\documentclass[default,iicol]{sn-jnl}

\usepackage{natbib}
\usepackage{tablefootnote}
\usepackage{graphicx}%
\usepackage{multirow}%
\usepackage{amsmath,amssymb,amsfonts}%
\usepackage{amsthm}%
\usepackage{mathrsfs}%
\usepackage[title]{appendix}%
\usepackage{xcolor}%
\usepackage{textcomp}%
\usepackage{manyfoot}%
\usepackage{booktabs}%
\usepackage{float}
\usepackage{listings}%
\usepackage[linesnumbered,lined,boxed,commentsnumbered,ruled]{algorithm2e}
\usepackage{xcolor}
\usepackage{blindtext}
\usepackage{hyperref}
\usepackage{xcolor}
\usepackage{array}
\usepackage{booktabs}       
\usepackage{longtable}
\usepackage{threeparttable}
\usepackage{multirow}
\usepackage{multicol}
\usepackage{graphicx}
\usepackage{bm}

\usepackage{amsthm}
\usepackage{dsfont}
\usepackage{lipsum} 

\usepackage{makecell,rotating,multirow,diagbox}
\usepackage{bm}
\usepackage{makecell}
\usepackage{cuted}

\usepackage{amssymb}
\usepackage{pifont}
\usepackage{booktabs}
\usepackage[export]{adjustbox}
\usepackage{amsmath}
\usepackage{amsfonts}
\usepackage{array}
\usepackage{graphicx}
\usepackage{booktabs}
\usepackage{bbding}
\usepackage{pifont}
\usepackage{wasysym}
\usepackage{utfsym}
\usepackage{fontawesome}
\usepackage{stfloats}
\usepackage{multirow}
\usepackage{threeparttable}
\usepackage{color}
\usepackage{tcolorbox}
\usepackage{tabularx}
\usepackage[misc]{ifsym}
\usepackage{bm}
\usepackage{url}

\theoremstyle{thmstyleone}%
\theoremstyle{thmstyletwo}%

\theoremstyle{thmstylethree}%

\begin{document}

\title[Article Title]{Retinal Vessel Segmentation via Neuron Programming}


\author[1]{\fnm{Tingting} \sur{Wu}}\email{wutt@njupt.edu.cn}

\author[1]{\fnm{Ruyi} \sur{Min}}\email{1222087522@njupt.edu.cn}

\author[1]{\fnm{Peixuan} \sur{Song}}\email{1023081914@njupt.edu.cn}

\author[2]{\fnm{Hengtao} \sur{Guo}}\email{hengtao.guo@outlook.com}

\author*[3]{\fnm{Tieyong} \sur{Zeng}}\email{zeng@math.cuhk.edu.hk}

\author[3]{\fnm{Feng-Lei} \sur{Fan}}\email{hitfanfenglei@gmail.com}

\affil[1]{\orgdiv{School of Science}, \orgname{Nanjing University of Posts and Telecommunication}, \orgaddress{\city{Nanjing}, \postcode{210023}, \country{China}}}

\affil[2]{\orgdiv{Independent Researcher}, \city{Seattle WA}, \postcode{98109}, \country{USA}}

\affil*[3]{\orgdiv{Department of Mathematics}, \orgname{The Chinese University of Hong Kong}, \orgaddress{\street{Shatin}, \city{Hong Kong}, \postcode{SAR}, \country{China}}}

\abstract{The accurate segmentation of retinal blood vessels plays a crucial role in the early diagnosis and treatment of various ophthalmic diseases. Designing a network model for this task requires meticulous tuning and extensive experimentation to handle the tiny and intertwined morphology of retinal blood vessels. To tackle this challenge, Neural Architecture Search (NAS) methods are developed to fully explore the space of potential network architectures and go after the most powerful one. Inspired by neuronal diversity which is the biological foundation of all kinds of intelligent behaviors in our brain, this paper introduces a novel and foundational approach to neural network design, termed ``neuron programming'', to automatically search neuronal types into a network to enhance a network's representation ability at the neuronal level, which is complementary to architecture-level enhancement done by NAS. Additionally, to mitigate the time and computational intensity of neuron programming, we develop a hypernetwork that leverages the search-derived architectural information to predict optimal neuronal configurations. Comprehensive experiments validate that neuron programming can achieve competitive performance in retinal blood segmentation, demonstrating the strong potential of neuronal diversity in medical image analysis.}



\maketitle

\section{Introduction}\label{sec1}
The retina is the only directly observable vascular network in the human body~\citep{yan2018joint}, and its health condition can provide a large amount of insight into numerous ophthalmic functions. Therefore, fundus imaging is currently the mainstream technology for the early diagnosis and treatment of various ophthalmic diseases, including diabetic retinopathy, hypertensive retinopathy, and age-related macular degeneration. In this vein, accurate segmentation of retinal blood vessels plays a crucial role. However, segmenting vessels in fundus images is challenging due to the tiny and intertwined morphology of retinal vessels and their sensitivity to factors such as image quality and lesions. Consequently, traditional model-based vessel segmentation methods~\citep{kalaie2017vascular} are often handicapped from achieving ideal outputs.  

Deep learning methods~\citep{liu2022full, shen2022scanet} have demonstrated exceptional potential in retinal vessel segmentation. Due to strong fitting abilities, these methods effectively overcome traditional limitations, enhancing the segmentation accuracy and robustness by automatically learning feature representations in images. However, despite these advancements, designing network architectures for retinal blood vessel segmentation requires meticulous tuning and extensive experimentation to handle the diverse image features. To tackle this challenge, Neural Architecture Search (NAS) methods~\citep{kucs2023evolutionary, wei2021genetic} were developed to fully explore the space of potential network architectures to identify the most suitable ones for specific retinal image features, which significantly boosts the automation level of network design and is particularly friendly to ophthalmologists who may not have much deep learning engineering expertise.

Recently, the field of NeuroAI has emerged, advocating that revealing the basic elements of brain intelligence can potentially catalyze the next-generation AI technologies \citep{zador2023catalyzing}. In our brain, neurons in the human brain display diverse forms and functions, and this diversity is crucial for the brain’s ability to handle complex tasks. Inspired by NeuroAI, recent studies have aimed to refine neuron types \citep{zoumpourlis2017non, micikevicius2017mixed, jiang2020nonlinear, mantini2021cqnn}, calling that artificial networks, like the brain, should be optimized along two dimensions: architecture and neuronal type, which complements the modification at the architecture level like NAS. Among these studies, quadratic neurons~\citep{r8} were proposed by replacing the inner product in traditional neurons with a quadratic term, enabling them to model more complex relationships and interactions in the data. Hereafter, we call neurons using the inner product as the aggregation function conventional neurons, while neurons using the quadratic function quadratic neurons. With neuronal types as a newly augmented dimension, previous studies have demonstrated that networks made of quadratic neurons possess greater expressive power and efficiency than networks of conventional neurons in some important tasks~\citep{r35}. Recently, theoretically, \citep{liao2024quadratic} proved there exists a function that can be approximated with a heterogeneous network of conventional and quadratic neurons using a polynomial number of neurons, while a homogeneous network of only one neuron type would require an exponential number of neurons. Encouraged by this remarked success in enhancing a model's ability, \textit{can we automatically design a network with heterogeneous neuron types for retinal vessel segmentation such that the resultant model not only is friendly to ophthalmologists but also has better feature representation power?} 

To answer this question, we propose a novel method called “neuron programming" to complement the architecture design by identifying the optimal composition of neuron types within the existing network. Imagine constructing a building where each room is uniquely built with different materials to suit its purpose best. Similarly, in neuron programming, we automatically customize the types of neurons at each location within a candidate network to maximize a network's performance. This approach, like our brain \citep{gupta2023visual}, moves beyond the traditional method of using a single type of neuron throughout the network. In addition to being a novel idea, neuron programming proves beneficial for retinal blood vessel segmentation. Due to the complexity of this task and the adoption of novel powerful neurons, neuron programming, used solely or combined with NAS, can identify the best neuronal types, which should outperform adopting the same architectures that rely solely on a single, universal neuron, as no single neuron type suits all cases. Thus, neuron programming offers a foundational and powerful method for enhancing neural networks, ensuring that networks are not just constructed, but carefully tuned for the optimal performance.

As the first step of neuron programming, we configure the network to utilize two neuron types: the conventional and quadratic. Quadratic neurons are integrated into the U-Net architecture, a widely-used framework in medical image segmentation. Similar to NAS, we use an evolutionary algorithm to determine the appropriate neuron type for each locations within the network. neuron programming can be done either simultaneously with NAS or exclusively. The latter approach is less complex than the former. However, the evolutionary algorithm to search for the optimal portfolio of neurons is still computationally expensive. To further simplify this process and lower computational costs, we explore introducing a network that maps the optimal neuronal configuration to the architecture, eliminating the need for exhaustive search and enabling direct prediction, which considerably reduces the time and resources needed to determine the ideal architecture. Such a network is referred to as the ``hypernetwork''~\citep{chauhan2024brief}. The hypernetwork method upgrades search, which is an advanced form of neuron programming, with the premise that a large amount of well-performing heterogeneous networks has been accumulated. To summarize, our contributions are threefold:

\begin{itemize}
    \item We introduce the concept of neuron programming, integrating quadratic neurons into the neural network design to create more sophisticated and adaptable models capable of capturing complex structures.

    \item We develop two search-based and one hypernetwork methodology to intelligently combine quadratic and conventional neurons to optimize network configurations for enhanced performance. Specifically, the hypernetwork efficiently predicts optimal network architectures and drastically reduces the computational demands of the search process.

    \item We validate the effectiveness of neuron programming in retinal vessel segmentation, highlighting its superior accuracy, efficiency, and generalizability across diverse datasets.
    
\end{itemize}

\section{Related Work}\label{sec2}
\subsection{Retinal Vessel Segmentation}\label{subsec2-1}
Accurate segmentation of retinal vessels is essential in medical image analysis, as it facilitates the detection and diagnosis of various ocular diseases. Over the years, numerous approaches have been proposed to address the challenges associated with accurate vessel segmentation. Early methods for retinal vessel segmentation primarily relied on traditional image processing techniques. These methods often utilized filtering and morphological operations. For example, using Gabor filters~\citep{giarratano2020automated} to enhance vessel structures and then applying thresholding techniques~\citep{jiang2003adaptive} for segmentation was prevalent. Additionally, mathematical morphology~\citep{zana2001segmentation} was applied to refine the segmented vessels by removing noise and enhancing connectivity. With the advent of machine learning, researchers began to explore more sophisticated methods for vessel segmentation. Techniques such as Support Vector Machines (SVM)~ \citep{ricci2007retinal} and Random Forests~\citep{annunziata2016accelerating} were employed to classify pixel-wise features derived from the retinal images. These methods demonstrated improved accuracy over traditional techniques, especially when combined with craft feature engineering.

Recent approaches leveraging deep learning, particularly convolutional neural networks (CNNs), have shown superior performance in retinal vessel segmentation. Among these, U-Net~\citep{r20} has become widely recognized and is now regarded as the benchmark model for biomedical image segmentation, including retinal vessels. Several U-Net variants have emerged to enhance its performance further. To name a few, residual U-Net \citep{alom2019recurrent} incorporates residual blocks to improve feature learning and gradient flow, addressing issues related to deep network training. Attention-based U-Net\citep{li2020accurate, guo2021sa} integrates attention mechanisms to focus on relevant features and suppress irrelevant ones, improving the segmentation accuracy of fine and thin vessels. SCS-Net \citep{wu2021scs} introduces multiple scales in the network to capture features at various resolutions, enhancing the segmentation of vessels with varying diameters. Hybrid approaches \citep{khan2020hybrid} that combine deep learning with traditional techniques or integrate multiple tasks, such as vessel segmentation and optic disc detection, have also been developed. The EXP-Net framework~\citep{shen2023expert} introduces an innovative approach to semi-supervised vessel segmentation that effectively utilizes limited annotations. EXP-Net incorporates an expert network to improve knowledge distillation. This expert network includes modules focused on knowledge enhancement and connectivity enhancement. All these methods consider the complex nature of retinal images in refining segmentation results.

\begin{table*}[htbp]
\centering
\caption{A summary of the recently-proposed quadratic neurons. $\sigma(\cdot)$ denotes the nonlinear activation function. $\odot$ denotes Hadamard product. Here, $\mathbf{W} \in \mathbb{R}^{n\times n}$, $\mathbf{w}_i \in \mathbb{R}^{n\times 1}$, and we omit the bias terms in these neurons.}
\scalebox{0.8}{
\begin{tabular}{l|l}
\hline
Authors           & Formulations          \\ \hline
\cite{zoumpourlis2017non, micikevicius2017mixed} & $\mathbf{y}=\sigma(\bm{x}^{\top}\mathbf{W}\bm{x}+\mathbf{w}^\top\bm{x})$               \\ \hline
 \cite{jiang2020nonlinear}      & \multirow{2}{*}{$\mathbf{y}=\sigma(\bm{x}^{\top}\mathbf{W}\bm{x}$)} \\ \cmidrule{1-1} 
\cite{mantini2021cqnn}     &                        \\ \hline
\cite{goyal2020improved}    & $\mathbf{y}=\sigma(\mathbf{w}^\top(\bm{x}\odot\bm{x}))$               \\ \hline
\cite{bu2021quadratic}       & $\mathbf{y}=\sigma((\mathbf{w}_1^\top\bm{x})(\mathbf{w}_2^\top\bm{x}))$             \\ \hline
\cite{xu2022quadralib} & $\mathbf{y}=\sigma((\mathbf{w}_1^\top\bm{x}) (\mathbf{w}_2^\top\bm{x})+\mathbf{w}_3^\top\bm{x})$ \\
\hline
\cite{fan2018new}      & $\mathbf{y}=\sigma((\mathbf{w}_1^\top\bm{x})(\mathbf{w}_2^\top\bm{x})+\mathbf{w}_3^\top(\bm{x}\odot\bm{x}))$        \\ \hline
\end{tabular}}
\label{tab:neurons}
\end{table*}

\subsection{Neuron Diversity}\label{subsec2-2}
Neuronal diversity \citep{r1} represents a fundamental aspect of neuroscience. Understanding neuronal diversity is crucial for explaining the internal workings of the nervous system and its role in generating intelligent behaviors. Artificial neural networks \citep{r2}, inspired by the biological nervous system, are computational models often designed by mimicking brain information processing. Despite significant advancements in foundational models in recent decades, neural networks exclusively rely on a single fundamental type of neuron. 

Incorporating neuronal diversity into neural networks is a fundamental approach to improve adaptability and capability in handling complex tasks. Researchers have developed new neuron designs such as polynomial neurons~\citep{r9,r4}, dendritic neurons~\citep{r5}, and spiking neurons~\citep{r6}. \citep{r7} proposed a parabolic neuron as an alternative to traditional hyperplane neurons, enhancing the flexibility of decision boundaries. \citep{r8} proposed a quadratic neuron with power and interactive terms, demonstrating its effectiveness in many downstream tasks such as low-dose CT image denoising. Furthermore, \citep{r9} tackled catastrophic forgetting with an active dendritic neuron design. These developments showcase the potential of integrating diverse neuron types into neural network architectures to enhance performance and address various challenges in computational modeling.

A quadratic neuron can be seen as a nonlinear adaptation of the conventional neuron, providing improved expressive capabilities compared to conventional neurons. It achieves this by calculating two inner products and applying a power operation to the input vector, subsequently aggregating them using a nonlinear activation function \citep{r8}. Mathematically, the quadratic neuron processes the $n$-dimensional input vector $\bm{x}=(x_1, x_2, ..., x_n)$ as outlined below:
\begin{equation}
y= \sigma((\mathbf{w}_1^\top \bm{x}+b_1)(\mathbf{w}_2^\top \bm{x}+b_2)+\mathbf{w}_3^\top(\bm{x}\odot \bm{x})+b_3),
\end{equation}
where $\sigma (\cdot)$ signifies a nonlinear activation function such as a rectified linear unit (ReLU), $\odot$ denotes the Hadamard product, $\bm{w}_1$,$\bm{w}_2$,$\bm{w}_3$ $\in$ $\mathbb{R}^n$ represent learnable weight vectors, while $b_1$,$b_2$,$b_3$  $\in$ $\mathbb{R}$ are bias terms. 

It is important to note that the design of quadratic neurons is not unique. Table \ref{tab:neurons} provides a summary of the existing quadratic neuron designs. It can be seen that neurons in \citep{zoumpourlis2017non, micikevicius2017mixed, jiang2020nonlinear, mantini2021cqnn} have the complexity of $\mathcal{O}(n^2)$, where $n$ denotes the dimension of inputs. Such a high complexity makes it quite hard to scale from these neurons to prototype large models. Therefore, these designs are excluded from consideration. Furthermore, the neuron designs presented in \citep{bu2021quadratic, xu2022quadralib} are primarily parabolic, while the design in \citep{goyal2020improved} adopts an elliptical form. They are special cases of our proposed neuron \citep{fan2018new}. Also, we argue that the fundamental quadratic neuron should encompass both elliptical and parabolic forms in order to extract a more comprehensive range of quadratic features. Thus, for scalability and completeness, we adopt \citep{fan2018new} as the alternative neuron design to evaluate the benefits of neuron programming.

\subsection{Neural Architecture Search}\label{subsec2-3}
Neural Architecture Search (NAS) \citep{r10} automates the neural network design by focusing on three key aspects: the definition of search space, the optimization of algorithms, and the evaluation of models. NAS has been applied to various visual tasks, including vessel image segmentation. For example, NAS-UNet \citep{weng2019unet} establishes a specialized search space tailored for medical image segmentation. GNAS-U$^2$Net \citep{sun2022gnas} employs a two-level nested U-structure for retinal image segmentation, addressing the limitations of single-level U-structures in incorporating contextual information for feature extraction. Genetic U-Net \citep{wei2021genetic} features a flexible search space with several enhancements to traditional genetic algorithms, achieving high-performance network architectures with reduced model parameters. MedUNAS \citep{kucs2023evolutionary} introduces a new cell-based search space and applies opposition-based differential evolution (ODE) to NAS for the first time, extending encoding strategies to continuous search spaces and reducing complexity compared to discrete methods. Additionally,  BTU-Net \citep{rajesh2023evolutionary} utilizes the Binary Teaching–Learning-Based Optimization (BTLBO) algorithm during the evolutionary process, offering an alternative to genetic algorithms for finding optimal node structures.

\begin{figure*}[t]
    \centering
    \includegraphics[width=1.0\textwidth]{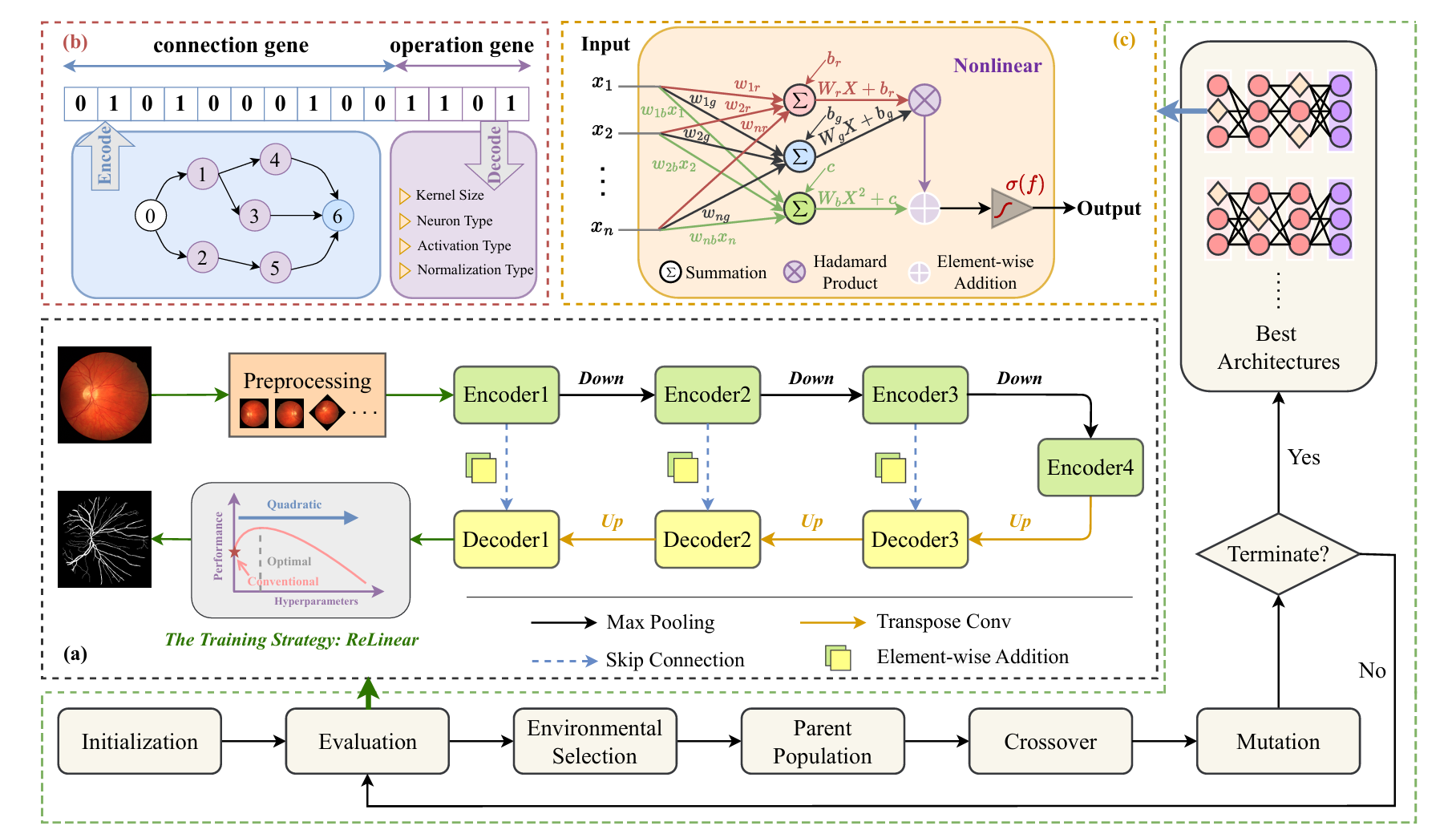}
    \caption{Overview of the proposed method framework. (a) The selected structure for retinal vessel segmentation. (b) The network architecture coding rules and block encoding example. (c) Our quadratic neuron}
    \label{fig: overview}
\end{figure*}

Hypernetworks represent an innovative paradigm in neural network design, where a secondary network (the hypernetwork) generates the weights for a primary network (the target network). Initially introduced by~\citep{ha2016hypernetworks}, this concept has been applied to various tasks, demonstrating its potential to improve model flexibility and efficiency~\citep{zhang2022implicit}. By decoupling the weight generation process from the target network, hypernetwork can produce more adaptive and scalable models. In recent years, hypernetwork has been increasingly integrated into NAS to mitigate its high computational costs. Several studies have proposed cost-effective alternatives to reduce NAS overhead~\citep{le2023generating}. One prominent approach is the one-shot methods~\citep{brock2017smash}, which involves training an auxiliary network (HyperNet) that dynamically generates weights for different architecture models. These approaches allow efficient exploration of multiple architectures within a single training process, significantly reducing the computational costs compared to traditional NAS methods. Our method stands out by constructing a hypernetwork that utilizes the optimal combination of the architecture and neuron types derived from neuron programming as its training data. After training, this hypernetwork then predicts the neuron types of unseen network structures, determining the operations to be applied at each position. Unlike traditional hypernetworks which primarily focus on weight generation for a single architecture, our approach leverages the optimal NAS-derived architecture to guide the design.

\section{Method}\label{sec3}
This study incorporates quadratic neurons into a U-shaped network structure as the first step of neuron programming for vessel image segmentation. The simultaneous integration of neuronal diversity and architecture is instrumental in offering basic yet unprecedented view to enhance a model's representation ability. The topological level allows for different network structures, and the neuronal level introduces quadratic neurons, optimizing each neuron for specific tasks to improve adaptability and generalization. To integrate different types of neurons, we use a genetic algorithm to address this problem. Neuron types are encoded as hyperparameters of the network structure for optimization, enabling automatic allocation of neurons and discovering the best network architecture. In the following sections, we will detail our methodology of neuron programming in terms of joint search, plug-and-play, and hypernet, towards reducing computational cost. Each includes the encoding of neuron types, the evolutionary algorithm used for architecture optimization, and the subsequent development of a hypernetwork for efficient and cost-effective network architecture adjustments.

\subsection{Neuron Programming: Joint Search with Architecture}\label{subsec3-2}

\textbf{Search Space Encoding}. Figure \ref{fig: overview} illustrates our overall flowchart. Like NAS, neuron programming can also be accomplished by the genetic algorithm. Thus, we can combine the search spaces of architecture and neuronal type to optimize both from scratch. We propose a block-based approach to the joint programming of architecture and neuronal types using a U-shaped encoder-decoder network as the backbone. Both the encoder and decoder consist of blocks, and the $i$-th block in an encoder or a decoder is denoted as \(\mathtt{Encoder}_i\) and \(\mathtt{Decoder}_i\), respectively. Then, skip connections that facilitate the flow of information between low-level and high-level features are added. In contrast to the original U-Net \citep{r20}, element-wise addition is used instead of concatenation to merge skip-connected and upsampled feature maps, thereby enhancing image features without increasing dimensionality and reducing the computational cost. 

\begin{table*}[htbp]
\caption{The node sequence operations for the joint search. Norm signifies Normalization, Conv represents Convolution Kernel Size, and IN stands for Instance Normalization.}\label{tab: operations}
\begin{tabular}{@{\extracolsep\fill}ccccc|ccccc}
\toprule%
\multicolumn{5}{@{}c@{}}{Pre-activation nodes} & \multicolumn{5}{@{}c@{}}{Post-activation nodes} \\\cmidrule{1-5}\cmidrule{6-10}%
Sequence & Gene & Norm & Conv & Neuron & Sequence & Gene & Norm & Conv & Neuron  \\
\midrule
1 & 0000 & - & 3×3 & Conventional & 5 & 0100  & - & 3×3 & Conventional  \\
2 & 0001 & - & 3×3 & Quadratic & 6 & 0101 & - & 3×3 & Quadratic  \\
3 & 0010 & IN & 3×3 & Conventional & 7 & 0110  & IN & 3×3 & Conventional  \\
4 & 0011 & IN & 3×3 & Quadratic & 8 & 0111  & IN & 3×3 & Quadratic  \\
9 & 1000 & - & 5×5 & Conventional & 13 & 1100 & - & 5×5 & Conventional  \\
10 & 1001 & - & 5×5 & Quadratic & 14 & 1101 & - & 5×5 & Quadratic  \\
11 & 1010 & IN & 5×5 & Conventional & 15 & 1110  & IN & 5×5 & Conventional \\
12 & 1011 & IN & 5×5 & Quadratic & 16 & 1111  & IN & 5×5 & Quadratic  \\ 
\botrule
\end{tabular}
\end{table*}

\begin{figure}[htbp]
    \centering
		\includegraphics[width=1\linewidth]{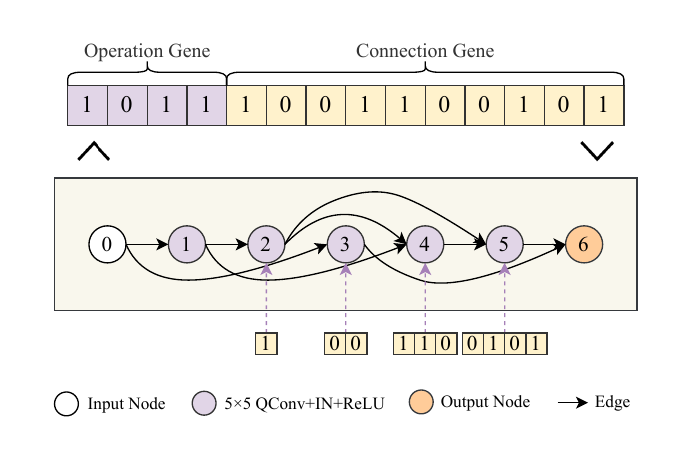}
        \small (a)
		\label{fig: encoding_a}
		\includegraphics[width=1\linewidth]{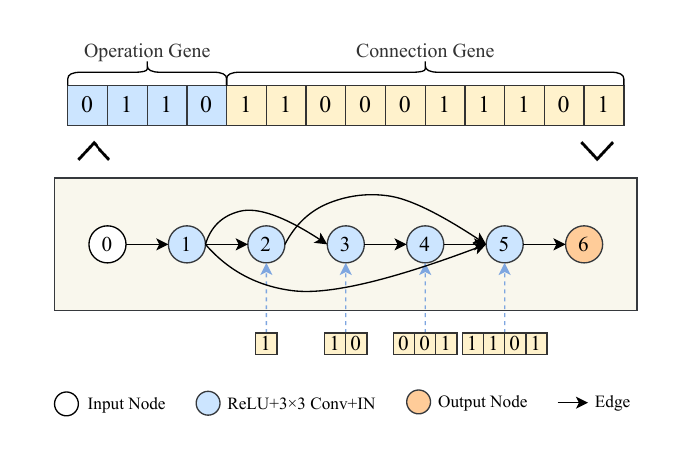}
        \small (b)
		\label{fig: encoding_b}
	\caption{Two examples of architecture encoding and their explanations. The intermediate nodes are colored to distinguish different operation genes, and the edge indicates the connection of internal nodes in the block, such as the 5×5 QConv+IN+ReLU, which is encoded as 1011 according to Table \ref{tab: operations}, where QConv indicates the use of quadratic neurons. Each edge shows whether nodes are connected; for example, node 1 is connected to node 2, coded as ``1''. Node 3 connects to neither node 1 nor node 2, hence node 3 is coded as ``00''. Node 4 connects to node 1 and node 2, but does not connect to node 3; therefore, the code for node 4 is ``110''. Node 5 connects both node 2 and node 4; so it is coded as ``0101''. Ultimately, the entire structure is encoded as ``1-00-110-0101''. The encoding scheme effectively represents the network structure, offering flexibility to encode various computational blocks within the network}
	\label{fig: encoding}
\end{figure}

We employ a directed acyclic graph \citep{xie2017genetic} and an adjacency list to encode a block in the U-Net architecture. Connection genes (0 and 1) denote if two nodes are connected, while operation genes represent sequences of basic operations such as convolution kernels and normalization types. Table \ref{tab: operations} shows the integration of quadratic neurons into the search space, resulting in 16 flexible operation sequences. Two concrete examples of encoding a block with one input node, one output node, and five intermediate nodes are shown in Figure \ref{fig: encoding}. The input node can only connect to subsequent nodes, but cannot receive connections from preceding nodes. Conversely, the output node can only be connected by preceding nodes, but cannot link subsequent nodes. Intermediate nodes can link both preceding and subsequent nodes, but connections must follow a lower-to-higher node sequence.

\textbf{Genetic Algorithm}.
With the encoded search space, we use the genetic algorithm to find the optimal network. First, an initial population of $N$ random neural networks is generated and evaluated based on their performance on training data, with performance metrics specially termed "fitness" in the evolutionary context. Two parents are selected via binary tournament selection and Hamming distance. In each iteration, crossover combines parts of these parent architectures to generate new network architectures. Random mutations are applied to some architectures to maintain diversity. These new offspring architectures then replace the worst-performing architectures in the population. This process of evaluation, selection, crossover, mutation, and replacement is iterated over $T$ generations, leading to a progressively better-performing network. We will now dive into some key steps during the evolutionary iterations.

\subsection{Neuron Programming: Plug-and-Play }\label{subsec4}
\begin{figure}[htbp]
    \centering
		\includegraphics[width=1\linewidth]{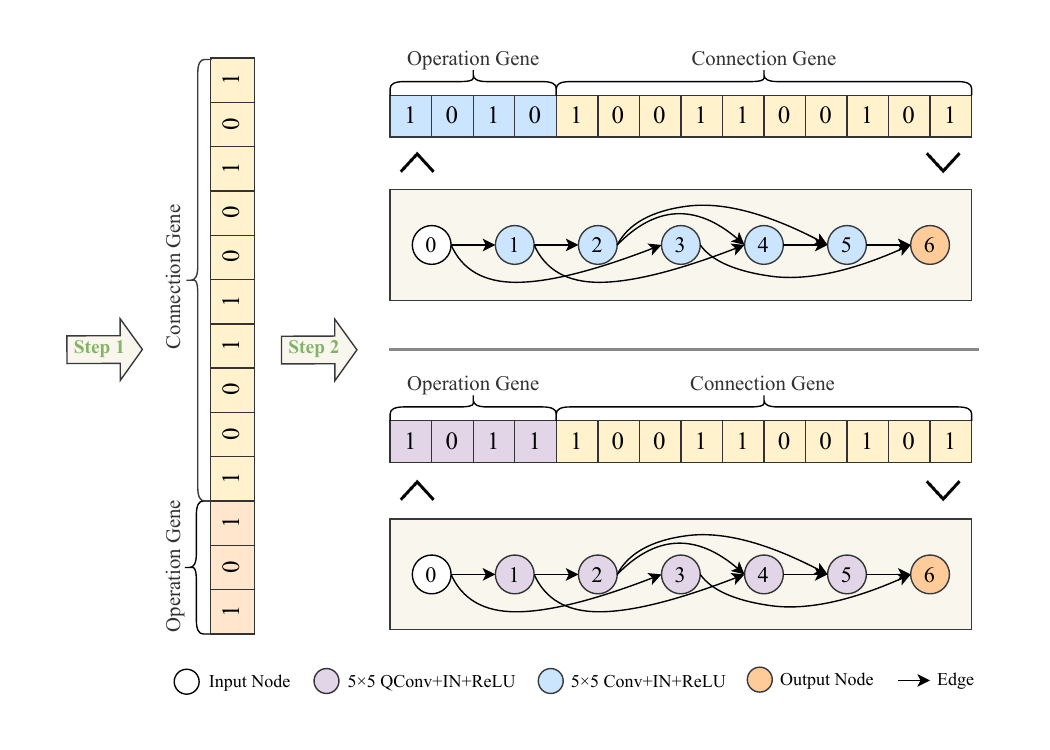}
        \
	\caption{Architecture encoding process during the plug-and-play search}
	\label{fig: encoding2}
\end{figure}

This work introduces “plug-and-play", a method that streamlines the search for an optimal network configuration by dividing it into two main steps: structural search and neuron type selection. In the first step, the search space encompasses parameters such as kernel size, the use of instance normalization, and the choice between pre-activation and post-activation, resulting in eight possible operation sequences. This step aims to find the optimal combination of these architectural components. Once the optimal architecture is identified in the first step, the second step then focuses on selecting the best neuron type for each position within the fixed architecture. Table \ref{tab: operations2} illustrates the two-part division of the search space. Figure \ref{fig: encoding2} illustrates the genetic encoding process in plug-and-play search, enabling the separate optimization of the architecture and neuron types and reducing the search complexity compared to joint search. In addition, the genetic algorithm used to explore the search space in the plug-and-play search remains the same as that in joint search.

Moreover, here, the network architecture is not necessarily obtained by search.  Since neuron programming can work in the plug-and-play manner, as long as the well-performing architecture is given, neuron programming can be enforced to further boost the model's performance. This is very likely, since the homogeneous network made of conventional neurons is a special case in the search space.

\begin{table*}[htbp]
\caption{The search space for the plug-and-play search. The first part details the search space for architecture parameters, while the second part focuses on neuron types. Norm signifies Normalization, Conv represents Convolution Kernel Size, and IN stands for Instance Normalization.}
\centering
\label{tab: operations2}
\begin{tabular}{@{\extracolsep\fill}cccc|ccc}
\toprule%
\multicolumn{4}{c|}{Step 1} & \multicolumn{3}{c}{Step 2} \\ \hline
Conv & Activation & Norm & Gene & Neuron & Gene & Sequence \\ \hline
3×3 & Pre-activation & - & 000 & Conventional/Quadratic & 0000/0001 & 1/2 \\
3×3 & Pre-activation & IN & 001 & Conventional/Quadratic & 0010/0011 & 3/4 \\
3×3 & Post-activation  & - & 010 & Conventional/Quadratic & 0100/0101 & 5/6 \\
3×3 & Post-activation  & IN & 011 & Conventional/Quadratic & 0110/0111 & 7/8 \\
5×5 & Pre-activation & - & 100 & Conventional/Quadratic & 1000/1001 & 9/10 \\
5×5 & Pre-activation & IN & 101 & Conventional/Quadratic & 1010/1011 & 11/12 \\
5×5 & Post-activation  & - & 110 & Conventional/Quadratic & 1100/1101 & 13/14 \\
5×5 & Post-activation  & IN & 111 & Conventional/Quadratic & 1110/1111 & 15/16 \\ \hline
\end{tabular}
\end{table*}

\subsection{Neuron Programming: Hypernetwork}
Searching often demands significant computational resources, and can struggle with deployment in practical applications. To overcome these challenges, we introduce neuron programming via a hypernetwork. Unlike traditional NAS, which searches for optimal architectures within a vast design space, neuron programming focuses on dynamically configuring networks. The hypernetwork acts as a meta-architect, predicting the optimal neuronal type for a target network, and effectively programming the neural configuration needed for specific tasks.

\textbf{Training Data and Strategy}.\label{subsubsec3-4-1}
The training data for the hypernetwork comes from our joint search results, evaluated using the F1 score as the performance metric. For each combination of network depth (encoder values of 2, 3, and 4) and channel sizes (ranging from 5 to 35 in increments of 5), the top 10 architectures with the highest F1 scores are selected, which can enhance the probability of finding the good network in practical applications. These high-performing configurations provide the gene codes for training the hypernetwork. This selective approach ensures that the hypernetwork is trained on the most promising architectures, enhancing its ability to predict optimal network structures efficiently. By learning from these optimal configurations, the hypernetwork can predict the neuronal types for the target network, streamlining the architecture search process. This strategy reduces the computational burden and improves the generalization capabilities of the resulting models, ensuring robust performance in practical applications.

\textbf{Inference Details}.\label{subsubsec3-4-2}
During inference, given a set of encoder and channel values, the hypernetwork predicts the necessary operations and neuron types, which are then decoded into readable network structures and trained for specific tasks like retinal vessel segmentation. This approach allows for the swift deployment of efficient, high-performing models, significantly reducing the time and resources required compared to traditional NAS methods. It also demonstrates the practicality and effectiveness of hypernetwork-driven architecture search in real-world applications.

In summary, rather than relying on an exhaustive and brute-force search like traditional NAS, the hypernetwork uses its training on well-performing architectures to understand the relationships between network parameters and performance metrics. By encoding this relation, the hypernetwork effectively ``programs'' in real-time and convenient manner. This process involves selecting the top 10 well-performing network structures from each parameter group to train a hypernetwork, which can then be used to predict the optimal combination of operations, neuron types, and layer configurations in terms of parameters like network depth and channel size. In this way, the hypernetwork transforms neural network design from a manual construction process into an adaptive and intelligent generation, towards a condensed neuron programming.

\section{EXPERIMENTS}\label{sec4}
\subsection{Experiment Setup}\label{subsec4-1}
\textbf{Datasets}.\label{subsubsec4-1-1}
We use four publicly available fundus image datasets: DRIVE \citep{r24}, STARE \citep{r25}, CHASE\_DB1 \citep{r28}, and HRF \citep{r30}. Each dataset is split into a training and a test set to evaluate the performance of our model. The DRIVE dataset contains 40 fundus images at a resolution of 565×584 pixels. This dataset is split into a training set of 20 images and a test set of 20 images. Two medical professionals annotated the test images, with the annotations provided by the first observer serving as the reference standard. The STARE dataset contains 20 images ($700\times605$ pixels), 10 of which show pathological changes. The leave-one-out method is employed \citep{r27}, with 20 iterations conducted. Each iteration consists of 19 images being used for training and 1 image for testing. This ensures that each image is tested once. The CHASE\_DB1 dataset comprises 28 images ($999\times960$ pixels), with 14 images obtained from the left eye and 14 from the right eye of 14 subjects. The initial 20 images are utilized for training, while the remaining 8 images are employed for testing~\citep{r29}. The HRF dataset comprises 45 high-resolution images ($3,504\times2,336$ pixels) from 15 healthy patients, 15 patients with diabetic retinopathy, and 15 patients with glaucoma, respectively. The initial five images from each category are utilized for training, with the remaining images reserved for testing. During the training phase, images and labels are downsampled by a factor of four to reduce the computational load.

\textbf{Evaluation Criterion}.\label{subsubsec4-1-2}
In this study, we employ a comprehensive set of evaluation metrics, including accuracy (ACC), sensitivity (SE), specificity (SP), F1 score (F1), and area under the ROC curve (AUC), to comprehensively assess the segmentation performance of our method. These metrics are used to evaluate the similarity between retinal images and ground truth, as well as the pixel-level classification performance. 

\textbf{Loss Function}.\label{subsubsec4-1-3}
The focal loss~\citep{r32} is chosen to address the class imbalance and to emphasize the importance of edge pixels in fundus image analysis. By reducing the influence of easy samples, it effectively corrects class imbalance, thereby enhancing the blood vessel segmentation accuracy. Furthermore, it enhances the learning of edge pixels, which is essential for precise vessel segmentation, addressing the limitations of traditional methods that treat all pixels equally. The focal loss is defined as follows:
\begin{equation}
\begin{aligned}
\mathcal{L}_{focal} = &-\sum_{n=1}^{N} (\alpha y_n (1-{\hat{y}}_n)^{\omega} \log{\hat{y}_n} 
\\&+ (1-\alpha)(1-y_n){{\hat{y}}_n}^{\omega} \log(1-{\hat{y}_n})).
\end{aligned}
\end{equation}
Here, \(y\) represents the ground truth, \(\hat{y}\) denotes the probability values generated by the model, \(N\) is the total number of samples, \(n\) is the sample index ranging from 1 to \(N\), \(\alpha\) is the balancing factor adjusting the importance of positive and negative samples, and \(w\) is the balancing factor in this power operation regulating the distinction between easy and challenging samples.

\textbf{Parameter Settings and Implementation Details}.\label{subsubsec4-2}
To mitigate overfitting caused by limited training data, we use image augmentation techniques to add variations such as horizontal/vertical flipping and random rotations (0-360 degrees). We apply pixel-wise normalization to standardize input data to a range of [-0.5, 0.5]. For the genetic algorithm, we set parameters with a population size of 20, running for 50 generations. Crossover and mutation probabilities are set to 0.9 and 0.7, respectively, with a 0.5 bit-flipping probability during mutation. In particular, the NAS and neuron programming are conducted exclusively on the DRIVE dataset, and the best-found network is then applied to the other three datasets. This approach not only conserves computational resources but also validates the generalization ability of the chosen network. We increase the training epochs to 1,000 to ensure the convergence of the model. The training process is conducted on PyTorch with two AMD100 GPUs for parallel processing.

\begin{table*}[htbp]
\caption{Comparison of the state-of-the-art networks and the proposed methods on DRIVE and CHASE\_DB1 dataset. The best and second-best results are bold-faced and underlined, respectively.}\label{tab: result1}
\scalebox{0.75}{
\begin{tabular}{@{\extracolsep\fill}cccccccc|ccccc}
\toprule%
& & & \multicolumn{5}{@{}c@{}}{DRIVE} & \multicolumn{5}{@{}c@{}}{CHASE\_DB1} \\
\cmidrule{4-8}\cmidrule{9-13}%
Methods & Year & Param (M) & ACC & SE & SP & F1 & AUC & ACC & SE & SP & F1 & AUC  \\
\midrule
U-Net & 2015 & 31.52 & 0.9536 & 0.7947 & 0.9803 & 0.8012 & 0.9860 & 0.9732 & 0.8041 & 0.9809 & 0.7933 & 0.9822 \\
DUNet & 2018 & 0.88 & 0.9566 & 0.7963 & 0.9800 & 0.8237 & 0.9802 & 0.9610 & 0.8155 & 0.9752 & 0.7883 & 0.9804 \\
Yan et al. & 2018 & 31.35 & 0.9542 & 0.7653 & 0.9818 & - & 0.9752 & 0.9610 & 0.7633 & 0.9809 & - & 0.9781 \\
R2U-Net & 2019 & 1.07 & 0.9556 & 0.7792 & 0.9813 & 0.8171 & 0.9784 & 0.9634 & 0.7756 & 0.9820 & 0.7928 & 0.9815 \\
Vessel-Net & 2019 & 1.70 & 0.9578 & 0.8038 & 0.9802 & - & 0.9821 & 0.9661 & 0.8132 & 0.9814 & - & 0.9860 \\
CS-Net & 2019 & 8.92 & 0.9632 & 0.8170 & \underline{0.9854} & - & 0.9798 & 0.9684 & 0.8471 & 0.9735 & - & 0.9384 \\
RVSeg-Net & 2020 & 5.20 & 0.9681 & 0.8107 & 0.9845 & - & 0.9817 & 0.9726 & 0.8069 & 0.9836 & - & 0.9833 \\
SA-UNet & 2020 & 0.54 & 0.9698 & 0.8212 & 0.9840 & 0.8263 & 0.9864 & \underline{0.9755} & \underline{0.8573} & 0.9835 & \underline{0.8153} & 0.9905 \\
CS$^2$-Net & 2021 & 5.96 & 0.9632 & 0.8218 & \textbf{0.9890} & - & 0.9825 & 0.9649 & 0.8316 & 0.9785 & 0.8133 & 0.9859 \\
FR-UNet & 2022 & 5.72 & \textbf{0.9705} & \underline{0.8356} & 0.9837 & \underline{0.8316} & \underline{0.9889} & 0.9748 & \textbf{0.8798} & 0.9814 & 0.8151 & \underline{0.9913} \\
EXP-Net & 2023 & 43.37 & 0.9667 & 0.8115 & 0.9837 & 0.8179 & 0.9816 & 0.9692 & 0.8086 & 0.9800 & 0.8010 & 0.9854 \\
SCANet$\_$U & 2023 & 29.64 & 0.9670 & 0.8076 & 0.9821 & - & - & 0.9722 & 0.8043 & \underline{0.9850} & - & - \\
Ours & 2024 & 0.84 & \underline{0.9701} & \textbf{0.8489} & 0.9825 & \textbf{0.8334} & \textbf{0.9901} & \textbf{0.9767} & 0.8545 & \textbf{0.9860} & \textbf{0.8224} & \textbf{0.9915}\\
\botrule
\end{tabular}}
\end{table*}

We utilize Lookahead \citep{r33} and ADAM \citep{r34} optimizers to stabilize training, using two sets of parameters to prevent oscillations (${ \alpha }$ = 0.05, ${k}$ = 6, ${\beta}_{1}$ = 0.9, ${\beta}_{2}$ = 0.999). For the optimization of quadratic networks, we apply the \textit{ReLinear} method~\citep{r35}, which initializes quadratic neurons weights as follows: $\mathbf{w}_2=0, b_2=1$ and $\mathbf{w}_3=0, b_3=0$, with regularization at adjusted learning rates ($\mathbf{w}_1$, $b_1$ at 0.001 and $\mathbf{w}_2=0, b_2=1$, $\mathbf{w}_3=0, b_3=0$ at 0.0001). This strategy ensures stable training by effectively controlling the learning of higher-order terms \citep{r35}.

\subsection{Segmentation Performance}
\textbf{Comparisons With State-of-the-Art Methods}.\label{subsubsec4-3}
We evaluate the effectiveness of our method by comparing it qualitatively and quantitatively with the state-of-the-art approaches across multiple retinal datasets. Table \ref{tab: result1} summarizes key metrics (ACC, SE, SP, F1, and AUC) along with parameter counts for each network. On the DRIVE dataset, our method achieves notably high metrics (SE: 0.8489, F1: 0.8334, AUC: 0.9901), indicating strong performance in vascular segmentation. The compared methods are U-Net~\citep{r20}, DUNet \citep{jin2019dunet}, Yan et al. \citep{yan2018joint}, R2U-Net \citep{alom2019recurrent}, Vessel-Net \citep{wu2019vessel}, CS-Net \citep{mou2019cs}, RVSeg-Net \citep{wang2020rvseg}, SA-UNet \citep{guo2021sa}, CS$^2$-Net \citep{mou2021cs2}, FR-UNet \citep{liu2022full}, EXP-Net \citep{shen2023expert}, SCANet \citep{shen2022scanet}, U-Net++ \citep{zhou2019unet++}, BSEResU-Net \citep{li2021bseresu} and Li et al. \citep{li2022global}. Our model achieves the highest SE and F1 among all models, with an ACC score second only to FR-UNet. Of note, our model is significantly smaller than the FR-UNet: the FR-UNet has 5.72M parameters, while our model has only 0.84M. Although our model's SP is slightly lower at 0.9825, the difference is marginal, within a magnitude of $0.001$.  On the CHASE\_DB1 dataset, our approach outperforms others in the ACC, SP, F1, and AUC metrics. Notably, our model surpasses other models by a large margin (at least over $1.8\%$). Tables \ref{tab: result2} and \ref{tab: result3} demonstrate our method's consistent excellence in ACC, F1, and AUC metrics on the STARE and HRF datasets, demonstrating the effectiveness of combining conventional and quadratic neurons.

\begin{figure*}[htbp]
\centering
        \multirow{6}{*}[3.96em]{\adjustbox{valign=m}{\rotatebox[origin=c]{90}{\fcolorbox{white}{cyan!10}{\parbox[c][0.3cm][c]{1.18cm}{\centering \footnotesize Images}}}}}%
            \vspace{1mm}
			\begin{minipage}[t]{0.148\linewidth}
			\centering
			\centerline{\includegraphics[width=1\linewidth]{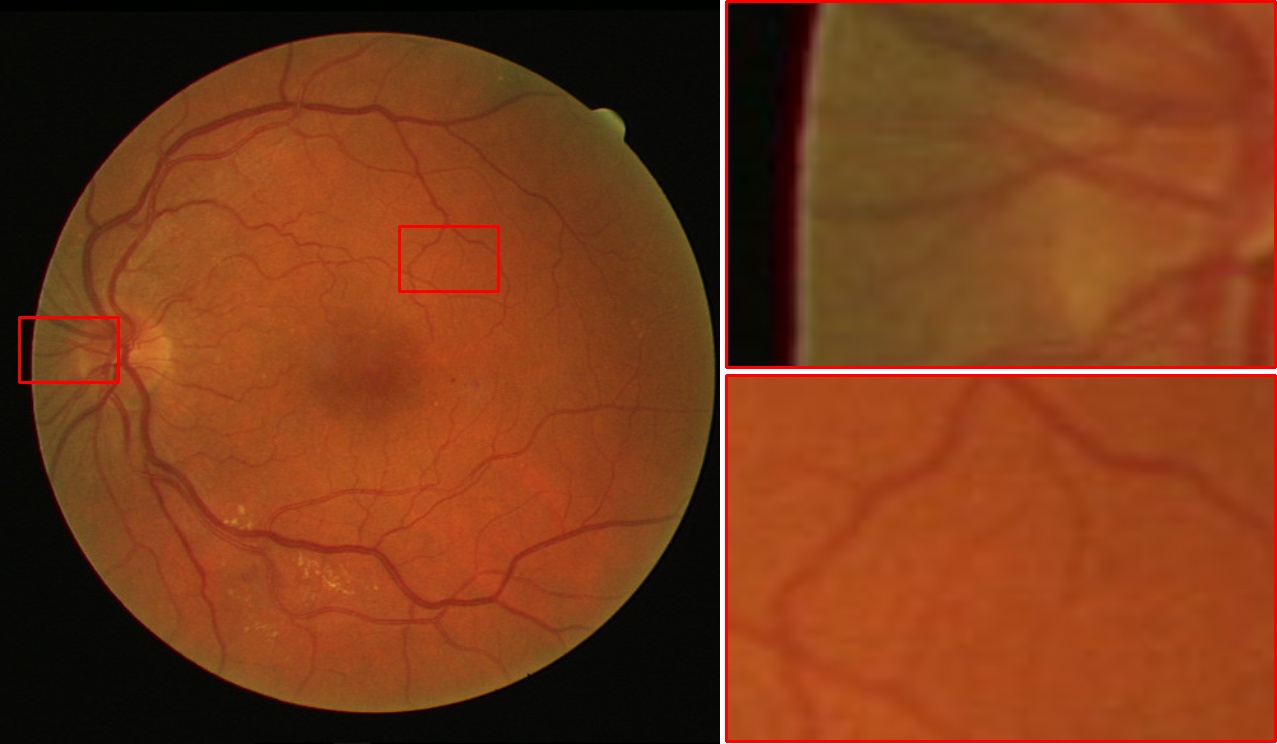}}
            \end{minipage}
            \begin{minipage}[t]{0.148\linewidth}
			\centering
			\centerline{\includegraphics[width=1\linewidth]{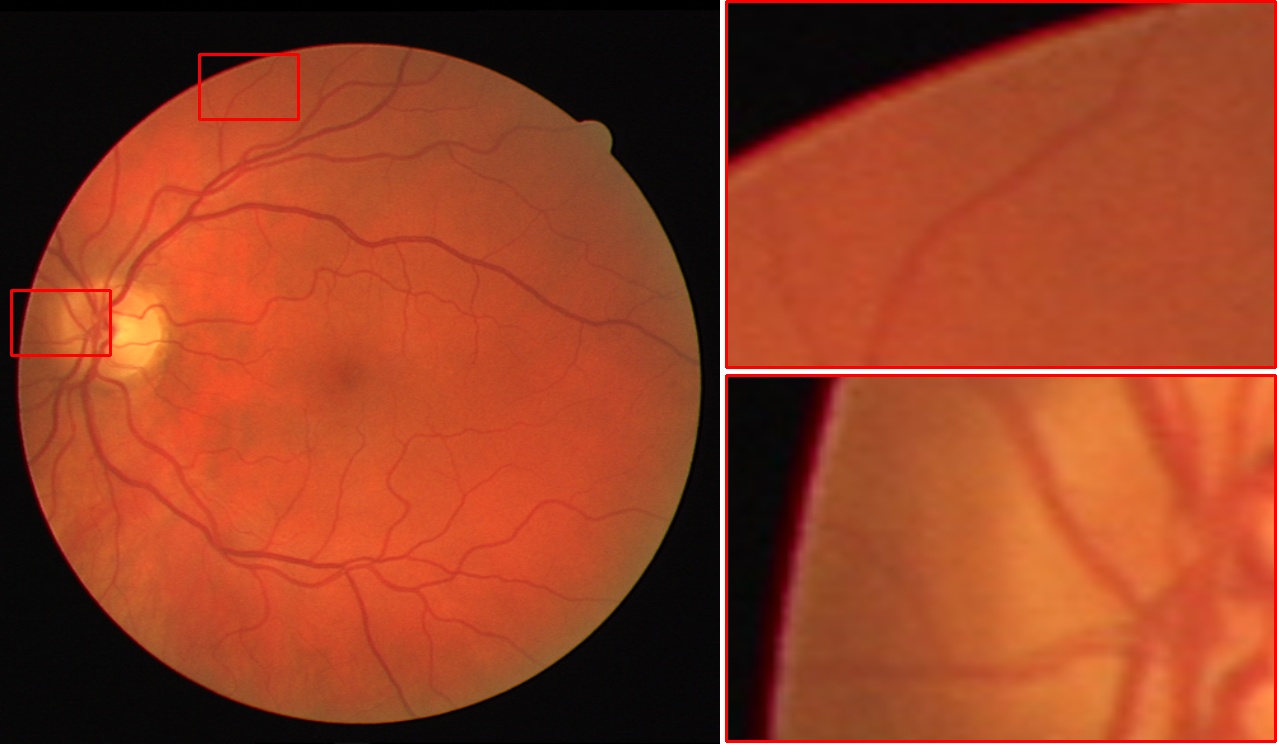}}
            \end{minipage}
            \begin{minipage}[t]{0.154\linewidth}
			\centering
			\centerline{\includegraphics[width=1\linewidth]{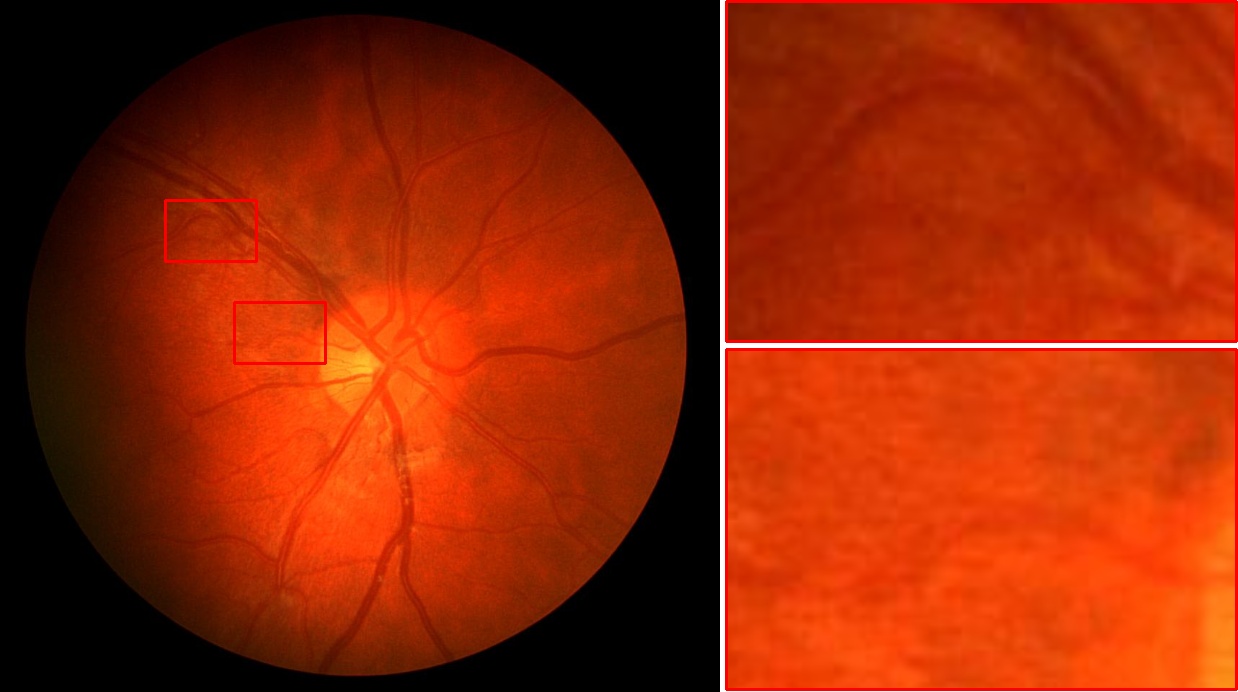}}
            \end{minipage}
            \begin{minipage}[t]{0.154\linewidth}
			\centering
			\centerline{\includegraphics[width=1\linewidth]{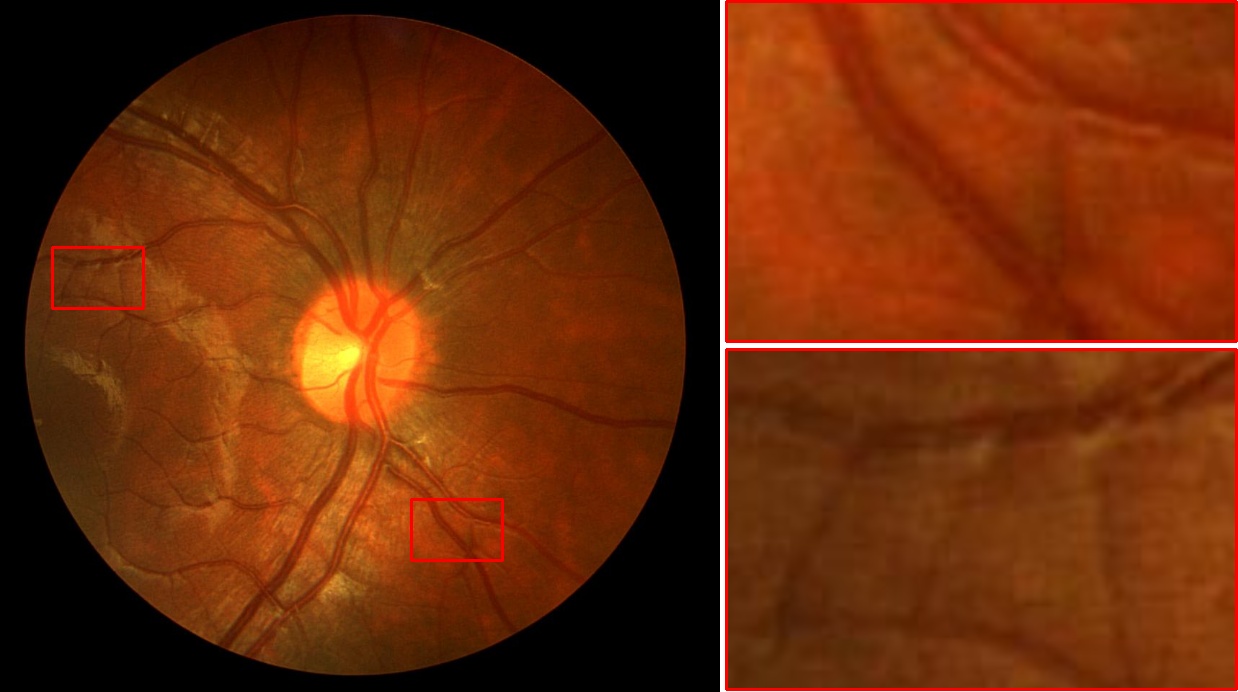}}
            \end{minipage}
            \begin{minipage}[t]{0.1635\linewidth}
			\centering
			\centerline{\includegraphics[width=1\linewidth]{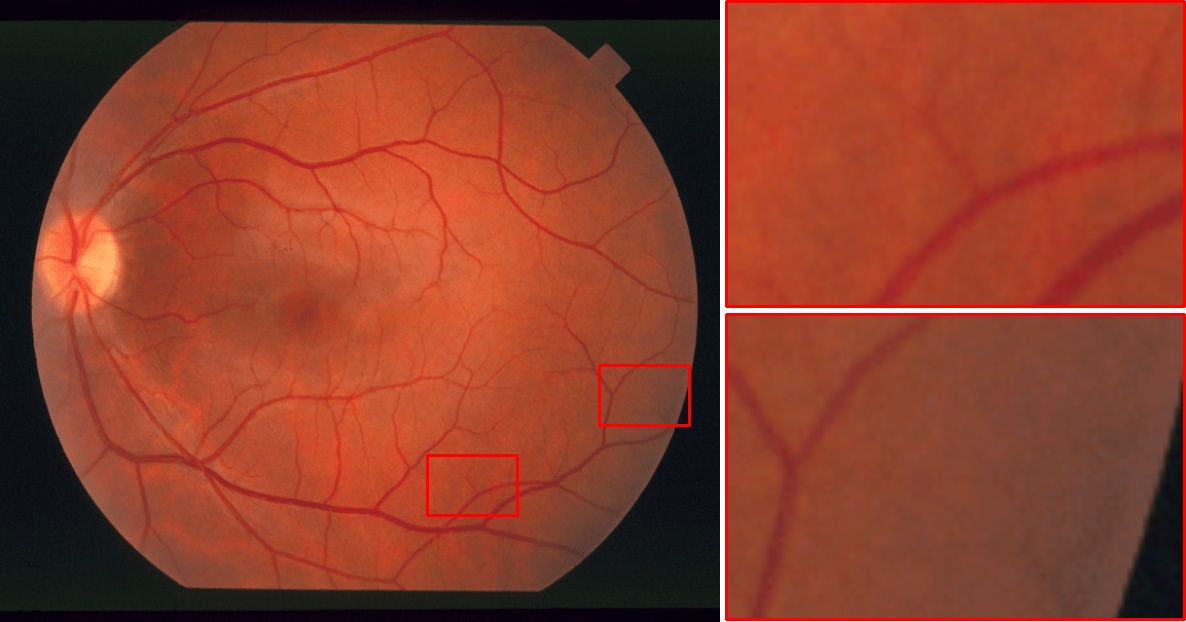}}
            \end{minipage}
            \begin{minipage}[t]{0.1635\linewidth}
			\centering
			\centerline{\includegraphics[width=1\linewidth]{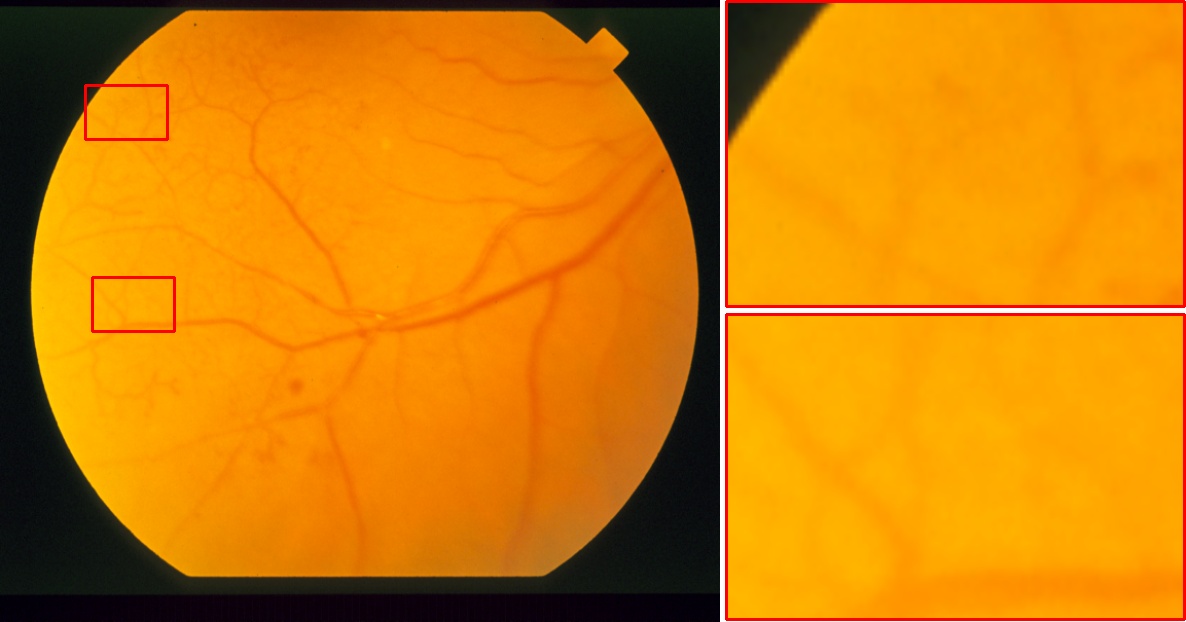}}
            \end{minipage}\\[0.1em]
            \vspace{-1mm}
             \multirow{6}{*}[3.96em]{\adjustbox{valign=m}{\rotatebox[origin=c]{90}{\fcolorbox{white}{cyan!10}{\parbox[c][0.3cm][c]{1.18cm}{\centering \footnotesize GT}}}}}%
            \vspace{1mm}
                \begin{minipage}[t]{0.148\linewidth}
			\centering
			\centerline{\includegraphics[width=1\linewidth]{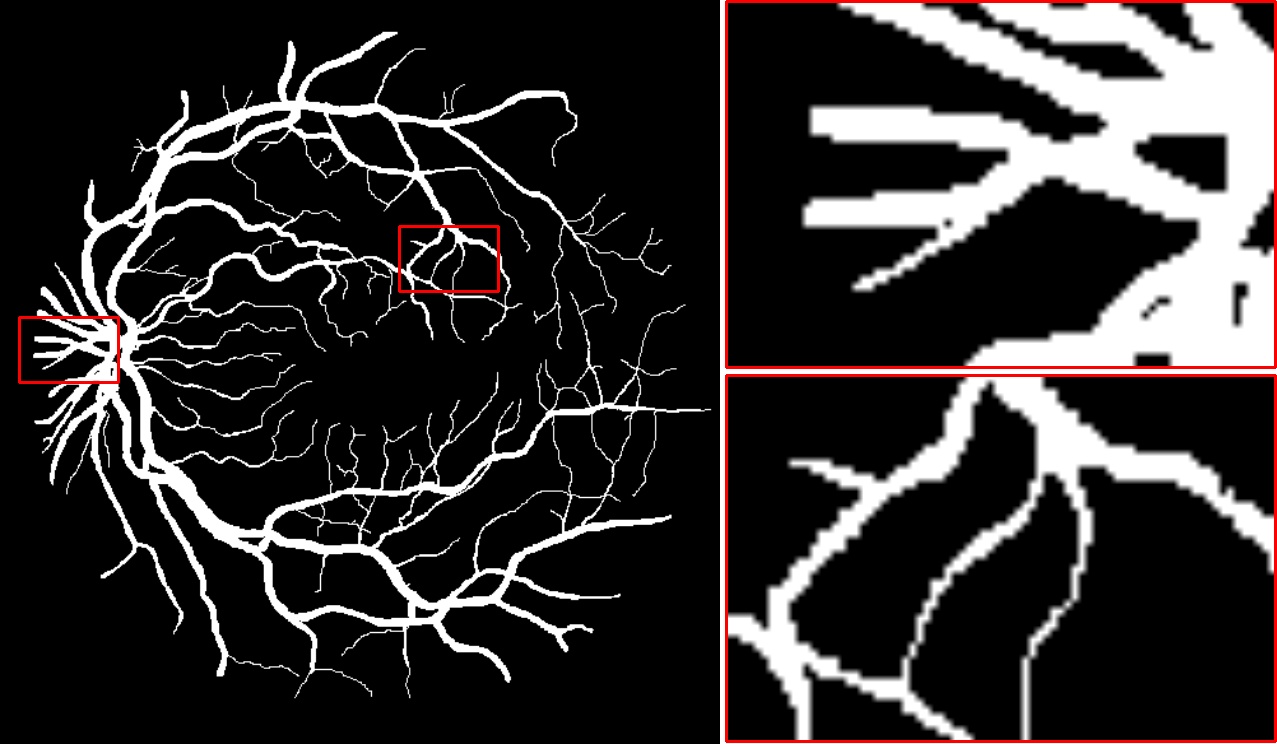}}
            \end{minipage}
            \begin{minipage}[t]{0.148\linewidth}
			\centering
			\centerline{\includegraphics[width=1\linewidth]{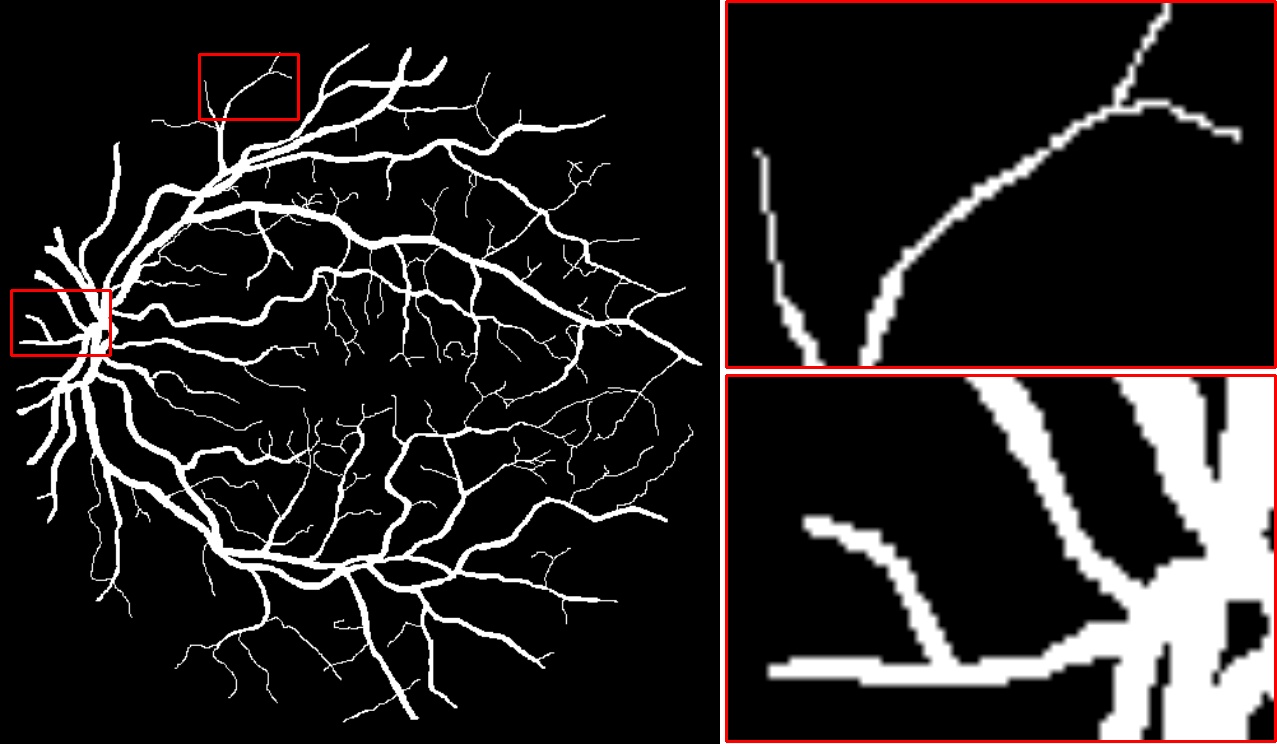}}
            \end{minipage}
            \begin{minipage}[t]{0.154\linewidth}
			\centering
			\centerline{\includegraphics[width=1\linewidth]{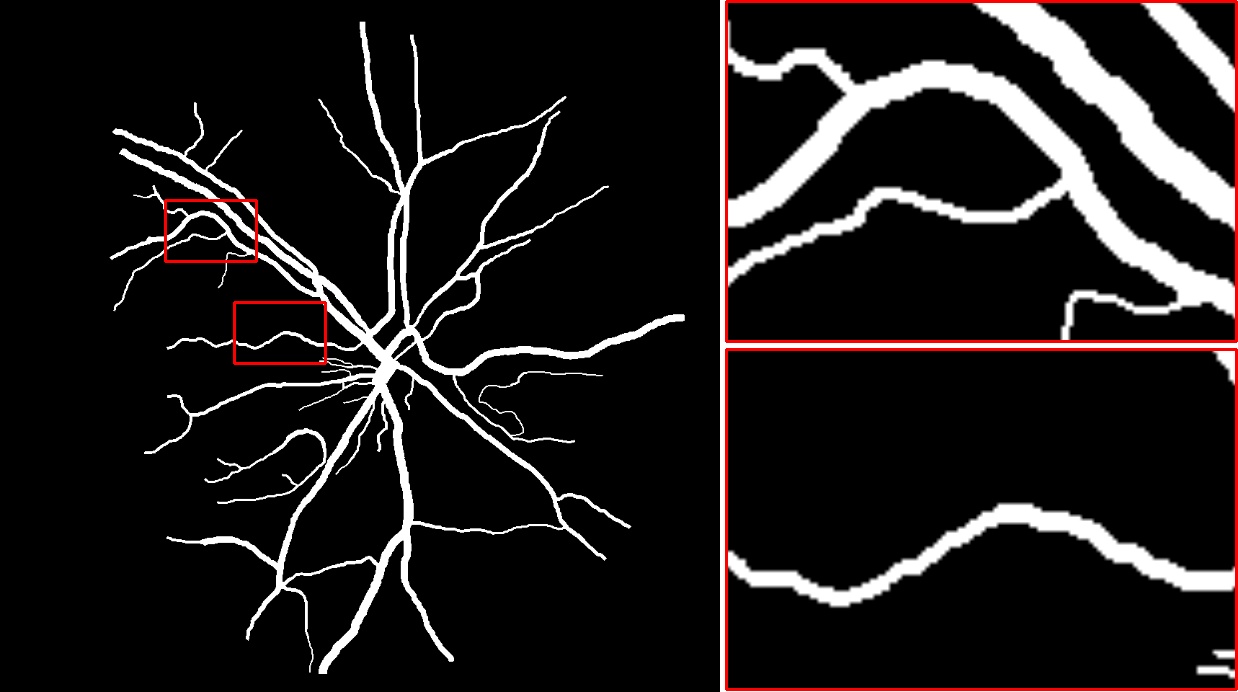}}
            \end{minipage}
            \begin{minipage}[t]{0.154\linewidth}
			\centering
			\centerline{\includegraphics[width=1\linewidth]{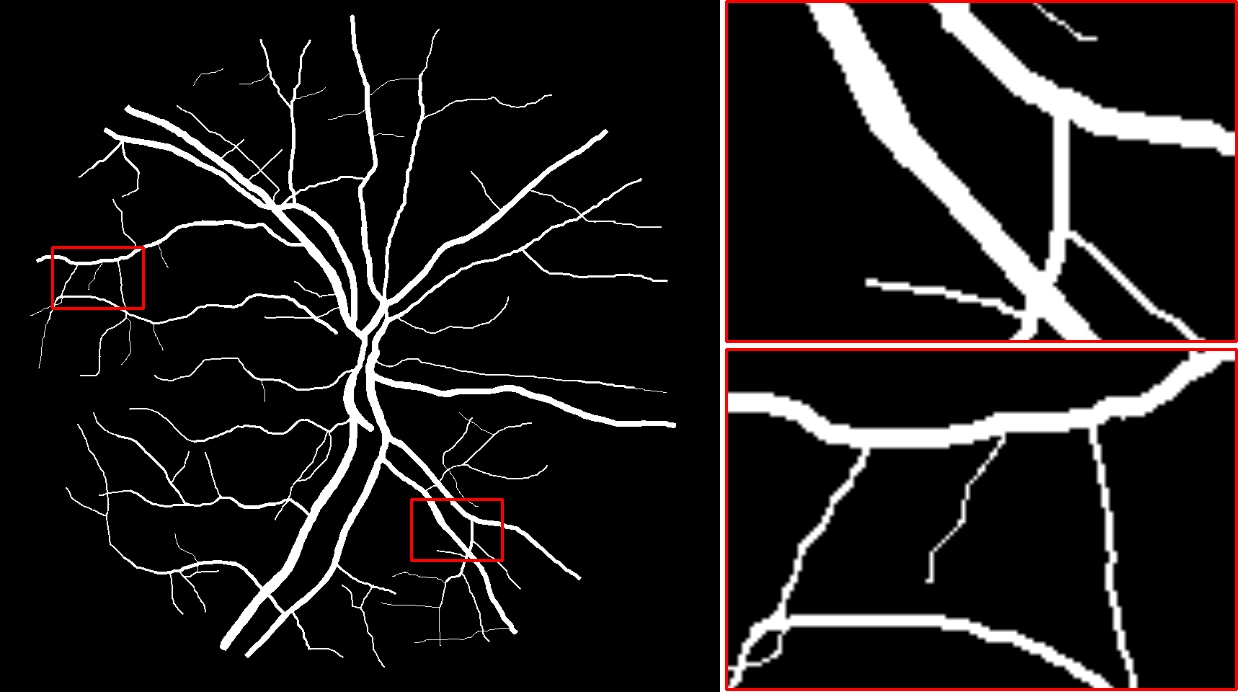}}
            \end{minipage}
            \begin{minipage}[t]{0.1635\linewidth}
			\centering
			\centerline{\includegraphics[width=1\linewidth]{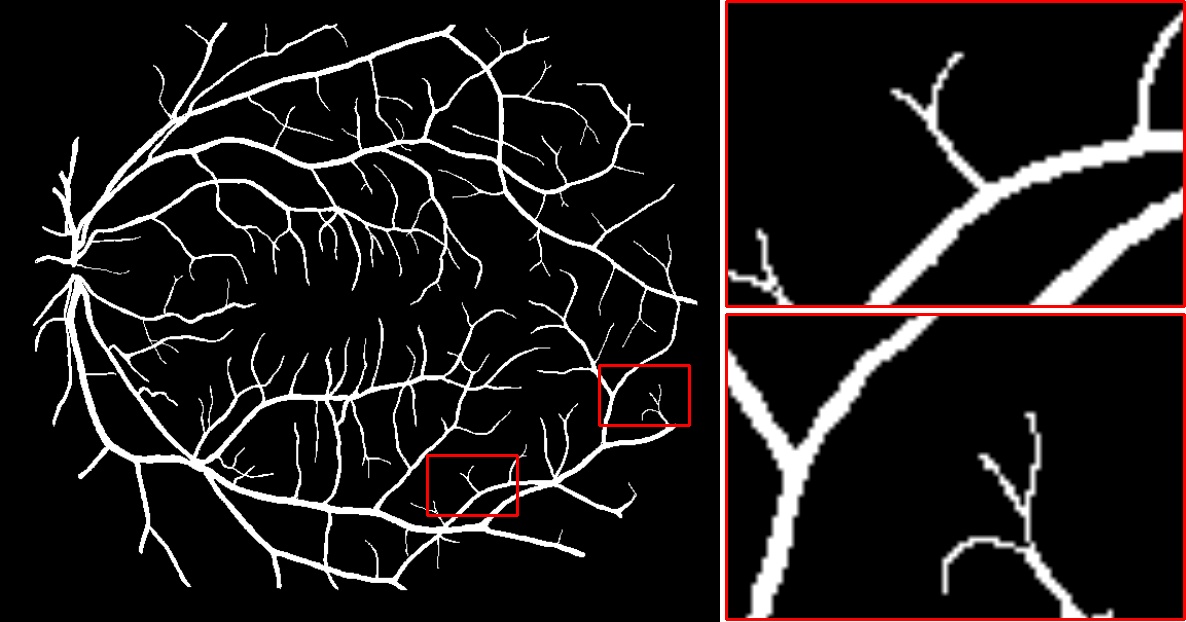}}
            \end{minipage}
            \begin{minipage}[t]{0.1635\linewidth}
			\centering
			\centerline{\includegraphics[width=1\linewidth]{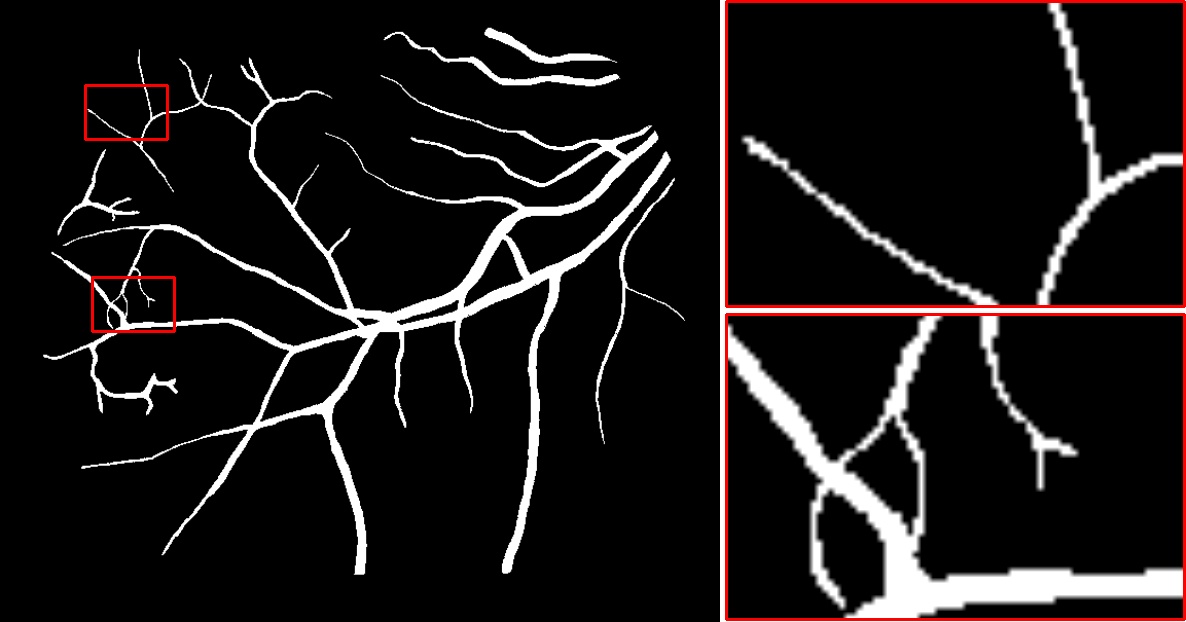}}
            \end{minipage}\\[0.1em]
            \vspace{-1mm}
            \multirow{6}{*}[3.96em]{\adjustbox{valign=m}{\rotatebox[origin=c]{90}{\fcolorbox{white}{cyan!10}{\parbox[c][0.3cm][c]{1.18cm}{\centering \footnotesize U-Net}}}}}%
            \vspace{1mm}
			\begin{minipage}[t]{0.148\linewidth}
			\centering
			\centerline{\includegraphics[width=1\linewidth]{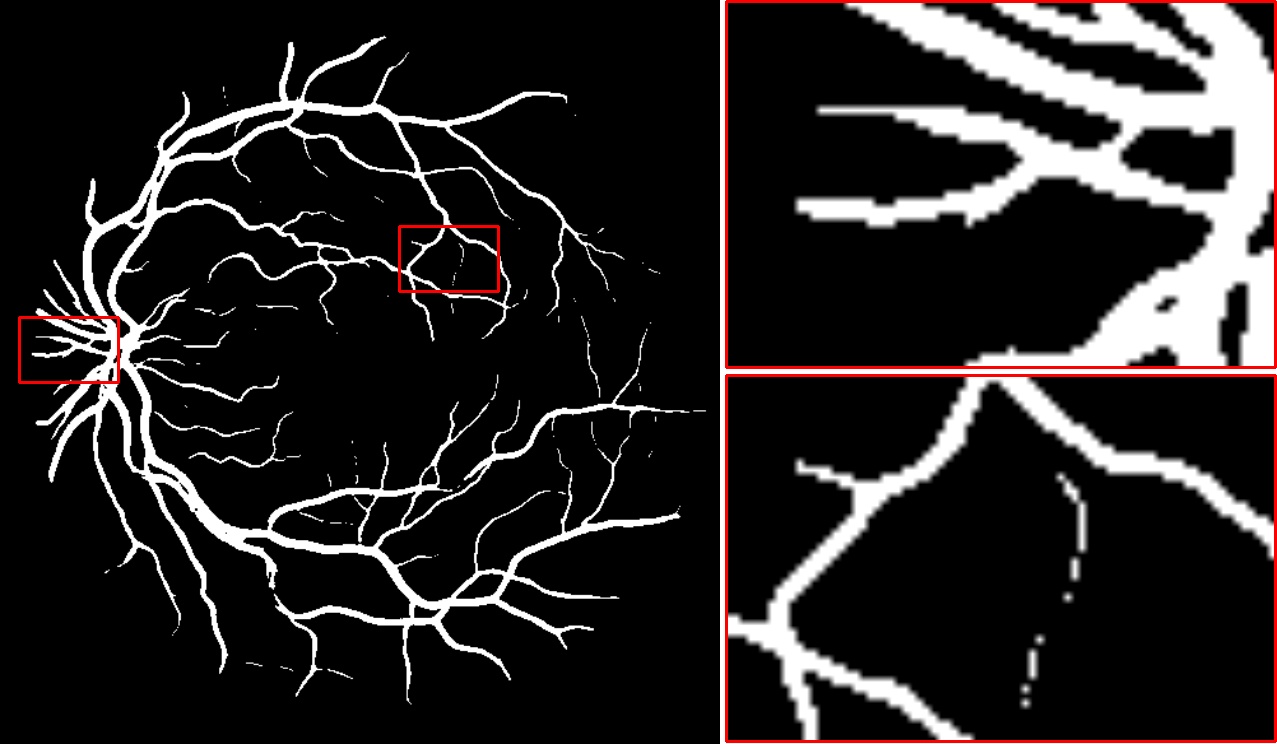}}
            \end{minipage}
            \begin{minipage}[t]{0.148\linewidth}
			\centering
			\centerline{\includegraphics[width=1\linewidth]{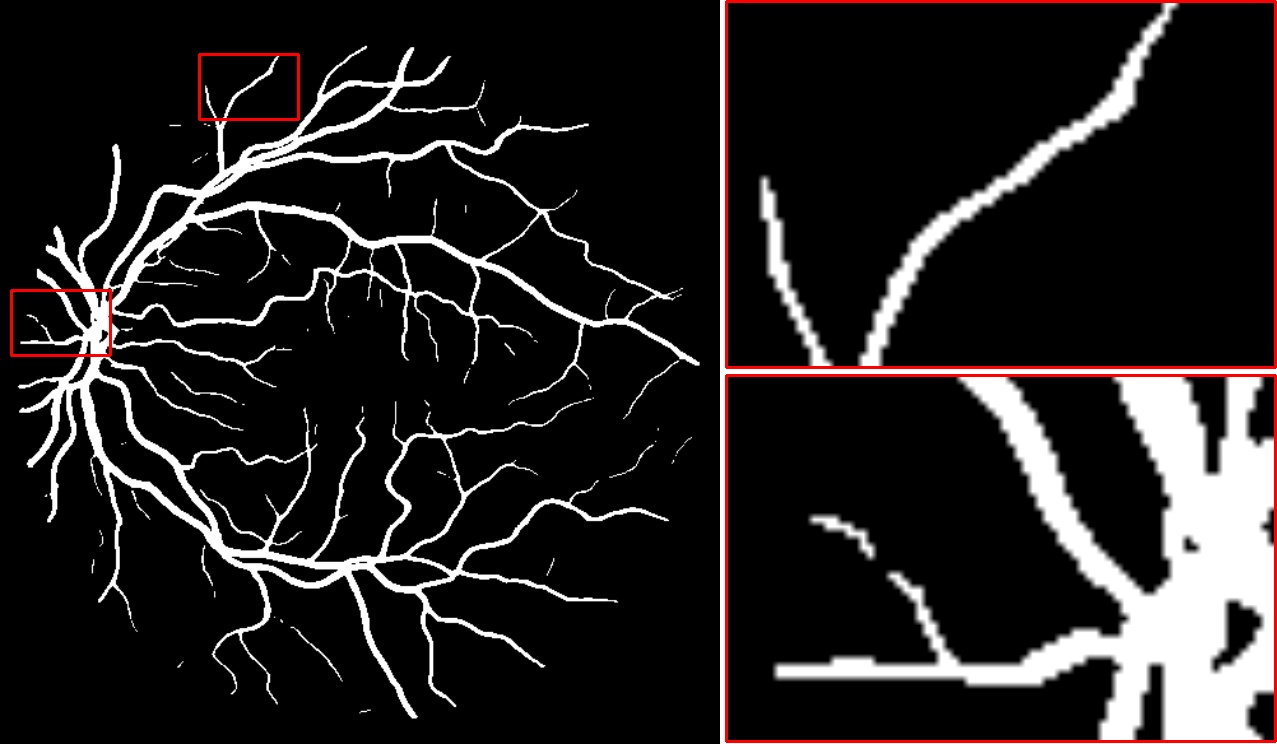}}
            \end{minipage}
            \begin{minipage}[t]{0.154\linewidth}
			\centering
			\centerline{\includegraphics[width=1\linewidth]{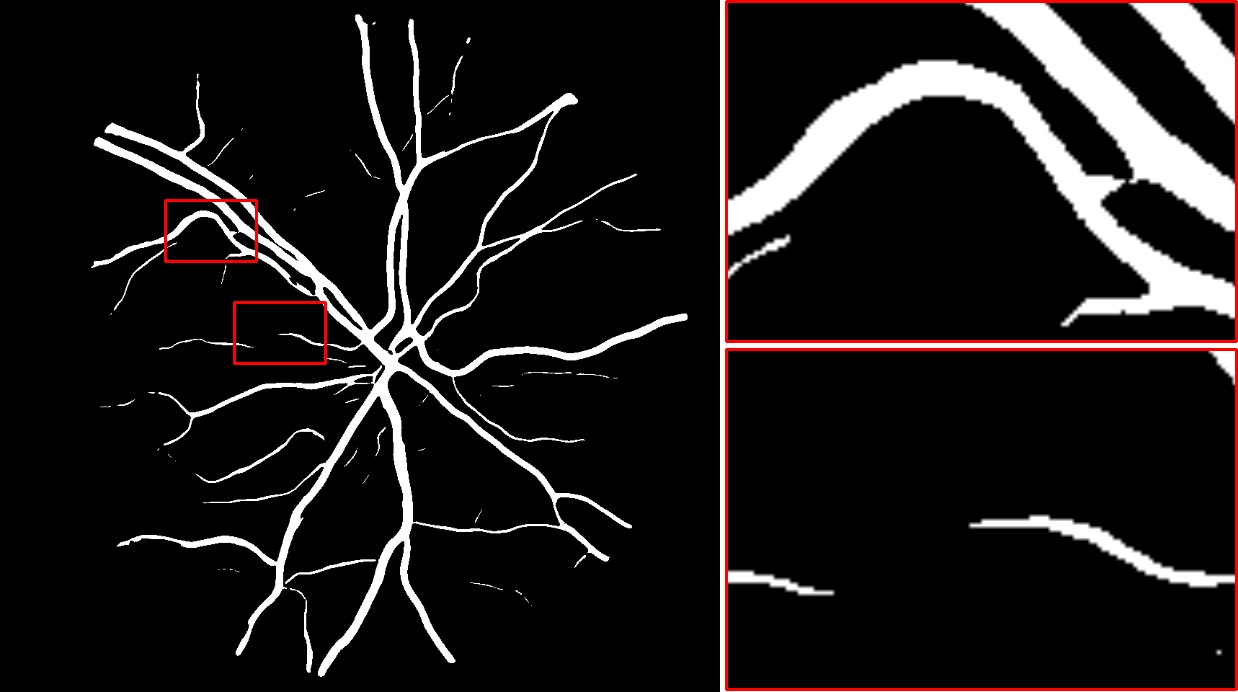}}
            \end{minipage}
            \begin{minipage}[t]{0.154\linewidth}
			\centering
			\centerline{\includegraphics[width=1\linewidth]{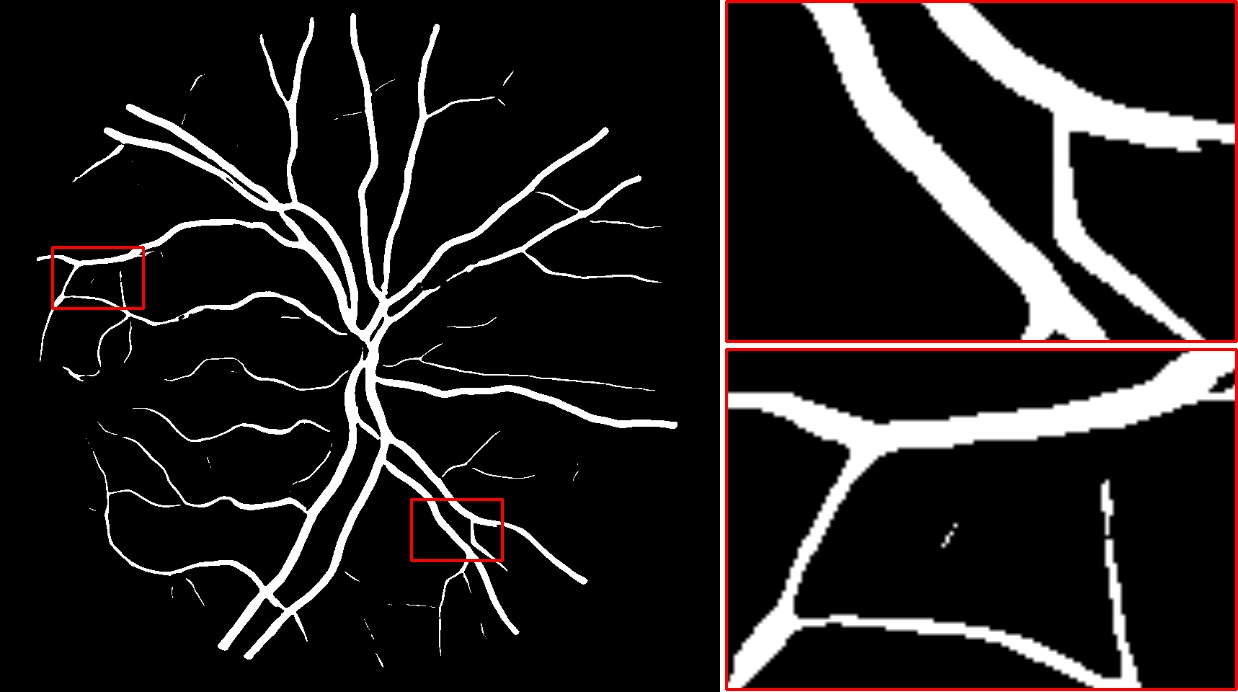}}
            \end{minipage}
            \begin{minipage}[t]{0.1635\linewidth}
			\centering
			\centerline{\includegraphics[width=1\linewidth]{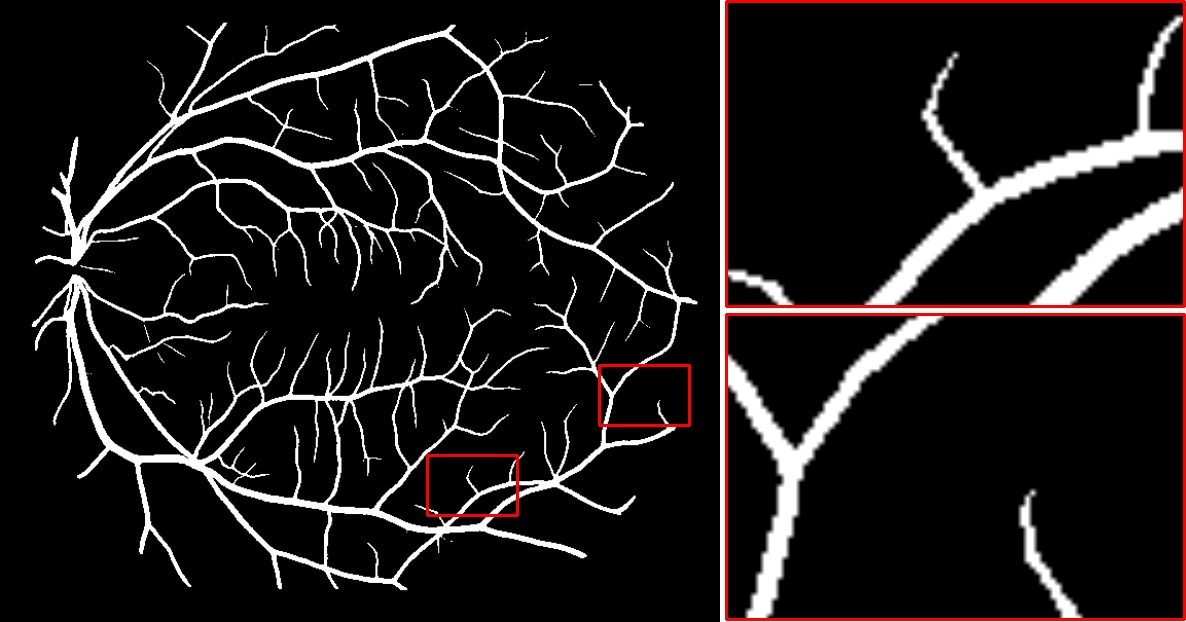}}
            \end{minipage}
            \begin{minipage}[t]{0.1635\linewidth}
			\centering
			\centerline{\includegraphics[width=1\linewidth]{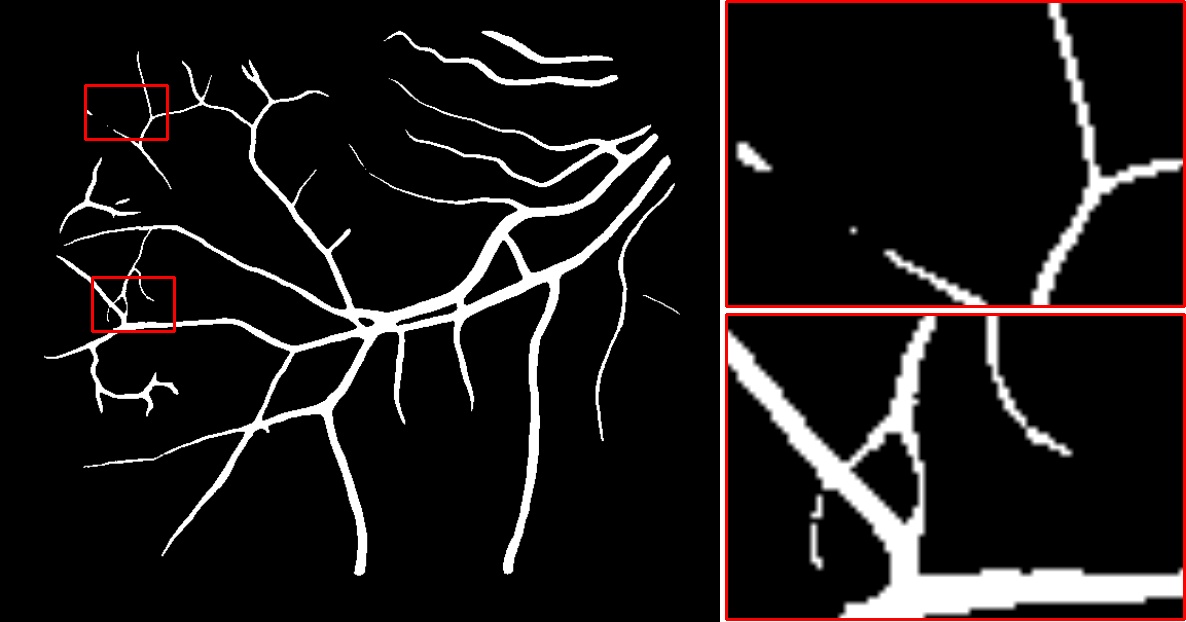}}
            \end{minipage}\\[0.1em]
            \vspace{-1mm}
           \multirow{6}{*}[3.96em]{\adjustbox{valign=m}{\rotatebox[origin=c]{90}{\fcolorbox{white}{cyan!10}{\parbox[c][0.3cm][c]{1.18cm}{\centering \footnotesize Yan et al.}}}}}%
            \vspace{1mm}
			\begin{minipage}[t]{0.148\linewidth}
			\centering
			\centerline{\includegraphics[width=1\linewidth]{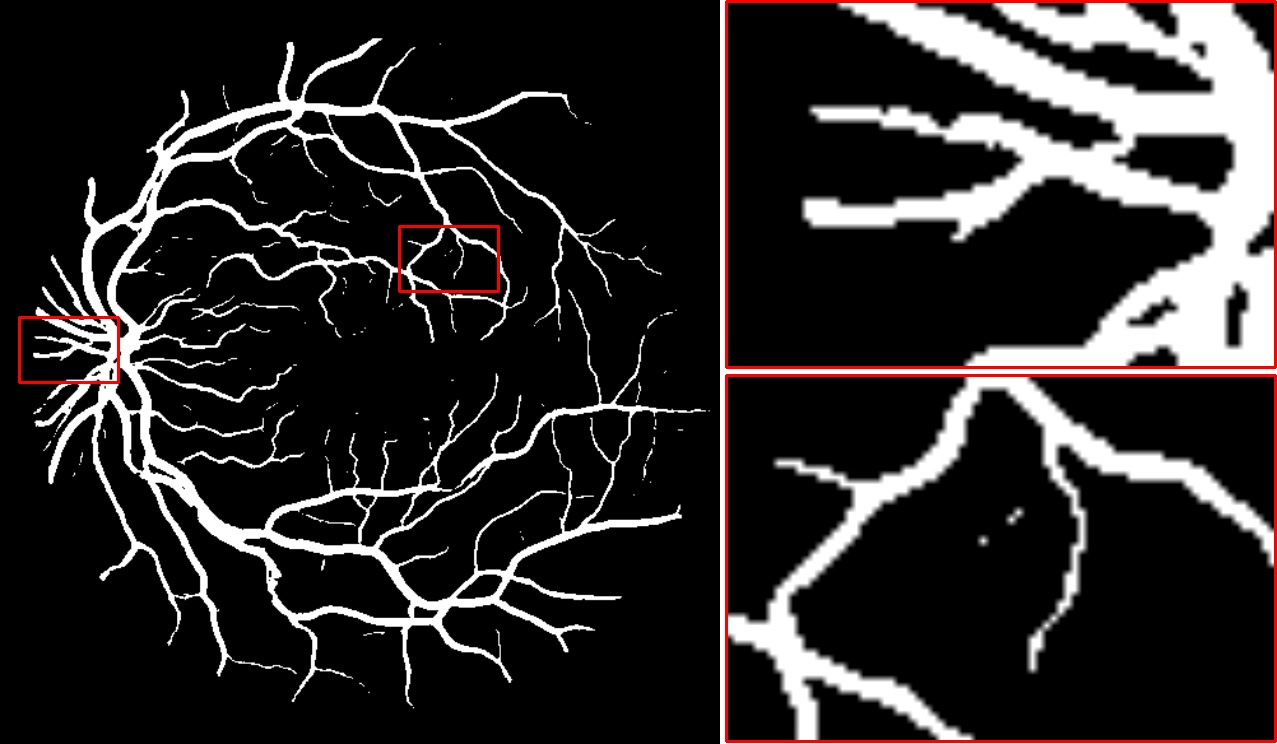}}
            \end{minipage}
            \begin{minipage}[t]{0.148\linewidth}
			\centering
			\centerline{\includegraphics[width=1\linewidth]{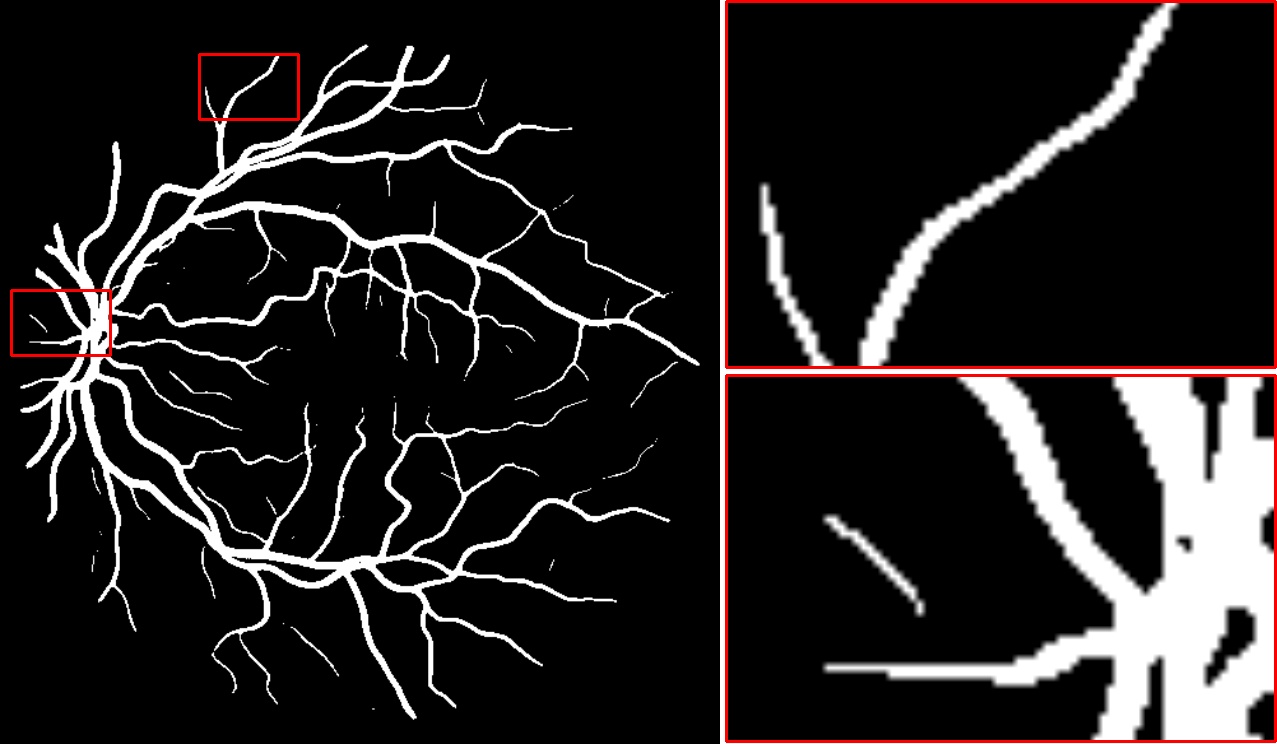}}
            \end{minipage}
            \begin{minipage}[t]{0.154\linewidth}
			\centering
			\centerline{\includegraphics[width=1\linewidth]{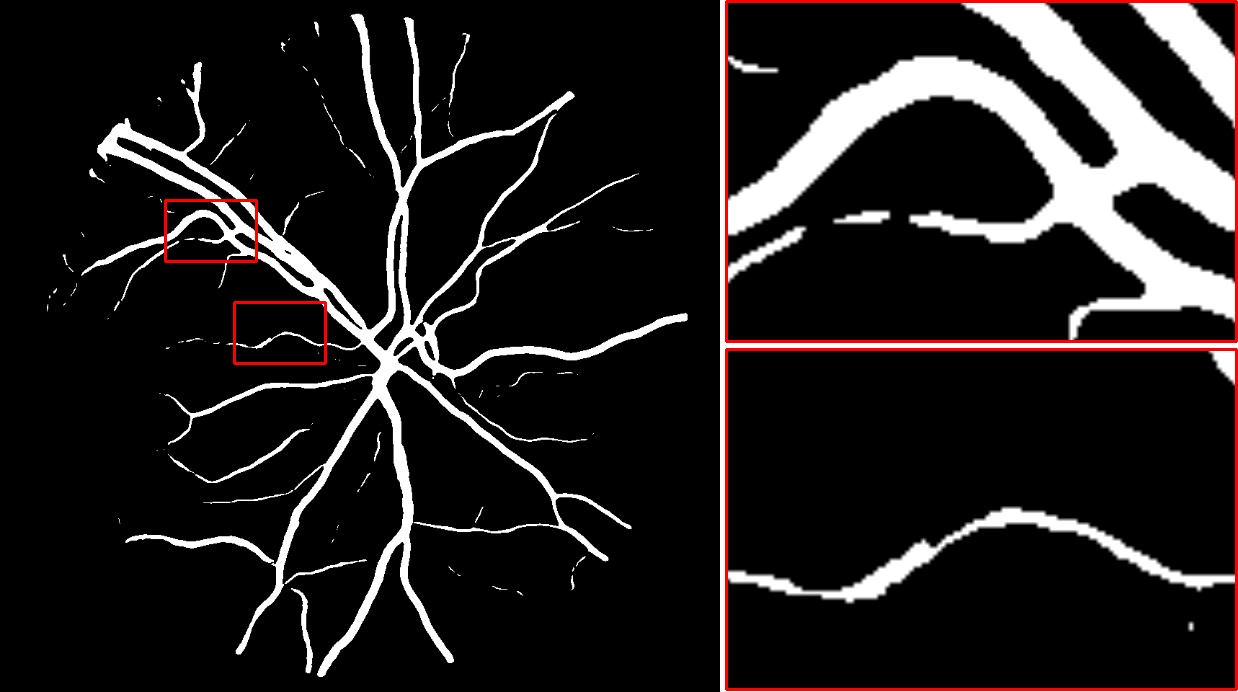}}
            \end{minipage}
            \begin{minipage}[t]{0.154\linewidth}
			\centering
			\centerline{\includegraphics[width=1\linewidth]{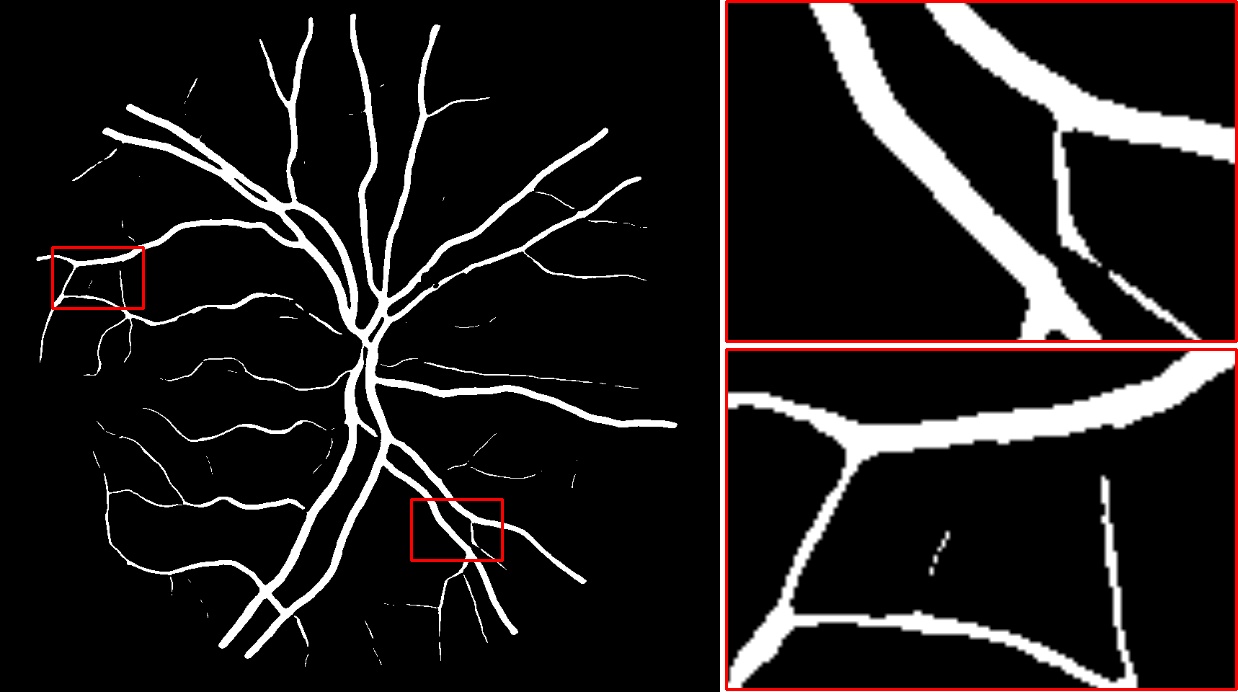}}
            \end{minipage}
            \begin{minipage}[t]{0.1635\linewidth}
			\centering
			\centerline{\includegraphics[width=1\linewidth]{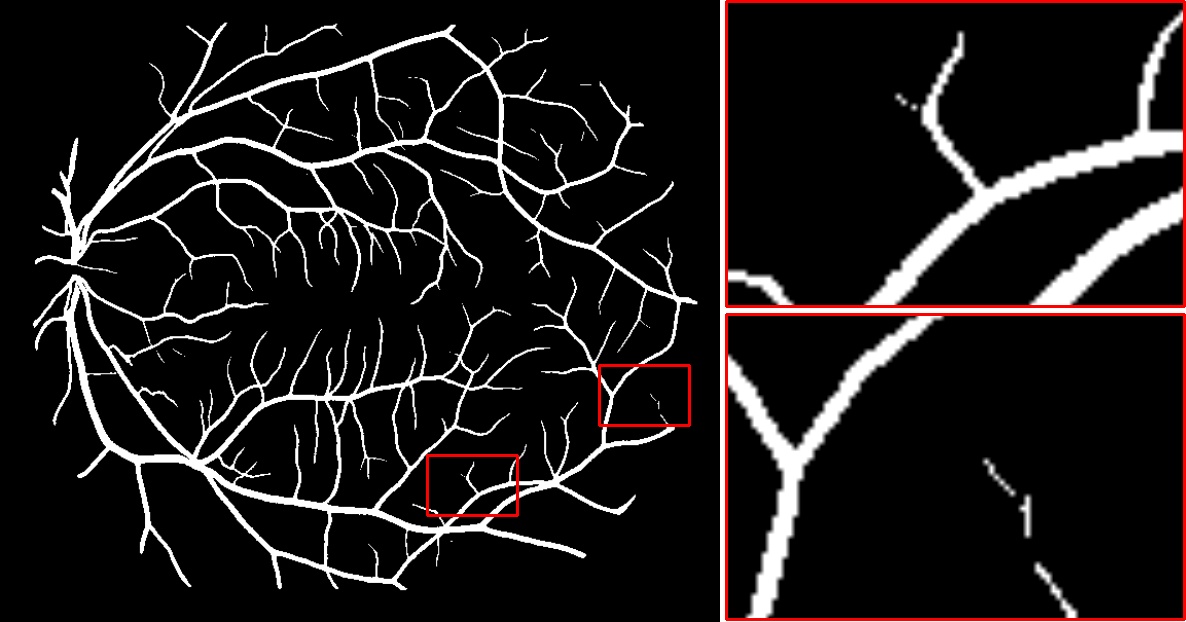}}
            \end{minipage}
            \begin{minipage}[t]{0.1635\linewidth}
			\centering
			\centerline{\includegraphics[width=1\linewidth]{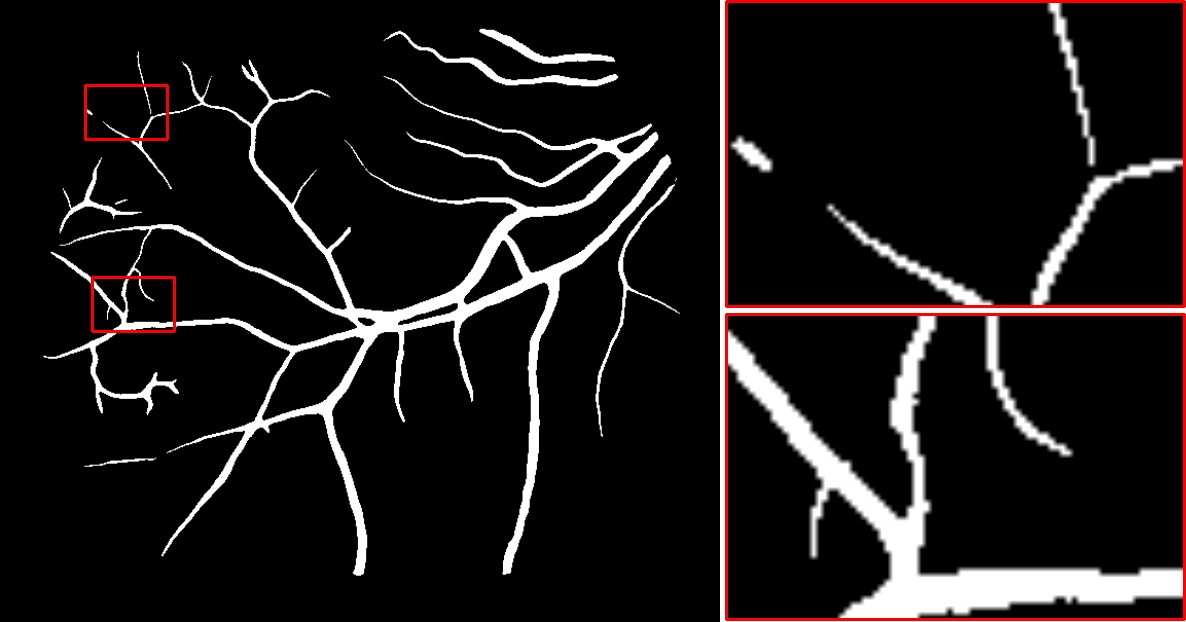}}
            \end{minipage}\\[0.1em]
            \vspace{-1mm}
          \multirow{6}{*}[3.96em]{\adjustbox{valign=m}{\rotatebox[origin=c]{90}{\fcolorbox{white}{cyan!10}{\parbox[c][0.3cm][c]{1.18cm}{\centering \footnotesize R2U-Net}}}}}%
            \vspace{1mm}
			\begin{minipage}[t]{0.148\linewidth}
			\centering
			\centerline{\includegraphics[width=1\linewidth]{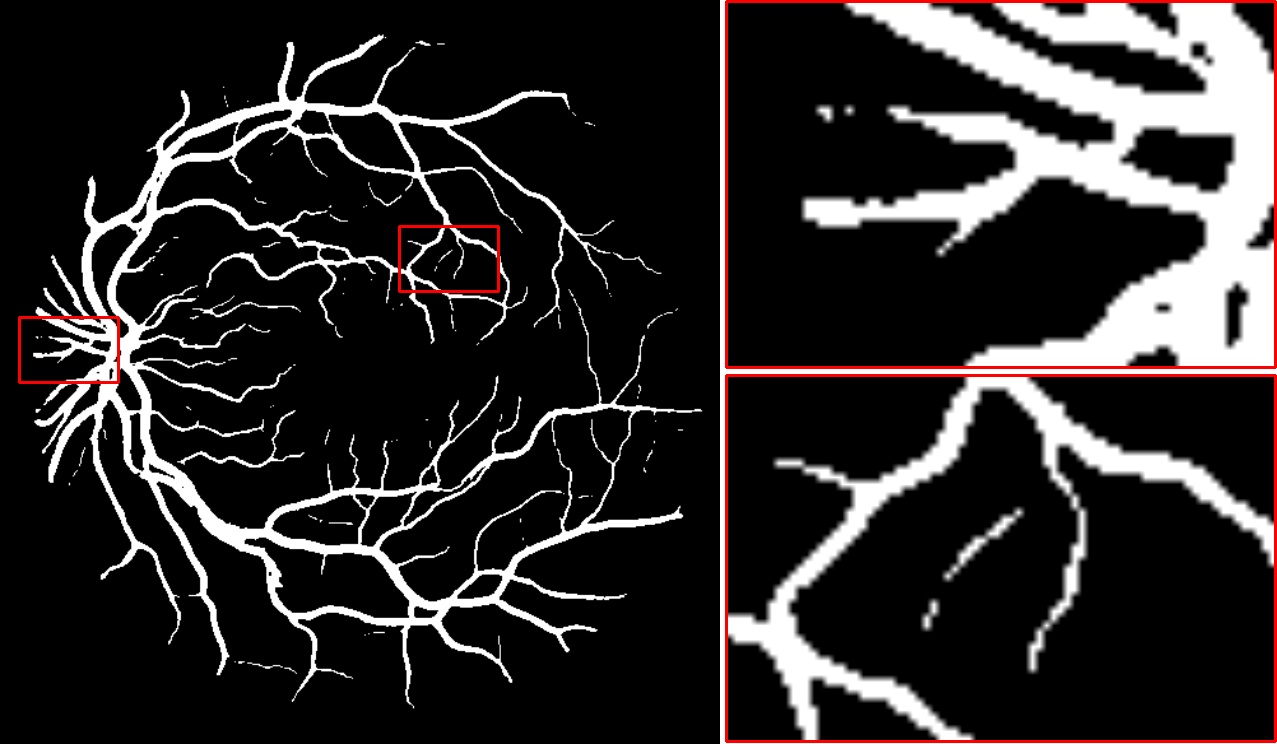}}
            \end{minipage}
            \begin{minipage}[t]{0.148\linewidth}
			\centering
			\centerline{\includegraphics[width=1\linewidth]{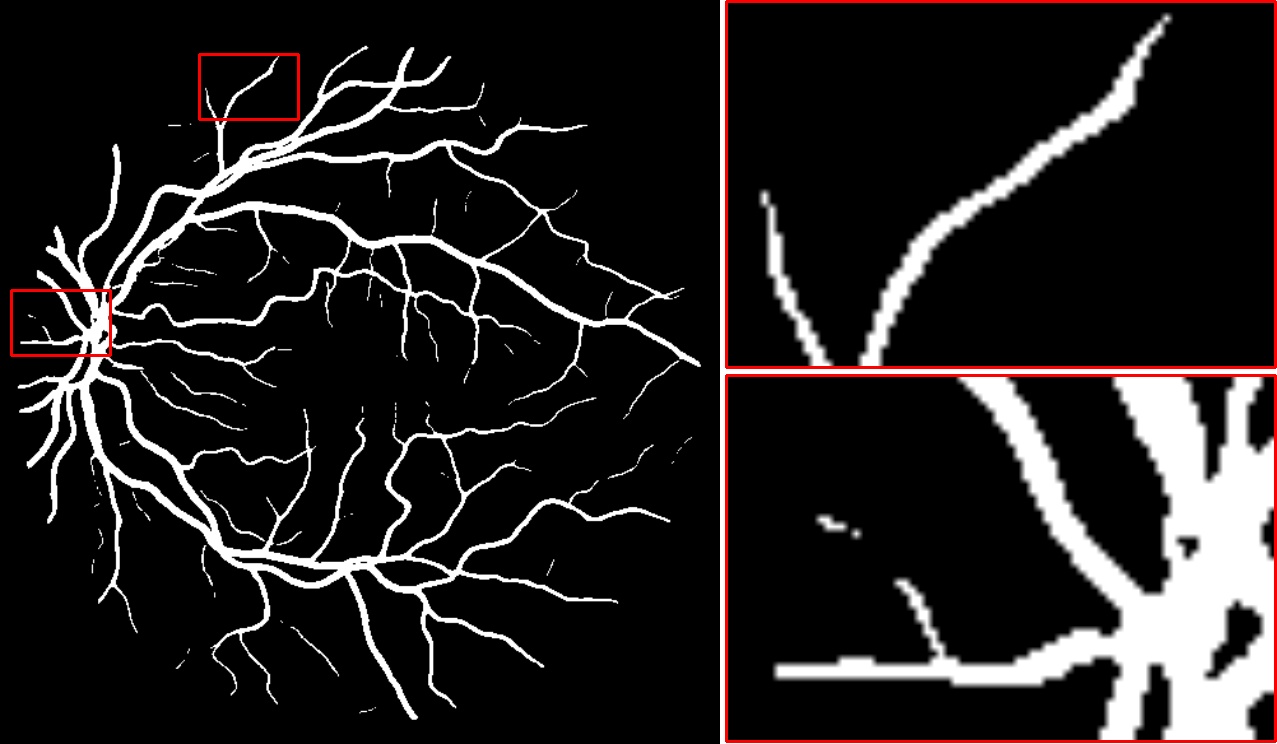}}
            \end{minipage}
            \begin{minipage}[t]{0.154\linewidth}
			\centering
			\centerline{\includegraphics[width=1\linewidth]{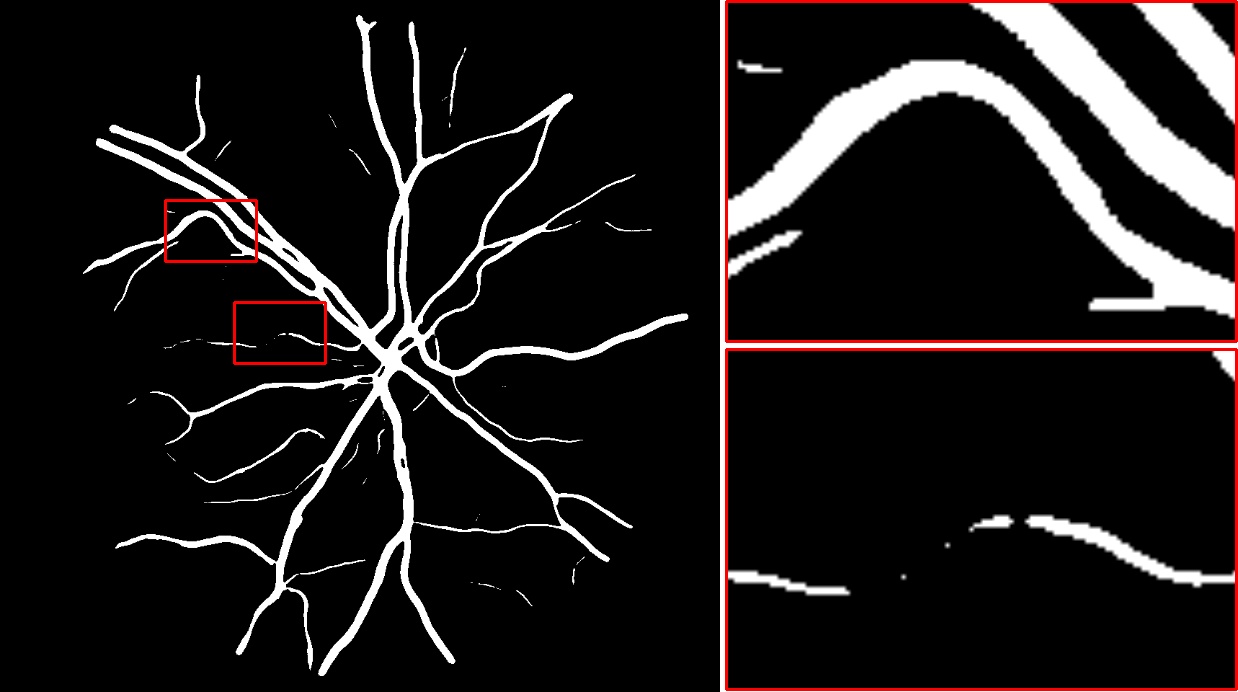}}
            \end{minipage}
            \begin{minipage}[t]{0.154\linewidth}
			\centering
			\centerline{\includegraphics[width=1\linewidth]{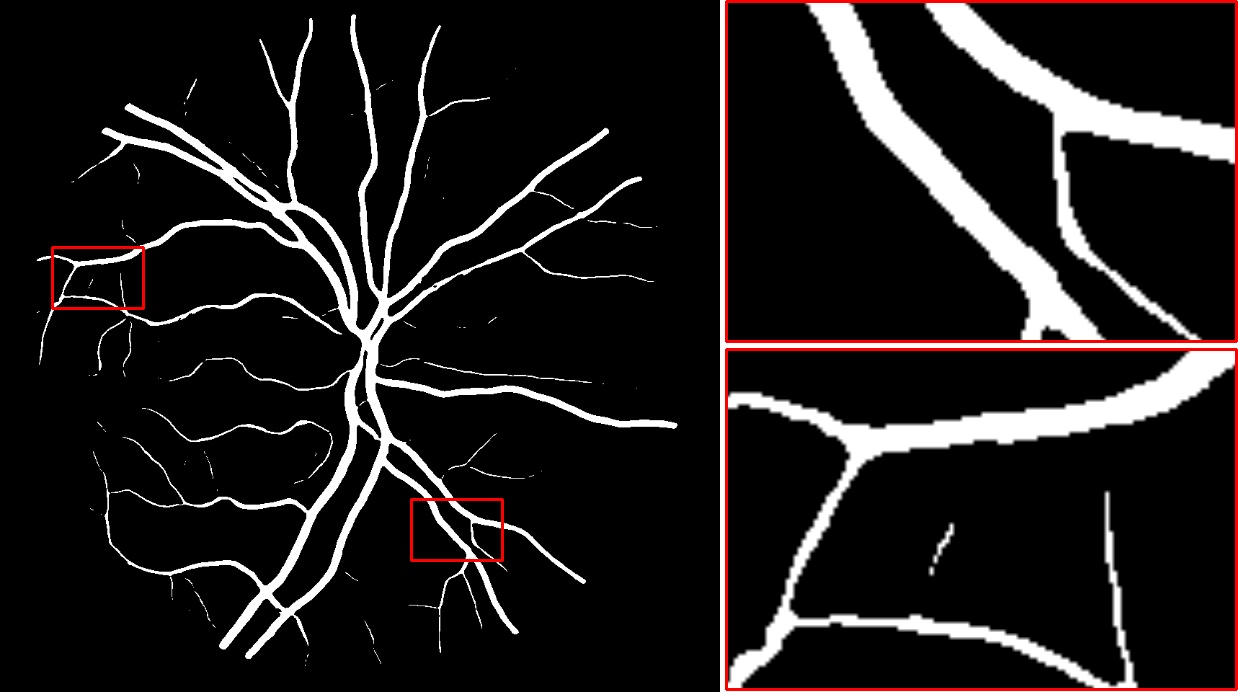}}
            \end{minipage}
            \begin{minipage}[t]{0.1635\linewidth}
			\centering
			\centerline{\includegraphics[width=1\linewidth]{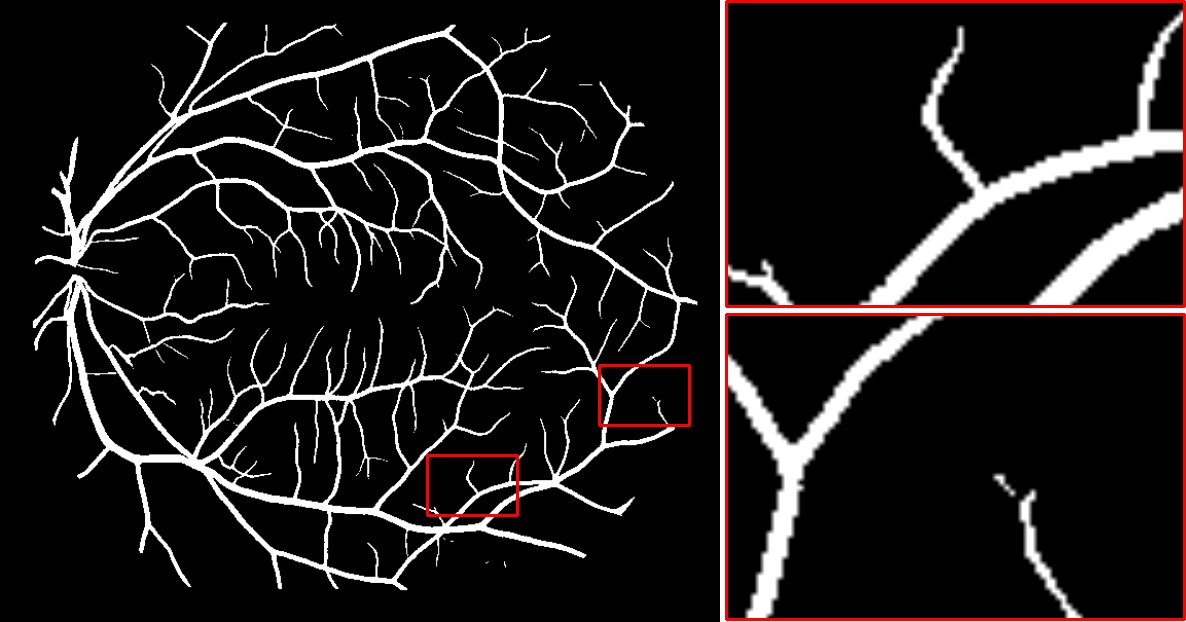}}
            \end{minipage}
            \begin{minipage}[t]{0.1635\linewidth}
			\centering
			\centerline{\includegraphics[width=1\linewidth]{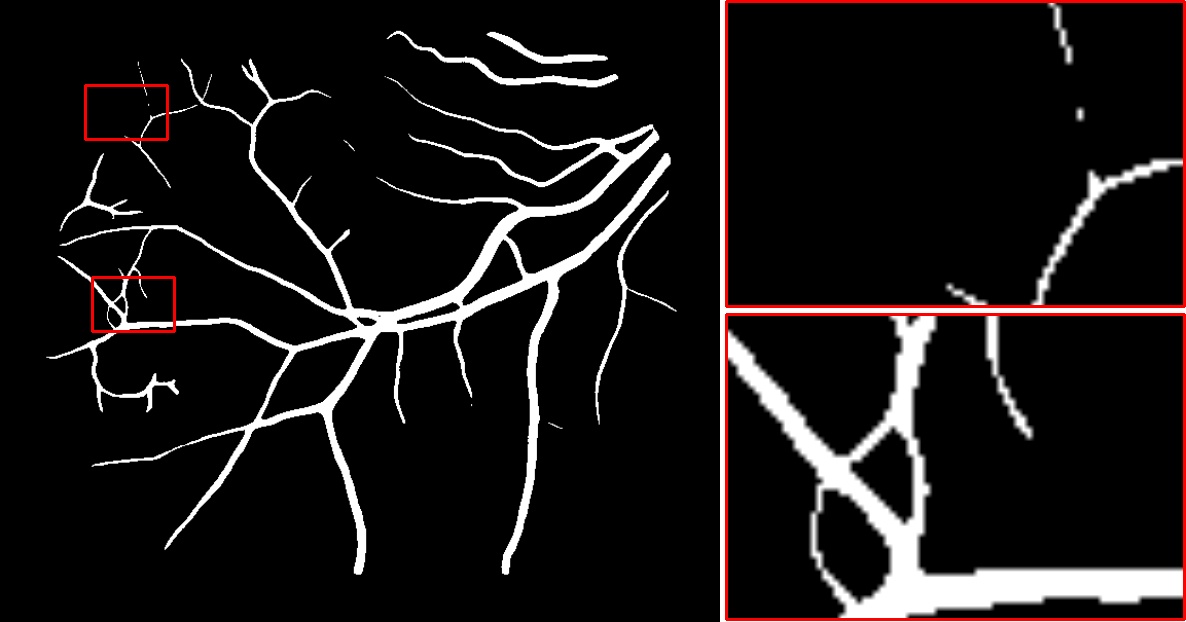}}
            \end{minipage}\\[0.1em]
            \vspace{-1mm}
          \multirow{6}{*}[3.96em]{\adjustbox{valign=m}{\rotatebox[origin=c]{90}{\fcolorbox{white}{cyan!10}{\parbox[c][0.3cm][c]{1.18cm}{\centering \footnotesize DUNet}}}}}%
            \vspace{1mm}
			\begin{minipage}[t]{0.148\linewidth}
			\centering
			\centerline{\includegraphics[width=1\linewidth]{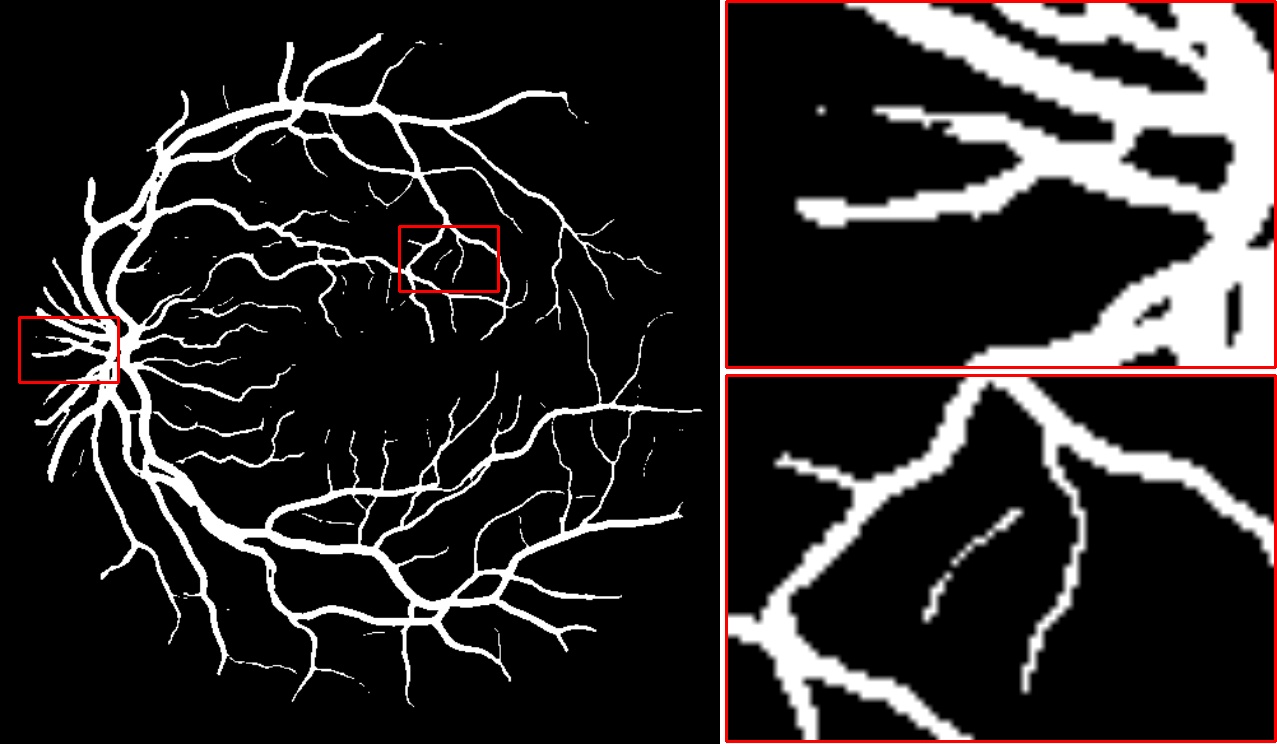}}
            \end{minipage}
            \begin{minipage}[t]{0.148\linewidth}
			\centering
			\centerline{\includegraphics[width=1\linewidth]{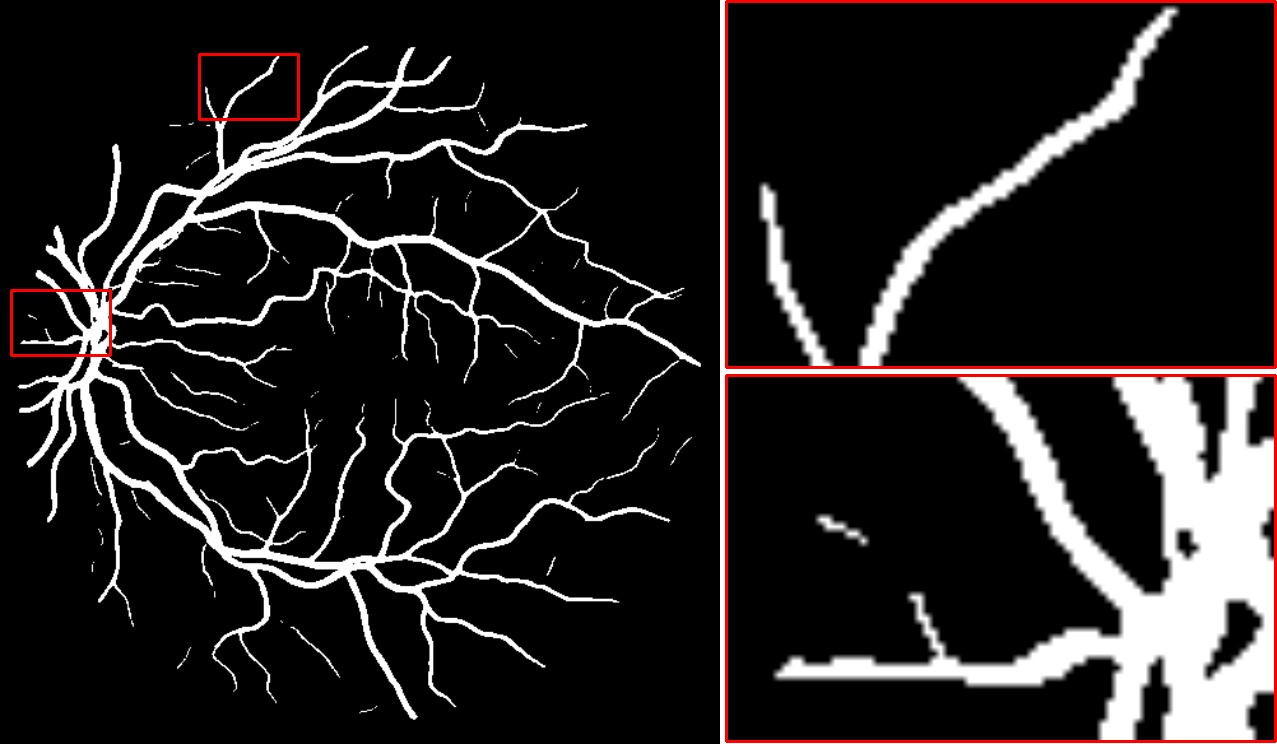}}
            \end{minipage}
            \begin{minipage}[t]{0.154\linewidth}
			\centering
			\centerline{\includegraphics[width=1\linewidth]{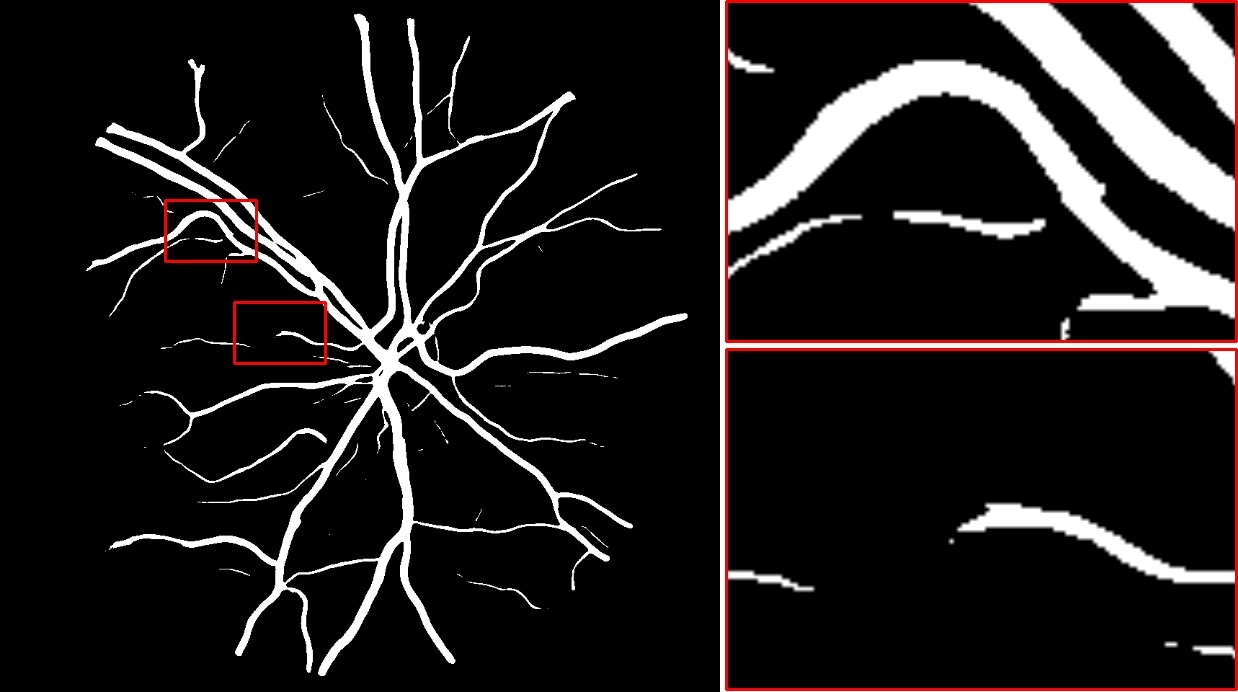}}
            \end{minipage}
            \begin{minipage}[t]{0.154\linewidth}
			\centering
			\centerline{\includegraphics[width=1\linewidth]{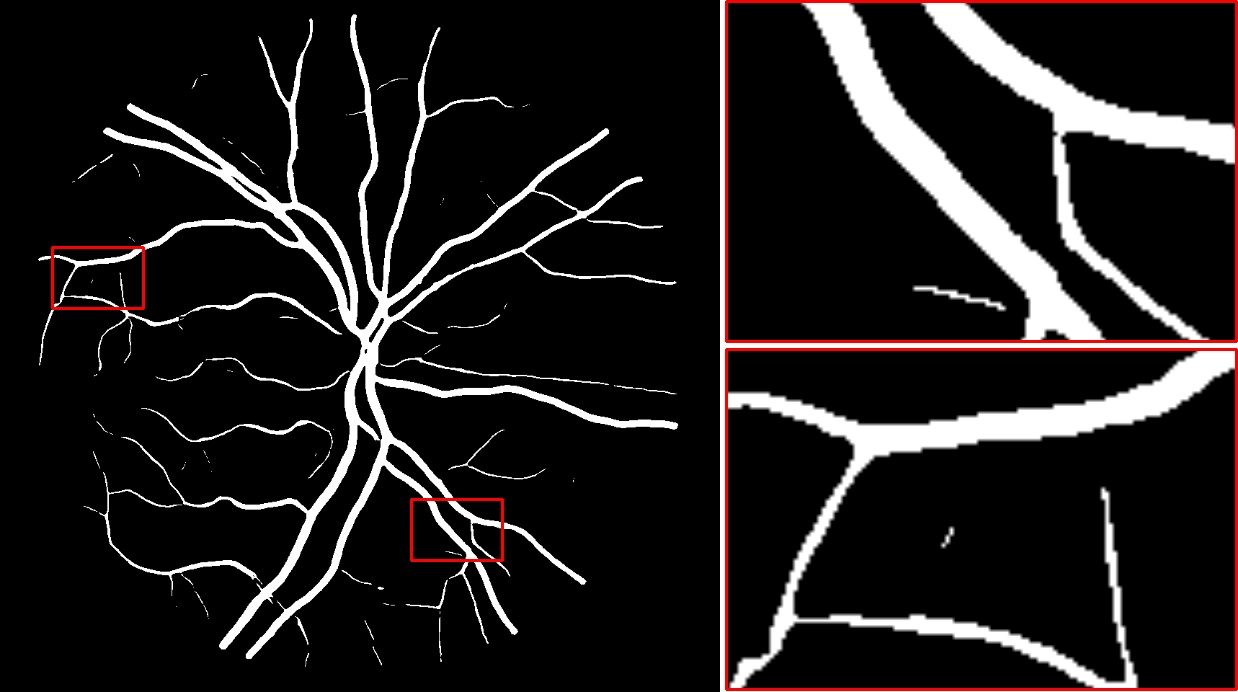}}
            \end{minipage}
            \begin{minipage}[t]{0.1635\linewidth}
			\centering
			\centerline{\includegraphics[width=1\linewidth]{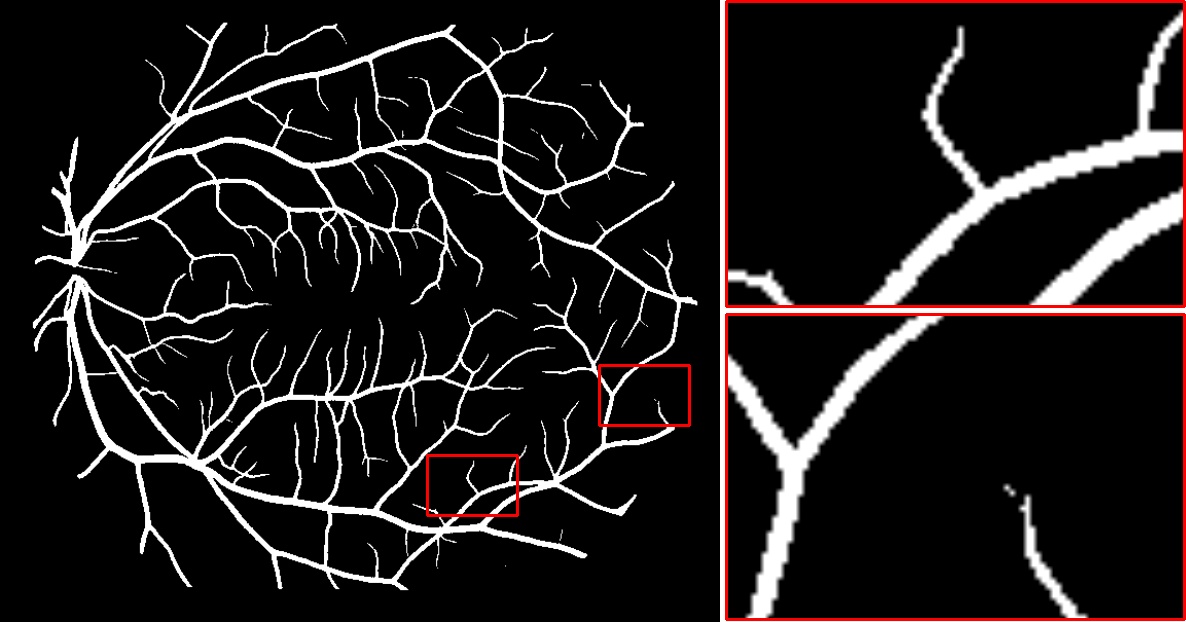}}
            \end{minipage}
            \begin{minipage}[t]{0.1635\linewidth}
			\centering
			\centerline{\includegraphics[width=1\linewidth]{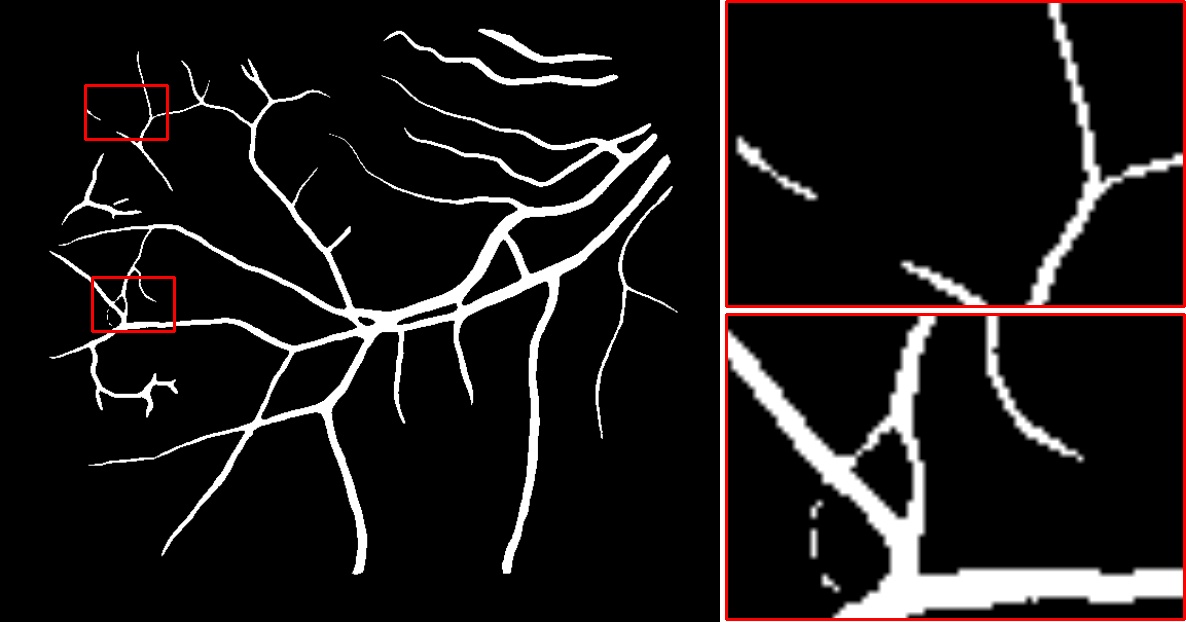}}
            \end{minipage}\\[0.1em]
            \vspace{-1mm}
           \multirow{6}{*}[3.96em]{\adjustbox{valign=m}{\rotatebox[origin=c]{90}{\fcolorbox{white}{cyan!10}{\parbox[c][0.3cm][c]{1.18cm}{\centering \footnotesize CS-Net}}}}}%
            \vspace{1mm}
			\begin{minipage}[t]{0.148\linewidth}
			\centering
			\centerline{\includegraphics[width=1\linewidth]{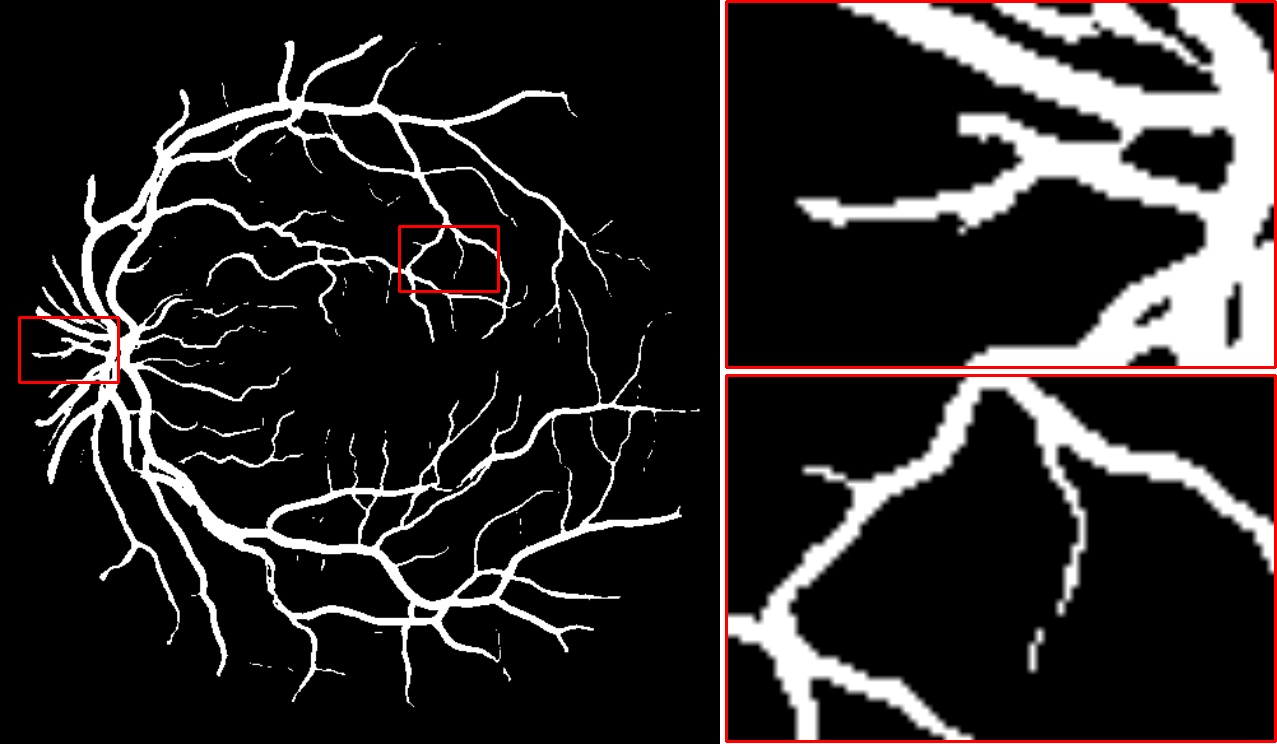}}
            \end{minipage}
            \begin{minipage}[t]{0.148\linewidth}
			\centering
			\centerline{\includegraphics[width=1\linewidth]{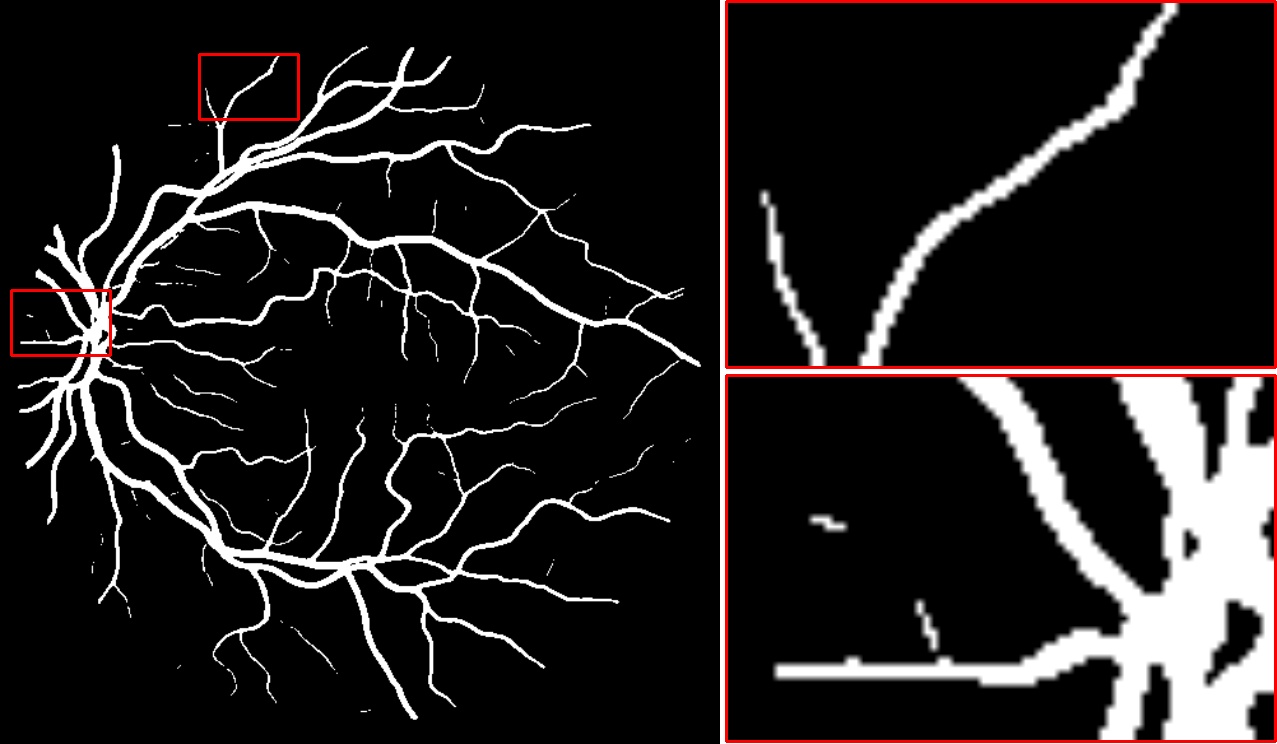}}
            \end{minipage}
            \begin{minipage}[t]{0.154\linewidth}
			\centering
			\centerline{\includegraphics[width=1\linewidth]{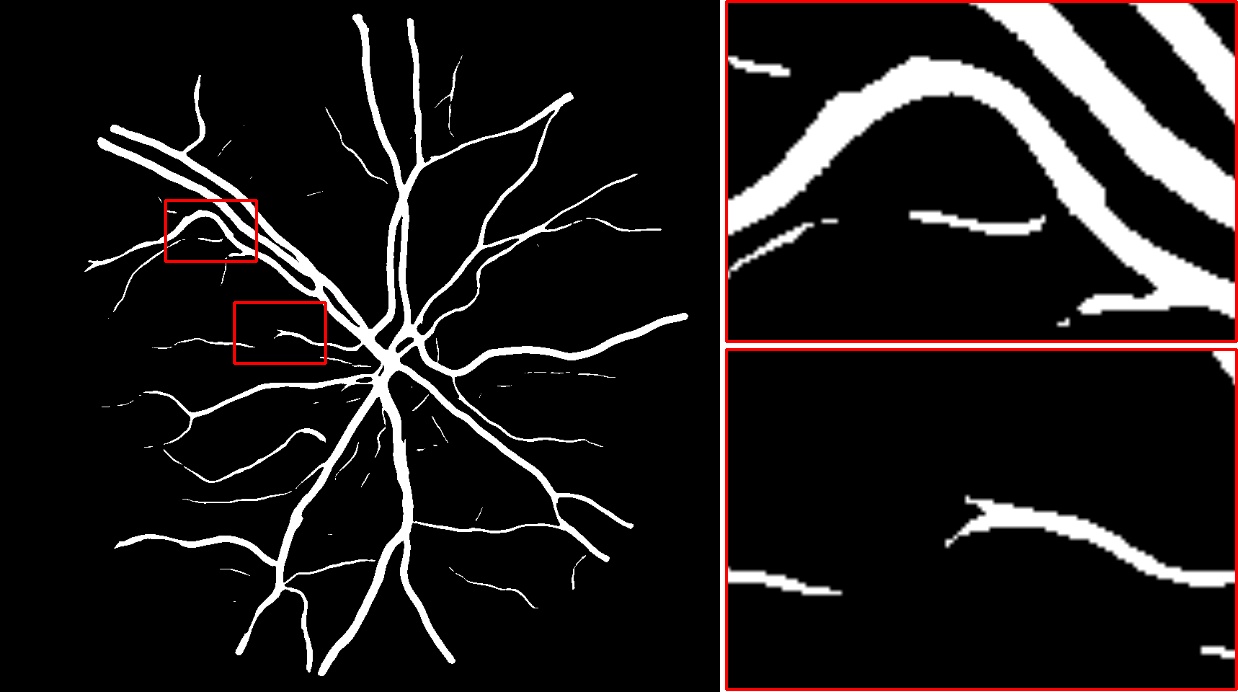}}
            \end{minipage}
            \begin{minipage}[t]{0.154\linewidth}
			\centering
			\centerline{\includegraphics[width=1\linewidth]{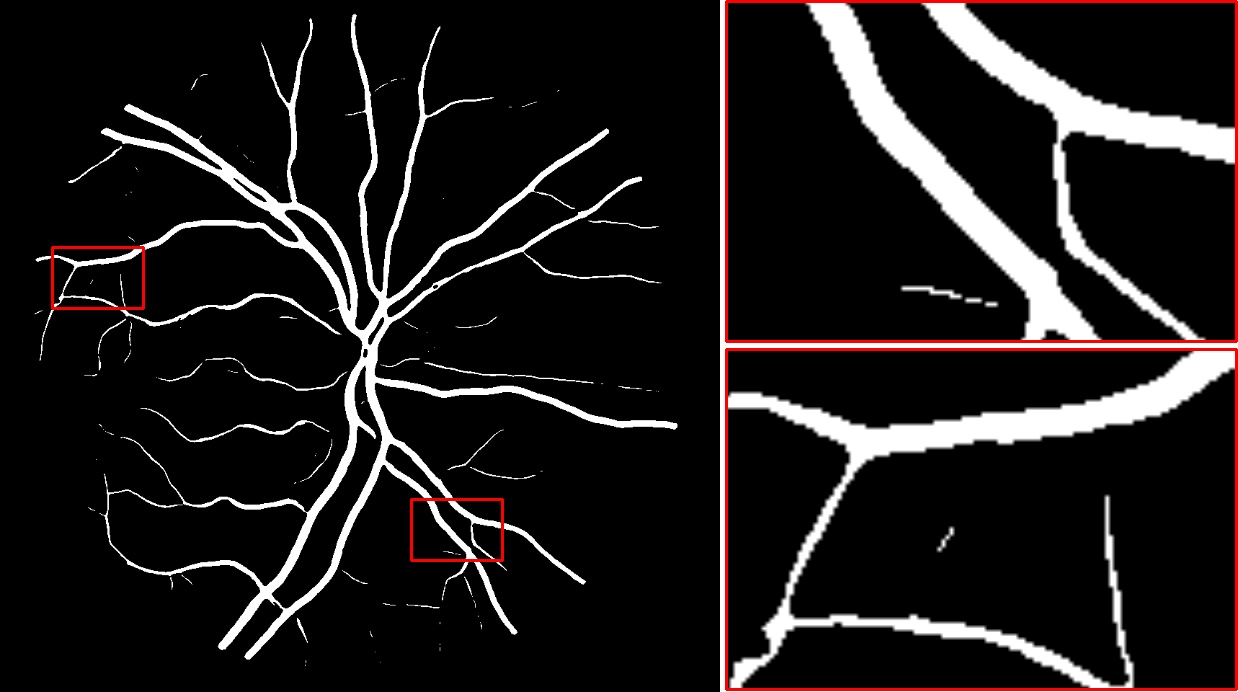}}
            \end{minipage}
            \begin{minipage}[t]{0.1635\linewidth}
			\centering
			\centerline{\includegraphics[width=1\linewidth]{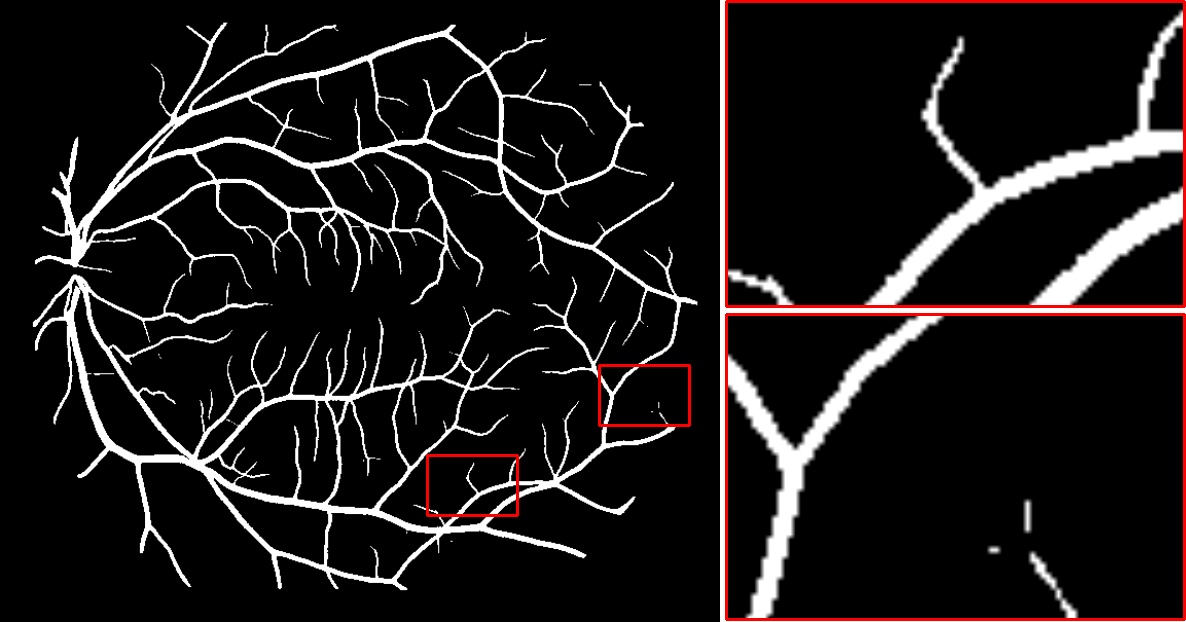}}
            \end{minipage}
            \begin{minipage}[t]{0.1635\linewidth}
			\centering
			\centerline{\includegraphics[width=1\linewidth]{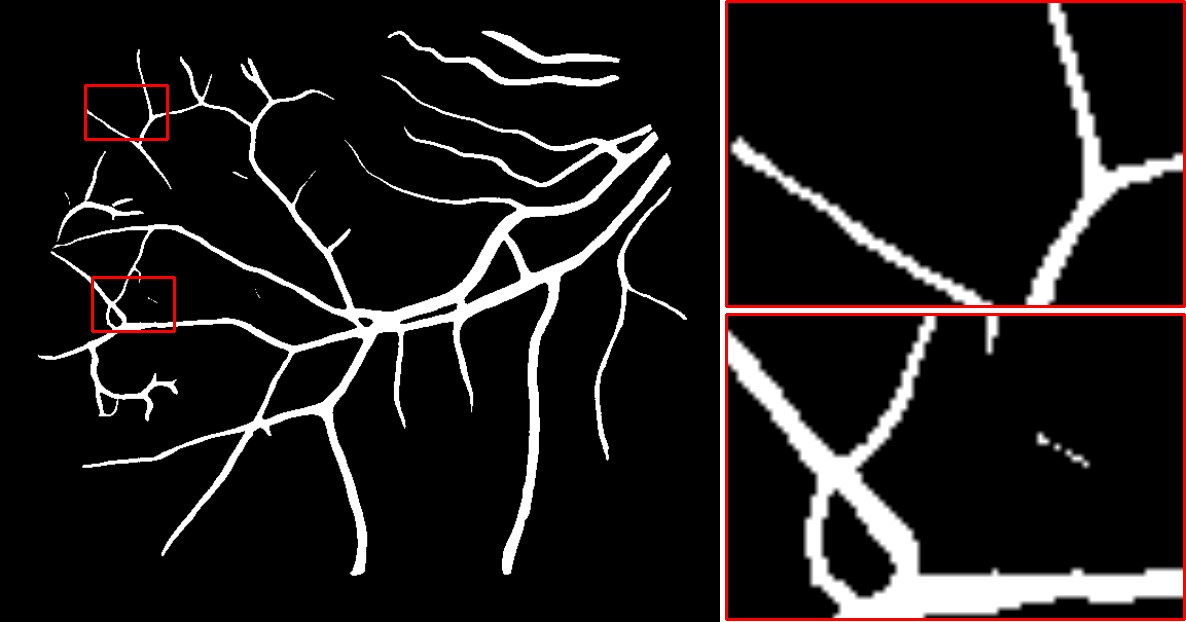}}
            \end{minipage}\\[0.1em]
               \vspace{-1mm}
            \multirow{6}{*}[3.96em]{\adjustbox{valign=m}{\rotatebox[origin=c]{90}{\fcolorbox{white}{cyan!10}{\parbox[c][0.3cm][c]{1.18cm}{\centering \footnotesize SA-UNet}}}}}%
            \vspace{1mm}
			\begin{minipage}[t]{0.148\linewidth}
			\centering
			\centerline{\includegraphics[width=1\linewidth]{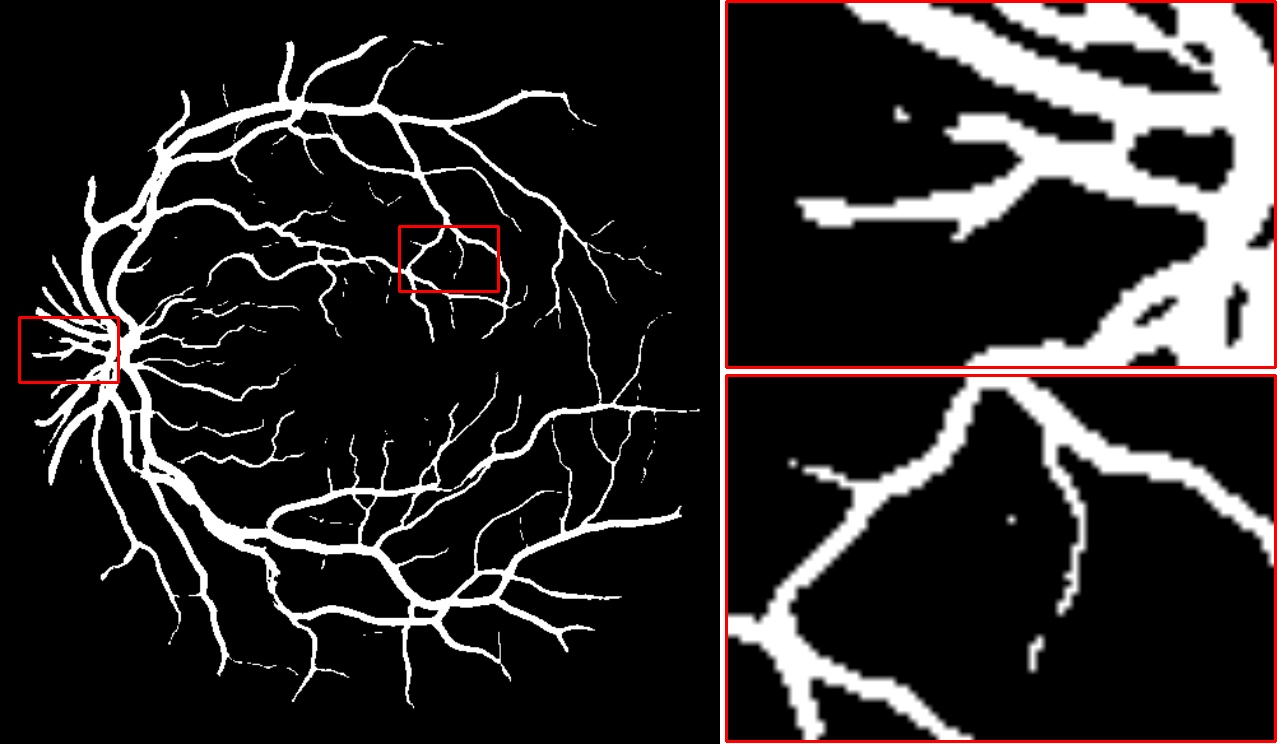}}
            \end{minipage}
            \begin{minipage}[t]{0.148\linewidth}
			\centering
			\centerline{\includegraphics[width=1\linewidth]{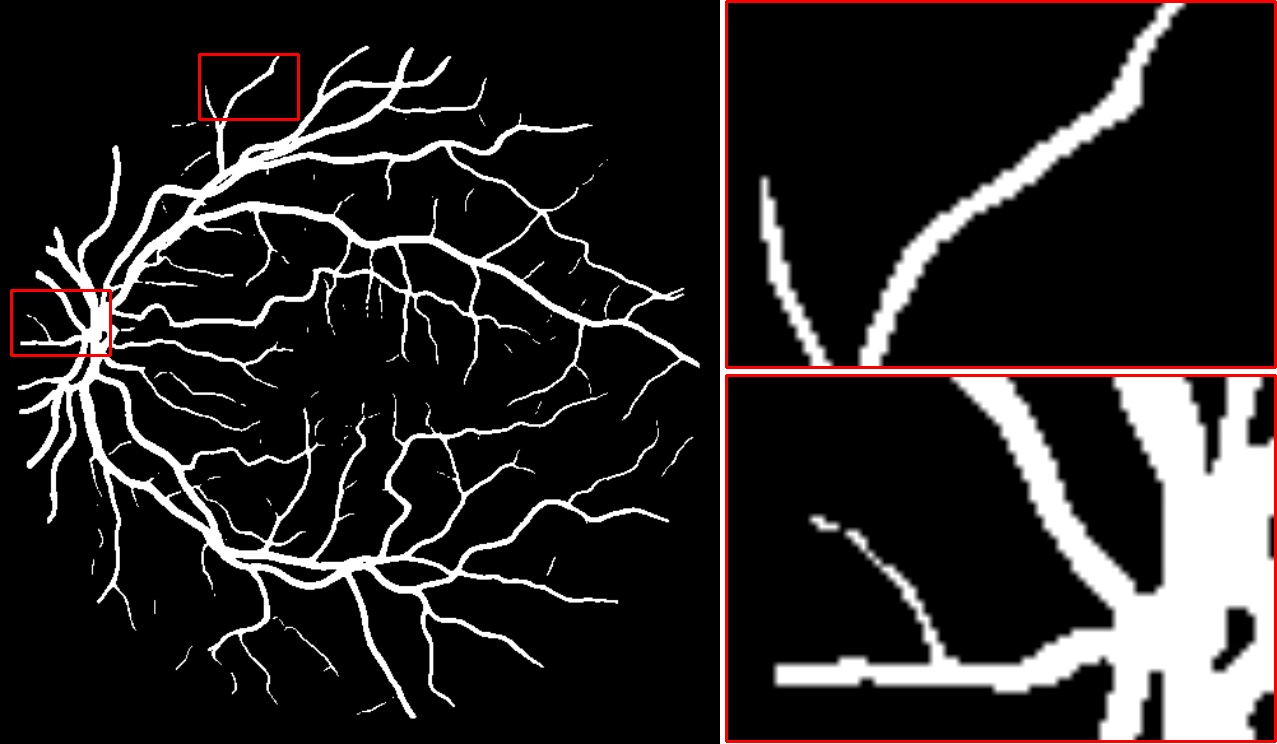}}
            \end{minipage}
            \begin{minipage}[t]{0.154\linewidth}
			\centering
			\centerline{\includegraphics[width=1\linewidth]{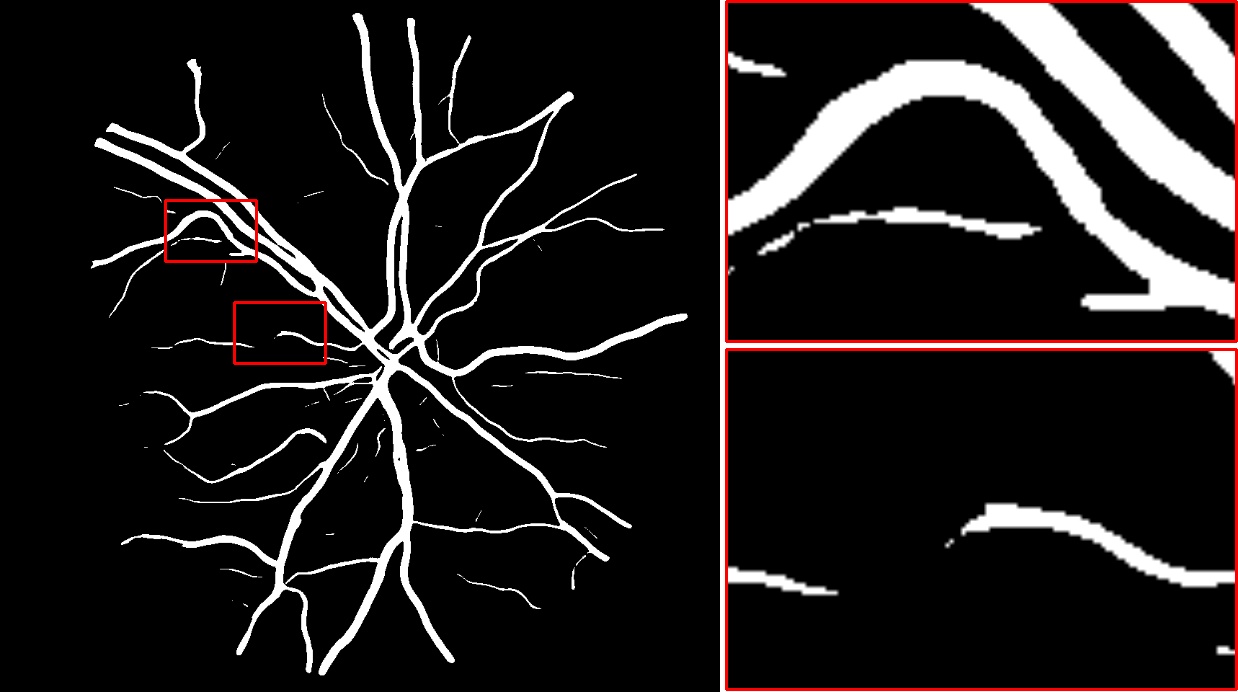}}
            \end{minipage}
            \begin{minipage}[t]{0.154\linewidth}
			\centering
			\centerline{\includegraphics[width=1\linewidth]{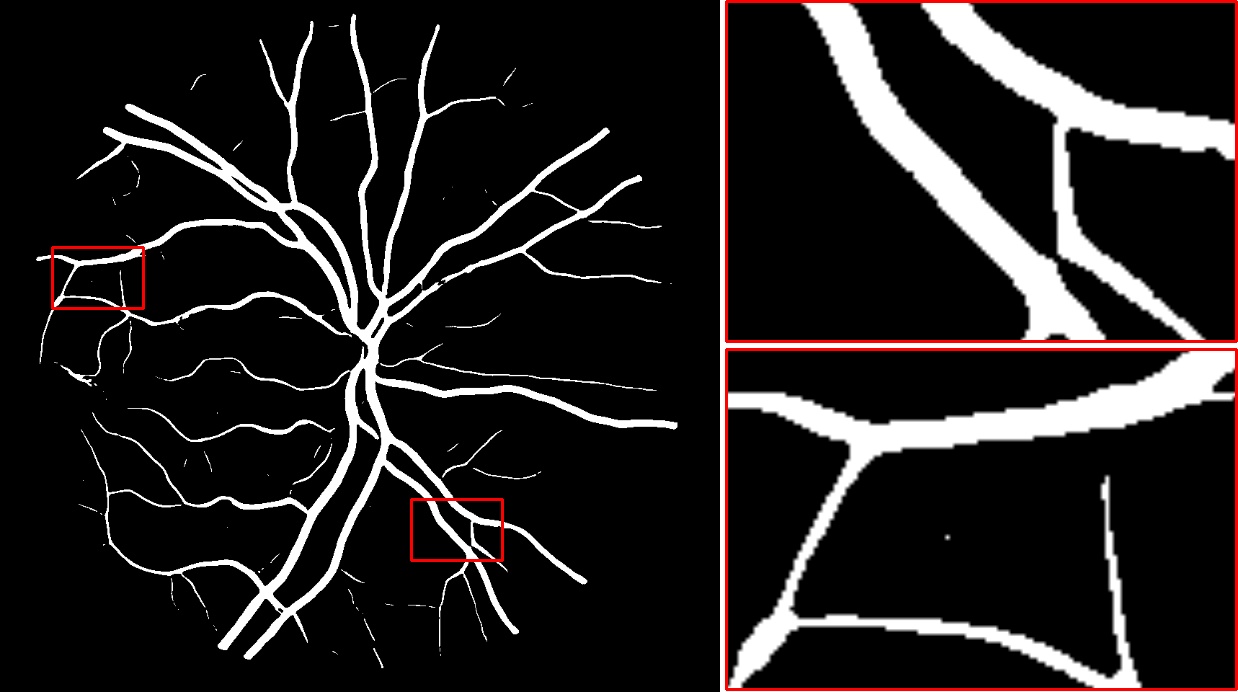}}
            \end{minipage}
            \begin{minipage}[t]{0.1635\linewidth}
			\centering
			\centerline{\includegraphics[width=1\linewidth]{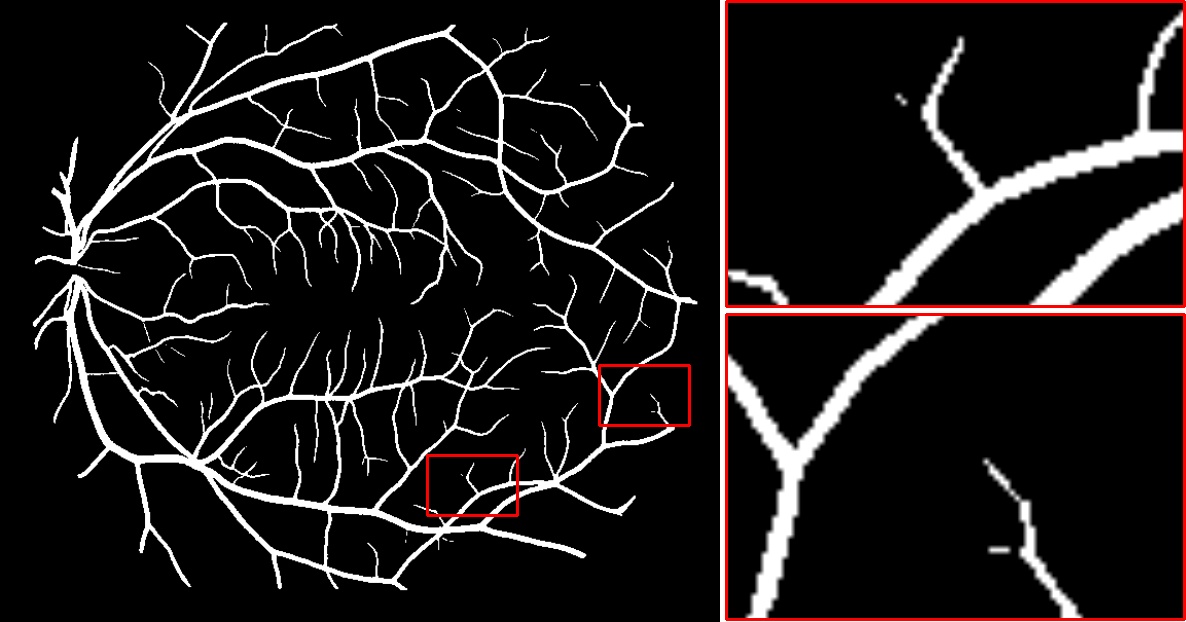}}
            \end{minipage}
            \begin{minipage}[t]{0.1635\linewidth}
			\centering
			\centerline{\includegraphics[width=1\linewidth]{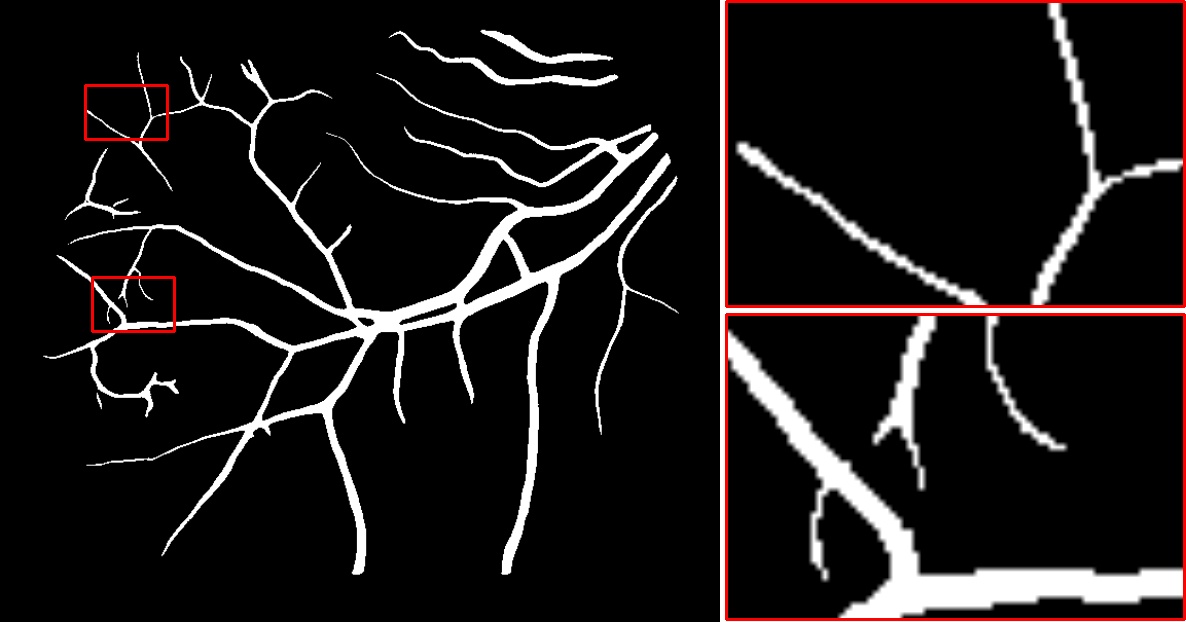}}
            \end{minipage}\\[0.1em]
               \vspace{-1mm}
             \multirow{6}{*}[3.96em]{\adjustbox{valign=m}{\rotatebox[origin=c]{90}{\fcolorbox{white}{cyan!10}{\parbox[c][0.3cm][c]{1.18cm}{\centering \footnotesize CS$^2$$-$Net}}}}}%
            \vspace{1mm}
			\begin{minipage}[t]{0.148\linewidth}
			\centering
			\centerline{\includegraphics[width=1\linewidth]{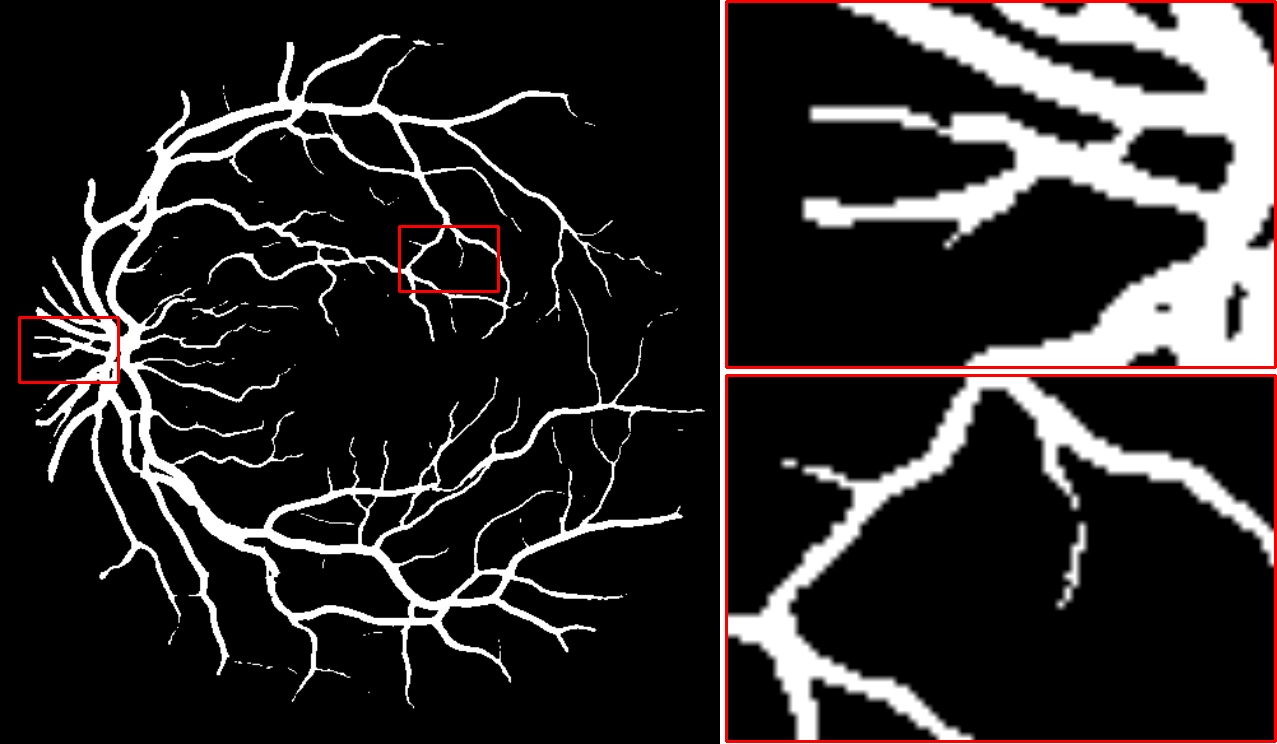}}
            \end{minipage}
            \begin{minipage}[t]{0.148\linewidth}
			\centering
			\centerline{\includegraphics[width=1\linewidth]{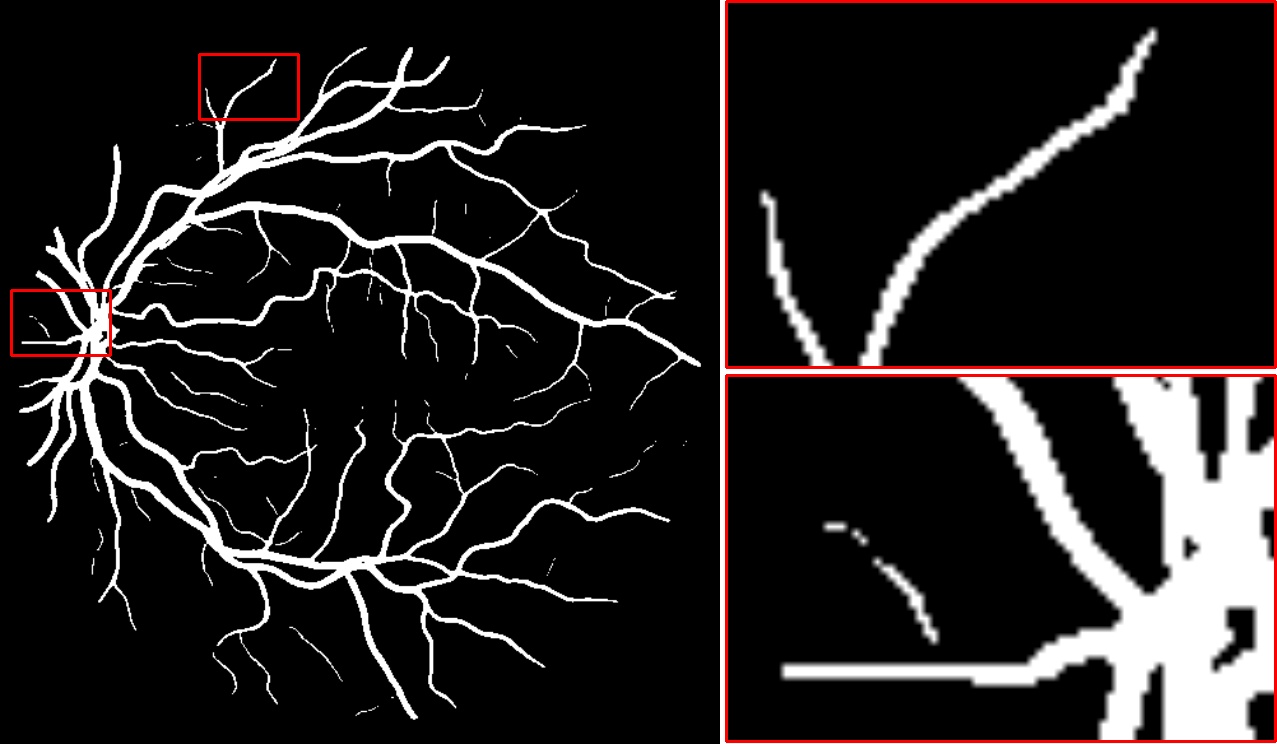}}
            \end{minipage}
            \begin{minipage}[t]{0.154\linewidth}
			\centering
			\centerline{\includegraphics[width=1\linewidth]{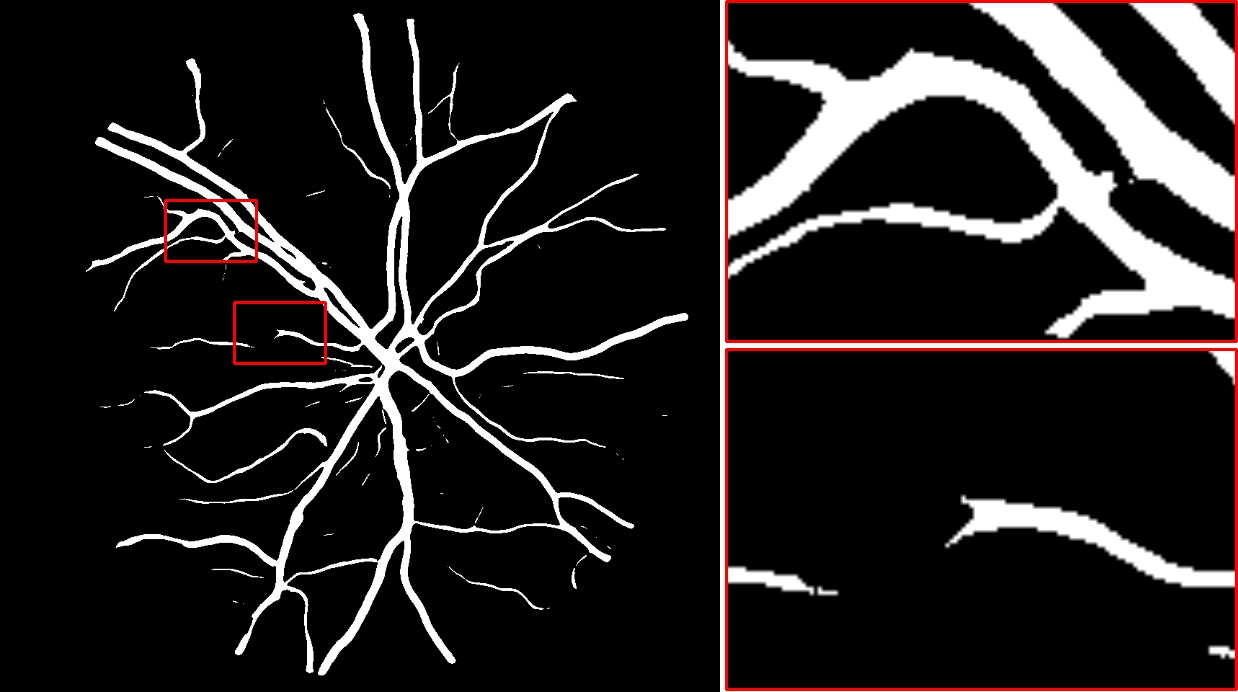}}
            \end{minipage}
            \begin{minipage}[t]{0.154\linewidth}
			\centering
			\centerline{\includegraphics[width=1\linewidth]{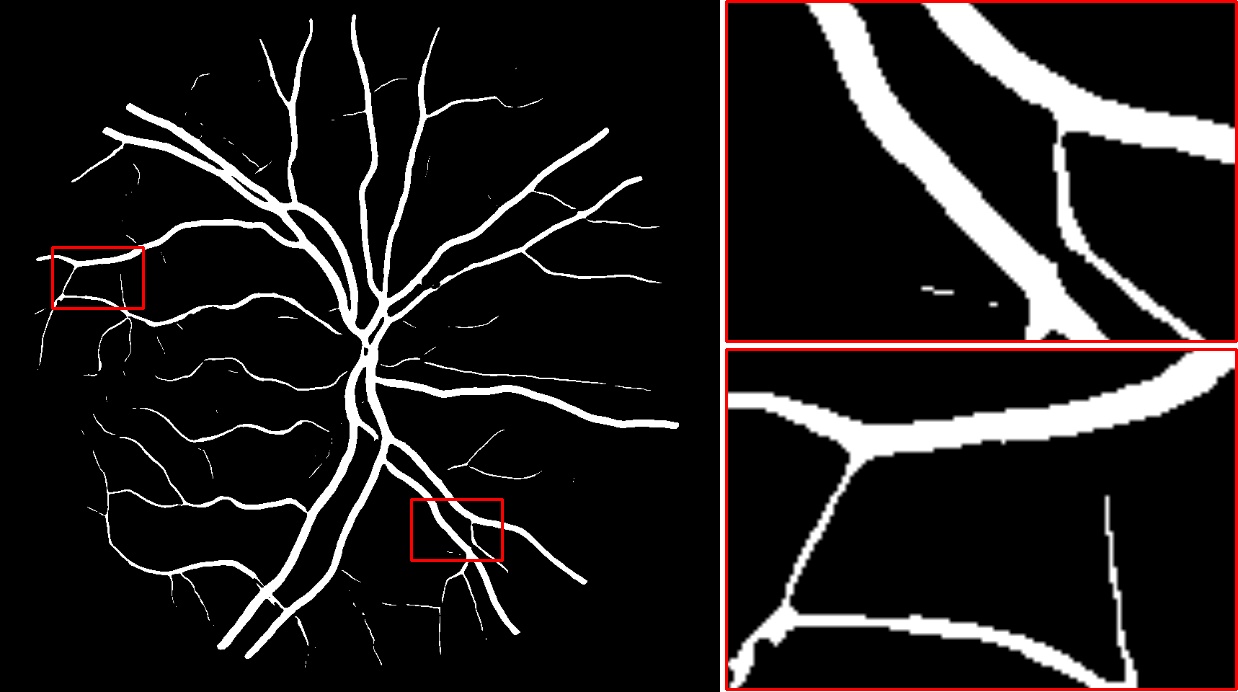}}
            \end{minipage}
            \begin{minipage}[t]{0.1635\linewidth}
			\centering
			\centerline{\includegraphics[width=1\linewidth]{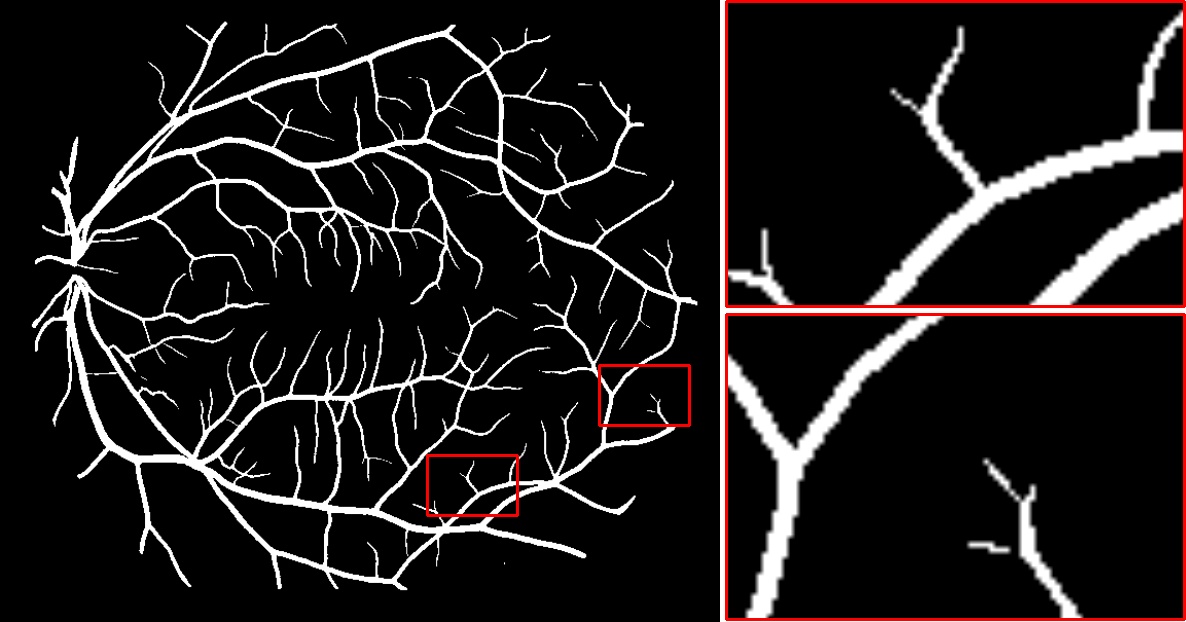}}
            \end{minipage}
            \begin{minipage}[t]{0.1635\linewidth}
			\centering
			\centerline{\includegraphics[width=1\linewidth]{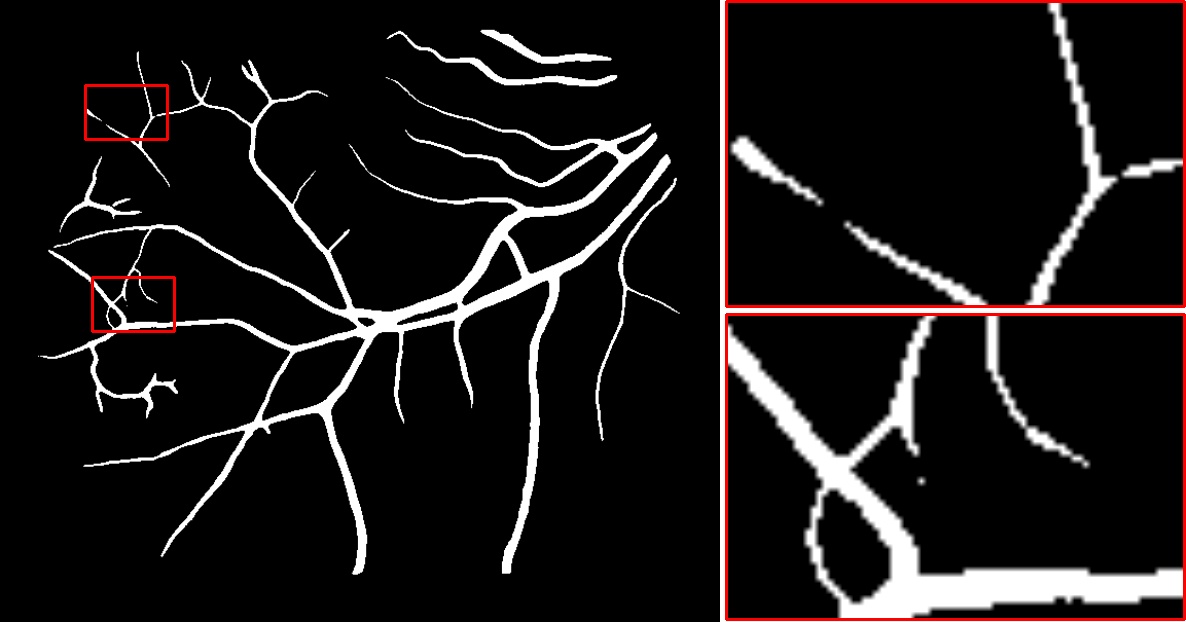}}
            \end{minipage}\\[0.1em]
               \vspace{-1mm}
           \multirow{6}{*}[3.96em]{\adjustbox{valign=m}{\rotatebox[origin=c]{90}{\fcolorbox{white}{cyan!10}{\parbox[c][0.3cm][c]{1.18cm}{\centering \footnotesize FR-UNet}}}}}%
            \vspace{1mm}
			\begin{minipage}[t]{0.148\linewidth}
			\centering
			\centerline{\includegraphics[width=1\linewidth]{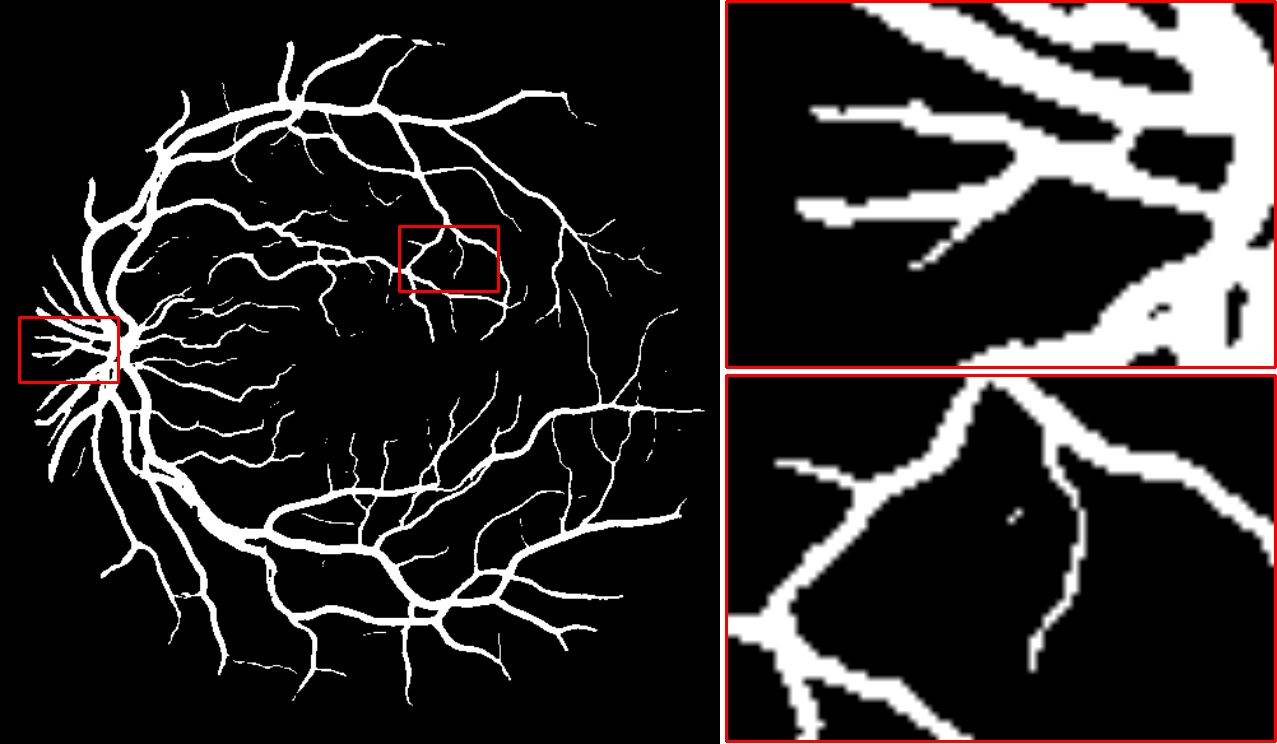}}
            \end{minipage}
            \begin{minipage}[t]{0.148\linewidth}
			\centering
			\centerline{\includegraphics[width=1\linewidth]{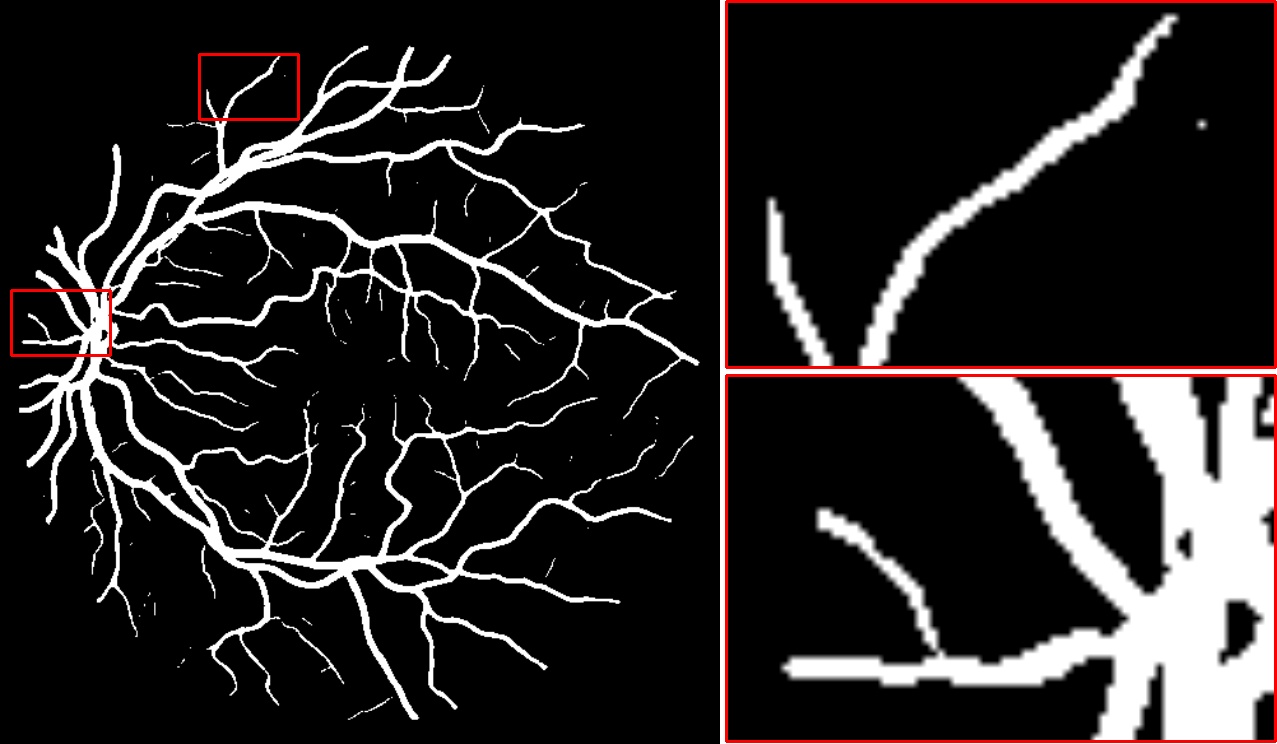}}
            \end{minipage}
            \begin{minipage}[t]{0.154\linewidth}
			\centering
			\centerline{\includegraphics[width=1\linewidth]{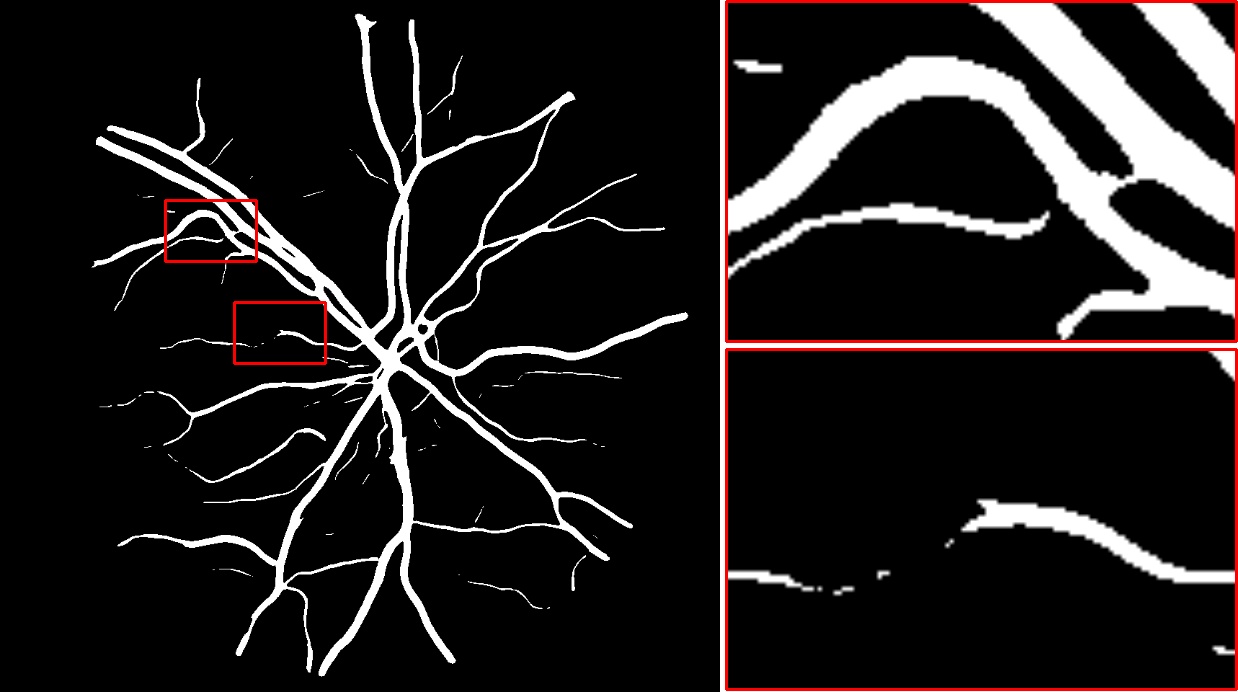}}
            \end{minipage}
            \begin{minipage}[t]{0.154\linewidth}
			\centering
			\centerline{\includegraphics[width=1\linewidth]{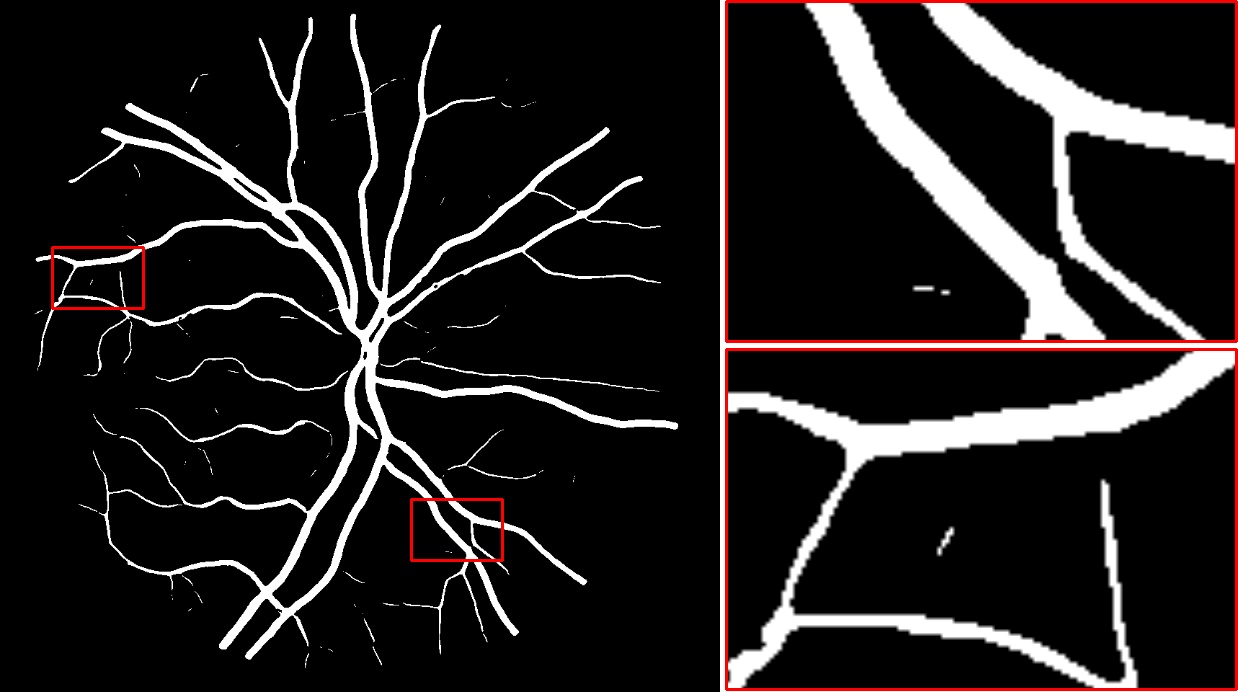}}
            \end{minipage}
            \begin{minipage}[t]{0.1635\linewidth}
			\centering
			\centerline{\includegraphics[width=1\linewidth]{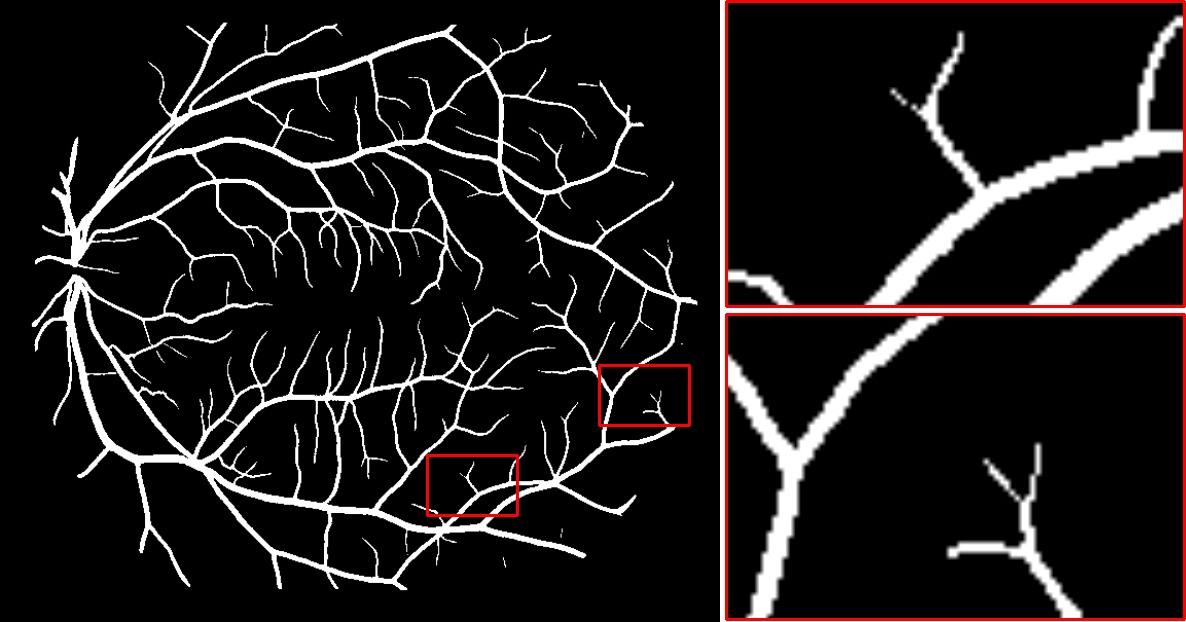}}
            \end{minipage}
            \begin{minipage}[t]{0.1635\linewidth}
			\centering
			\centerline{\includegraphics[width=1\linewidth]{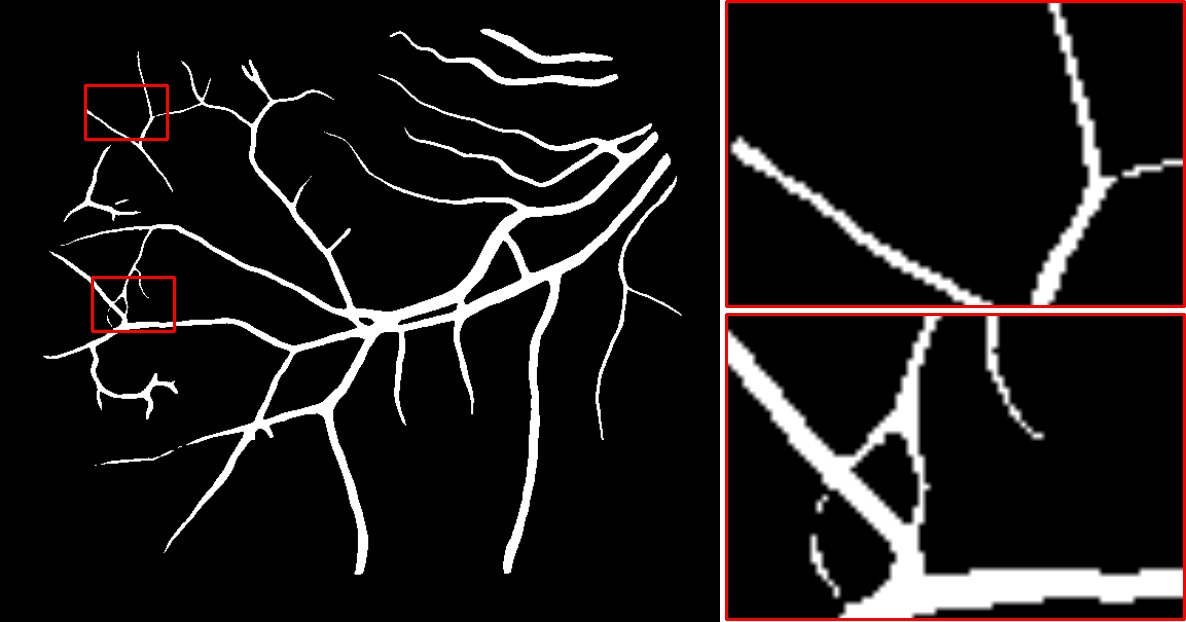}}
            \end{minipage}\\[0.1em]
            \vspace{-1mm}
            \multirow{6}{*}[3.96em]{\adjustbox{valign=m}{\rotatebox[origin=c]{90}{\fcolorbox{white}{cyan!10}{\parbox[c][0.3cm][c]{1.18cm}{\centering \footnotesize EXP-Net}}}}}%
            \vspace{1mm}
			\begin{minipage}[t]{0.148\linewidth}
			\centering
			\centerline{\includegraphics[width=1\linewidth]{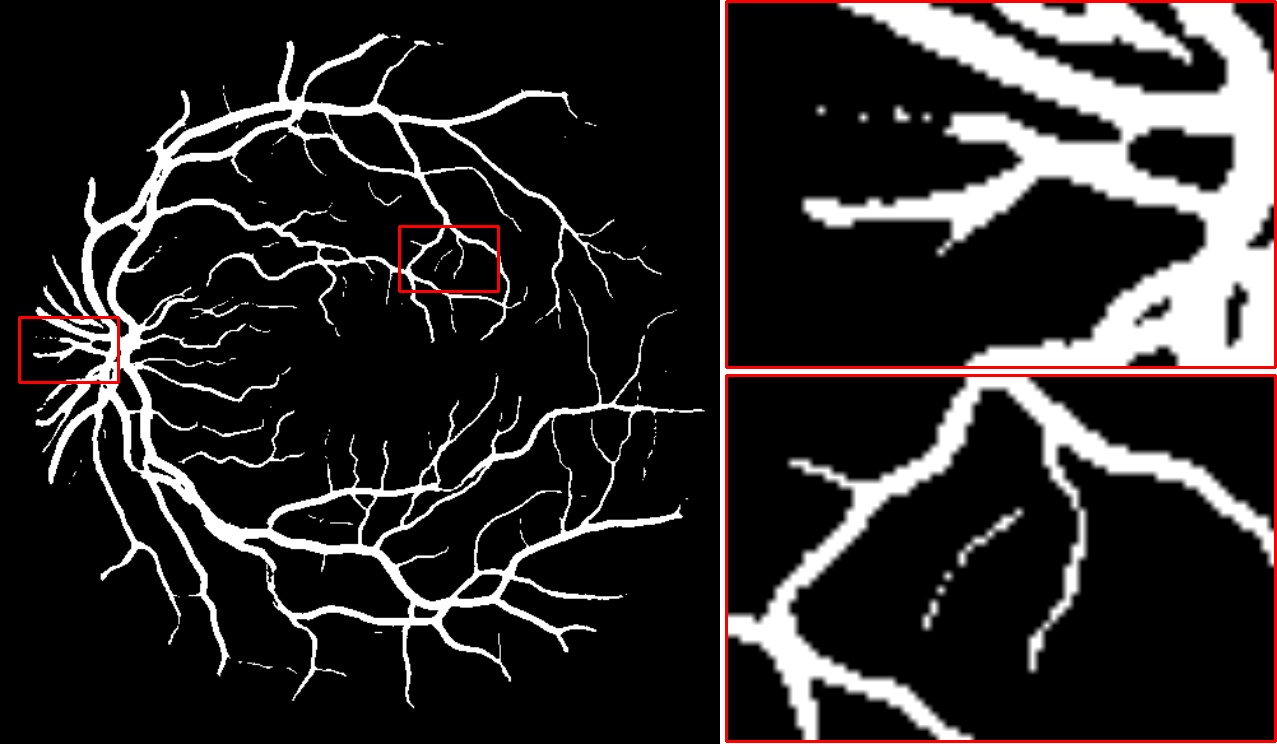}}
            \end{minipage}
            \begin{minipage}[t]{0.148\linewidth}
			\centering
			\centerline{\includegraphics[width=1\linewidth]{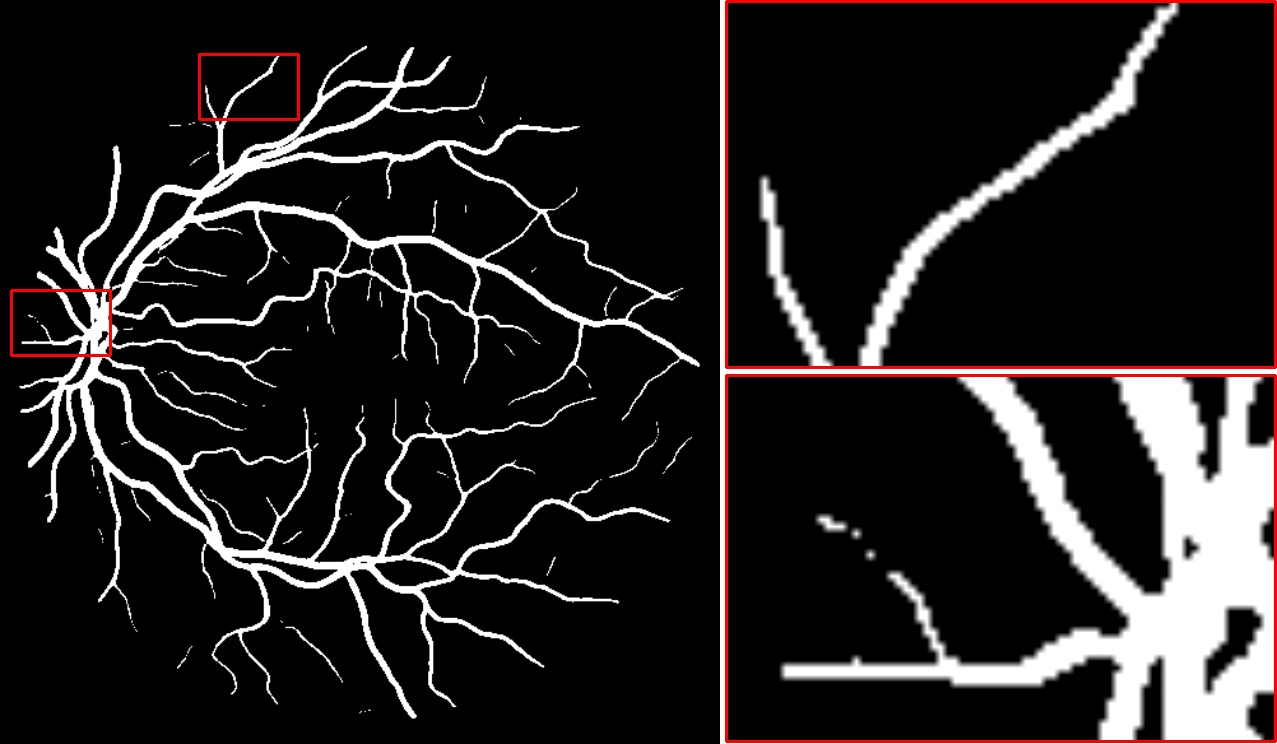}}
            \end{minipage}
            \begin{minipage}[t]{0.154\linewidth}
			\centering
			\centerline{\includegraphics[width=1\linewidth]{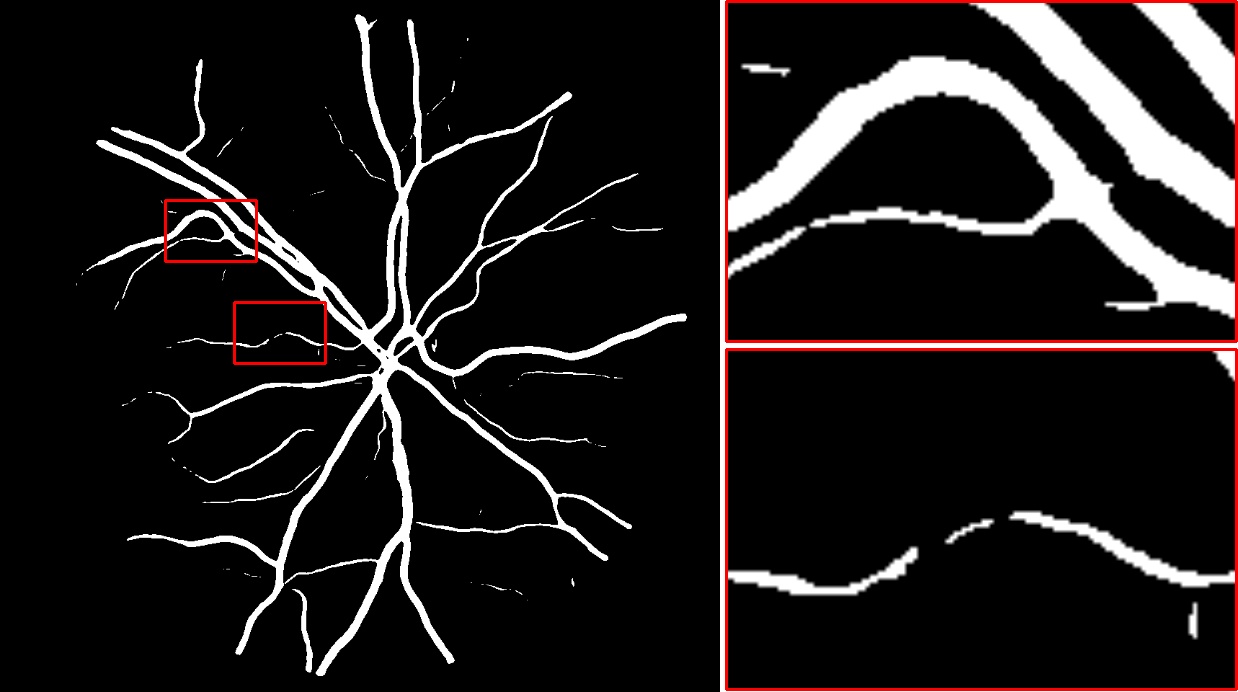}}
            \end{minipage}
            \begin{minipage}[t]{0.154\linewidth}
			\centering
			\centerline{\includegraphics[width=1\linewidth]{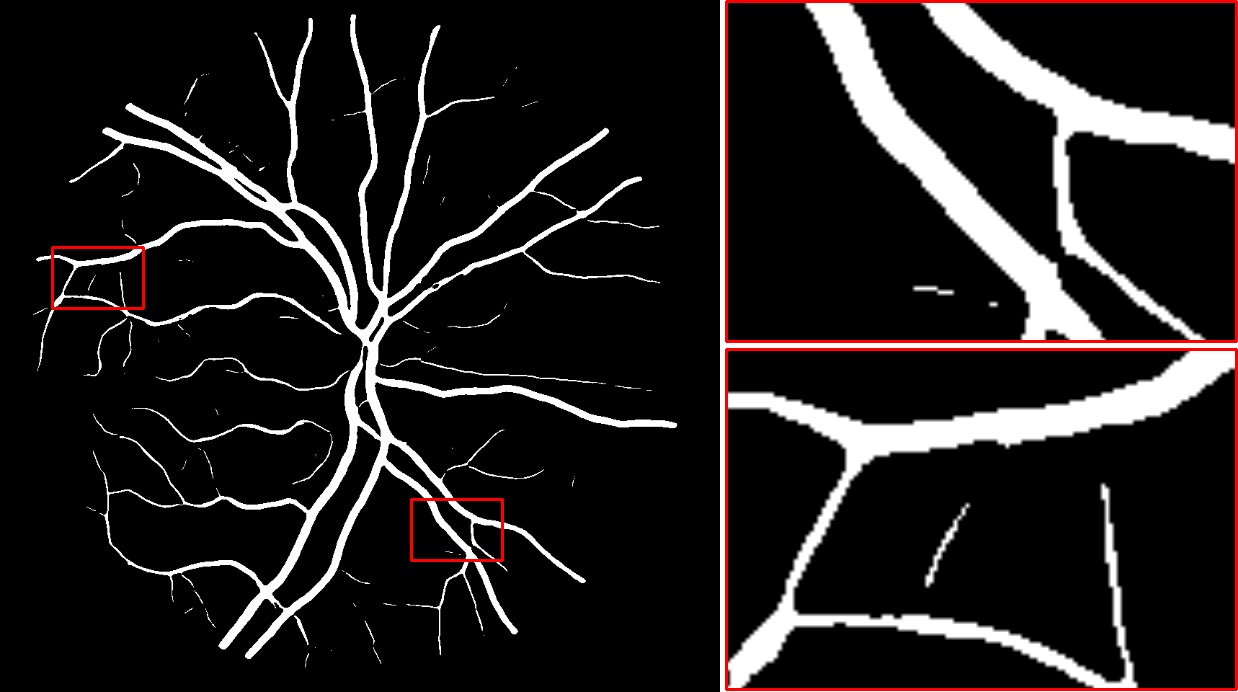}}
            \end{minipage}
            \begin{minipage}[t]{0.1635\linewidth}
			\centering
			\centerline{\includegraphics[width=1\linewidth]{STARE1_mag_saunet.jpg}}
            \end{minipage}
            \begin{minipage}[t]{0.1635\linewidth}
			\centering
			\centerline{\includegraphics[width=1\linewidth]{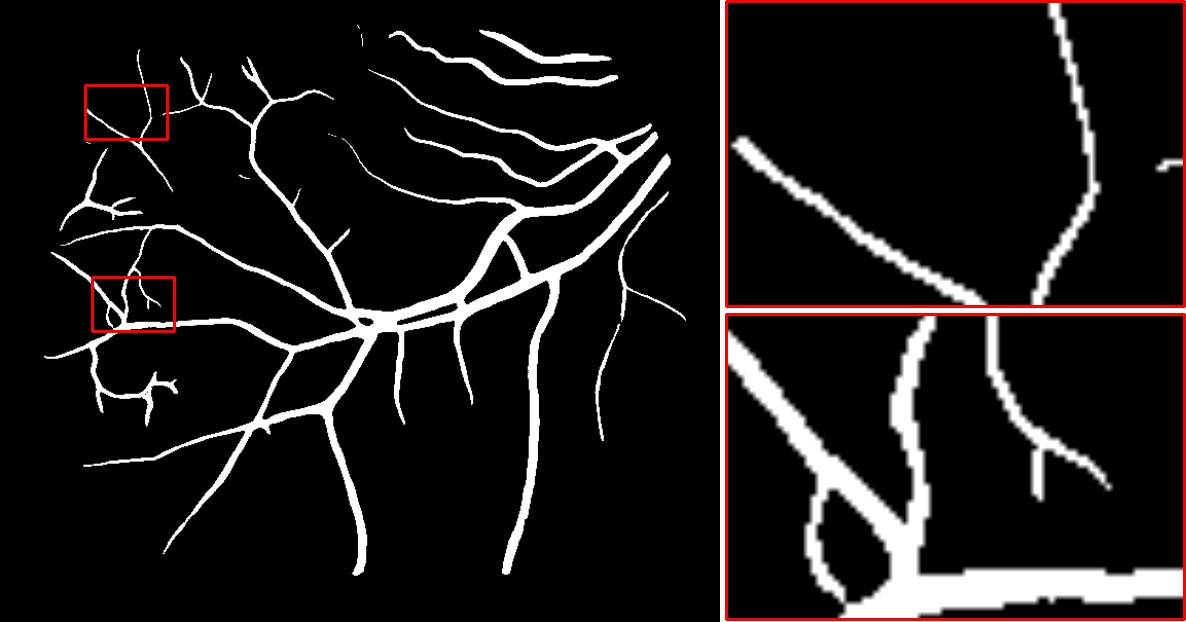}}
            \end{minipage}\\[0.1em]
            \vspace{-1mm}
           \multirow{6}{*}[3.96em]{\adjustbox{valign=m}{\rotatebox[origin=c]{90}{\fcolorbox{white}{cyan!10}{\parbox[c][0.3cm][c]{1.18cm}{\centering \footnotesize Ours}}}}}%
            \vspace{1mm}
			\begin{minipage}[t]{0.148\linewidth}
			\centering
			\centerline{\includegraphics[width=1\linewidth]{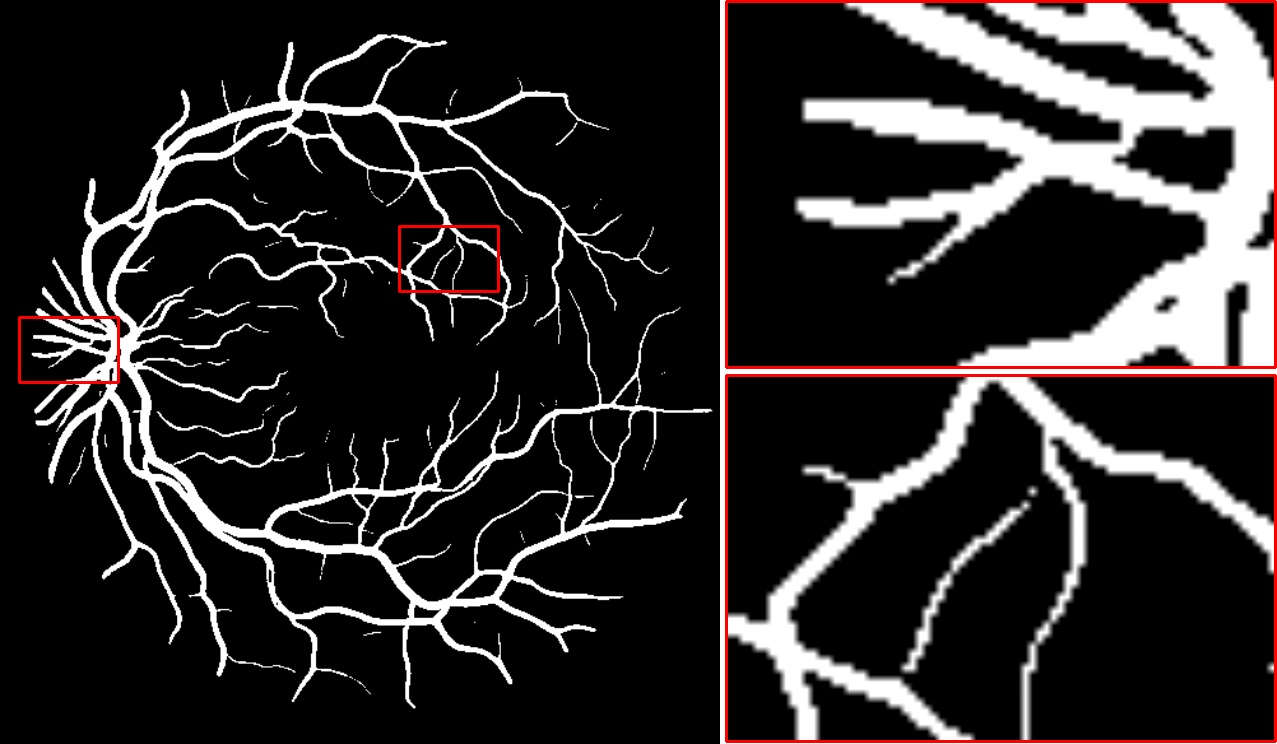}}
            \end{minipage}
            \begin{minipage}[t]{0.148\linewidth}
			\centering
			\centerline{\includegraphics[width=1\linewidth]{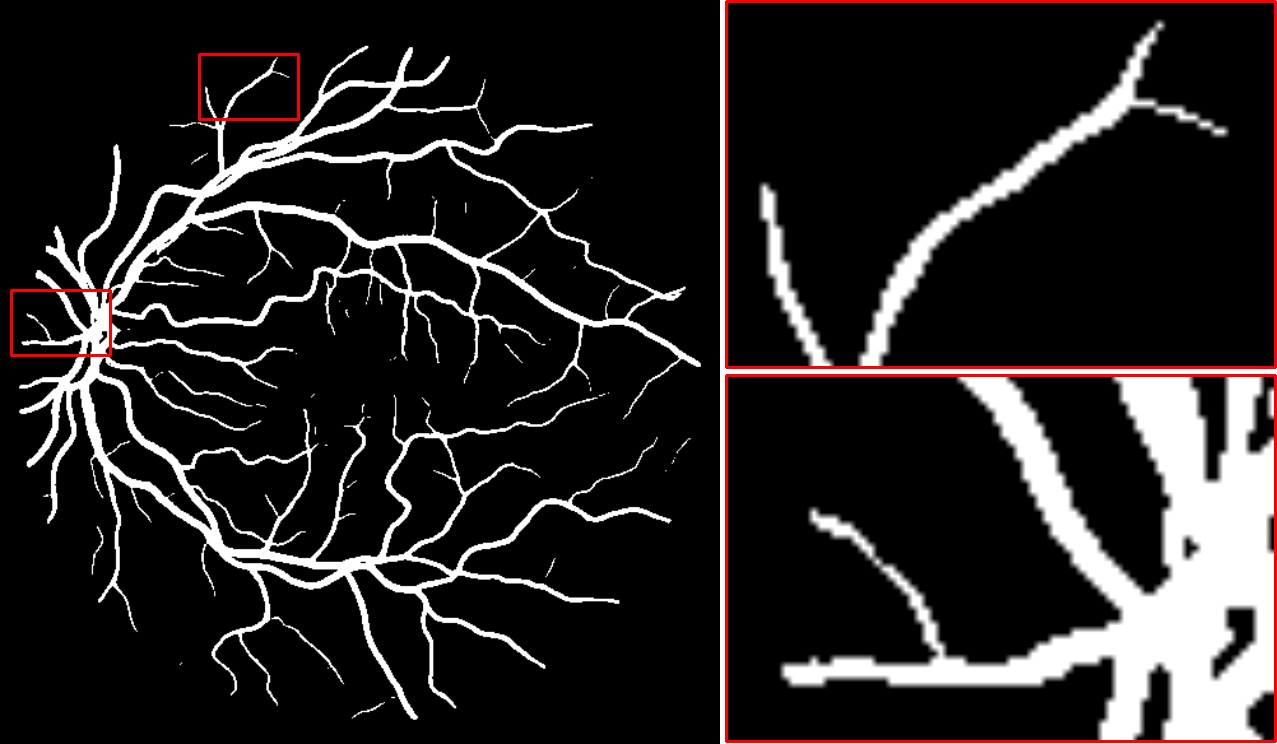}}
            \end{minipage}
            \begin{minipage}[t]{0.154\linewidth}
			\centering
			\centerline{\includegraphics[width=1\linewidth]{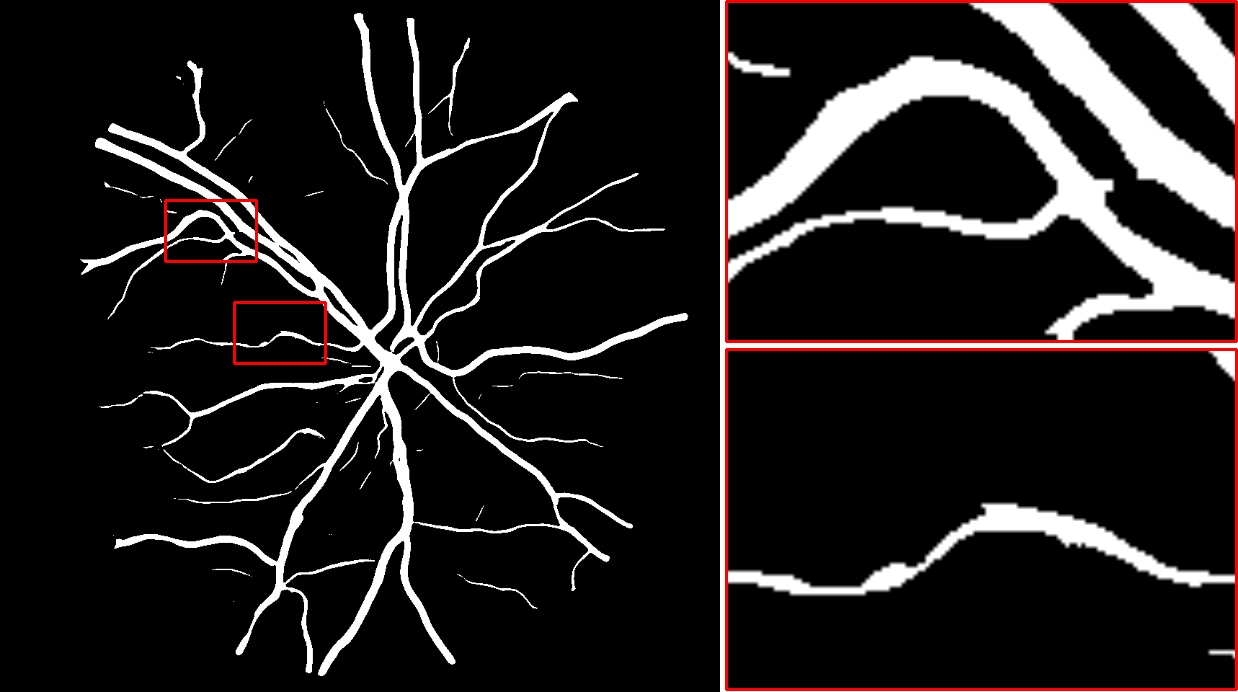}}
            \end{minipage}
            \begin{minipage}[t]{0.154\linewidth}
			\centering
			\centerline{\includegraphics[width=1\linewidth]{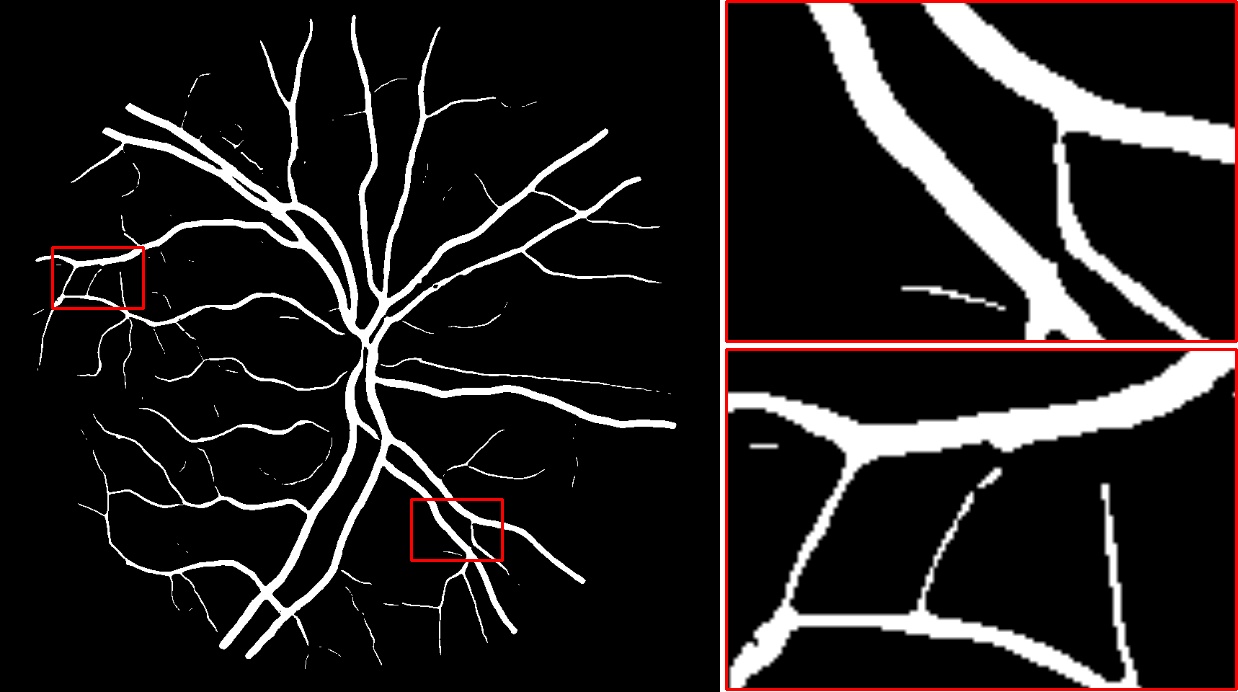}}
            \end{minipage}
            \begin{minipage}[t]{0.1635\linewidth}
			\centering
			\centerline{\includegraphics[width=1\linewidth]{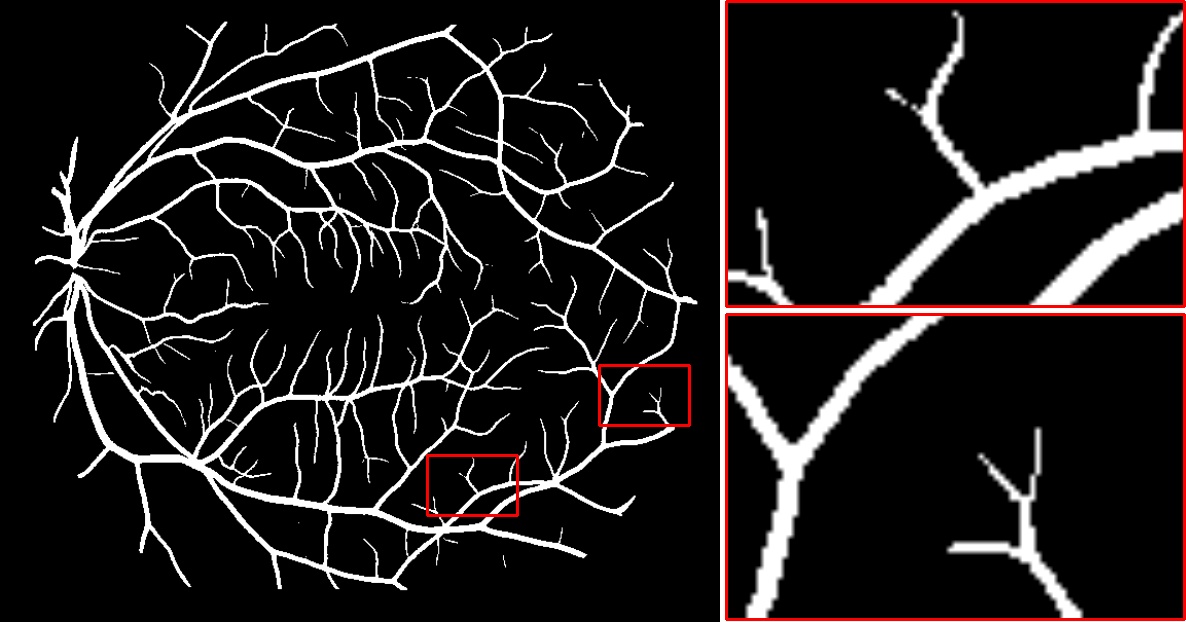}}
            \end{minipage}
            \begin{minipage}[t]{0.1635\linewidth}
			\centering
			\centerline{\includegraphics[width=1\linewidth]{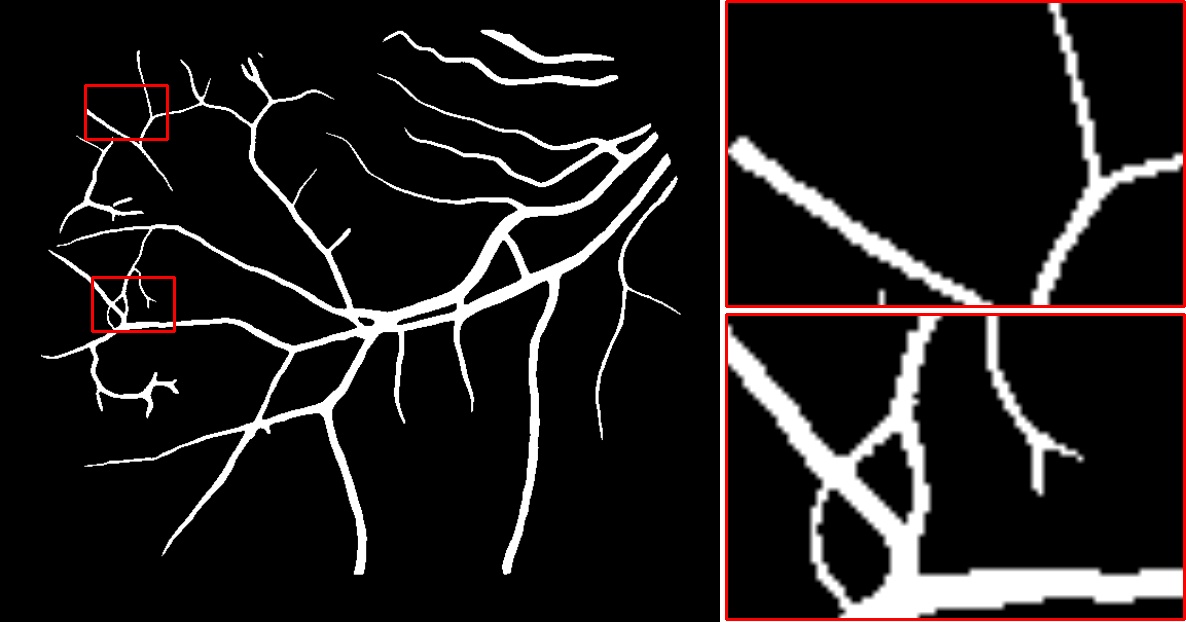}}
            \end{minipage}\\[0.1em]
		\caption{Visualization of vessel segmentation results on DRIVE (rows 1-2),  CHASE\_DB1 (rows 3-4) and STARE (rows 5-6)}
  \label{img: visualization}
\end{figure*}

\begin{figure}[ht]
\begin{tcolorbox}[colback=white, colframe=cyan!30, colbacktitle=white, coltitle=black, boxrule=0.5mm, left=0.05mm, right=0.05mm, sharp corners]
  \vspace{-2mm}
  \scriptsize \textbf{Segmentation of low-contrast tiny vessels} \\
  \begin{minipage}[t]{0.191\linewidth}
			\centering
			\centerline{\includegraphics[width=1\linewidth]{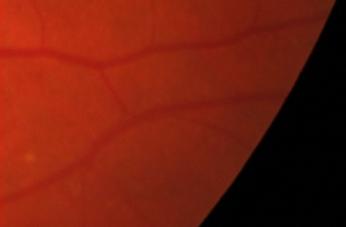}}
		\end{minipage}
			\begin{minipage}[t]{0.191\linewidth}
			\centering
			\centerline{\includegraphics[width=1\linewidth]{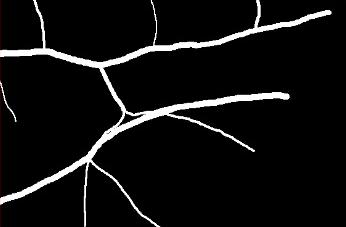}}
		\end{minipage}
			\begin{minipage}[t]{0.191\linewidth}
				\centerline{\includegraphics[width=1\linewidth]{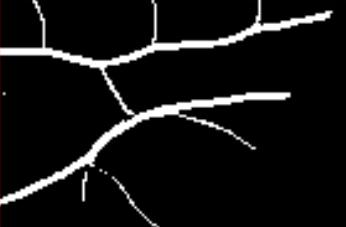}}
			\end{minipage}
				\begin{minipage}[t]{0.191\linewidth}
				\centering
				\centerline{\includegraphics[width=1\linewidth]{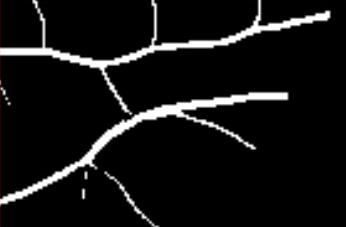}}
			\end{minipage}
                \begin{minipage}[t]{0.191\linewidth}
				\centering
				\centerline{\includegraphics[width=1\linewidth]{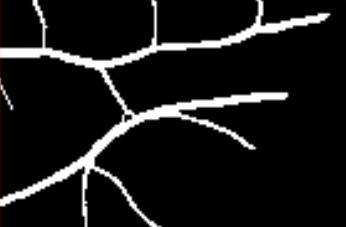}}
			\end{minipage}
   \vspace{-3mm} 
   \\ 
   \begin{minipage}[t]{0.191\linewidth}
			\centering
			\centerline{\includegraphics[width=1\linewidth]{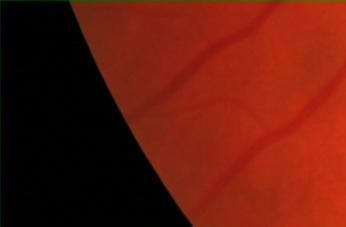}}
			\vspace{-3pt}
		\end{minipage}
			\begin{minipage}[t]{0.191\linewidth}
			\centering
			\centerline{\includegraphics[width=1\linewidth]{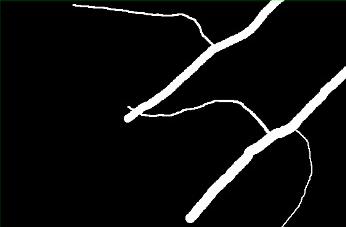}}
			\vspace{-3pt}
		\end{minipage}
			\begin{minipage}[t]{0.191\linewidth}
				\centerline{\includegraphics[width=1\linewidth]{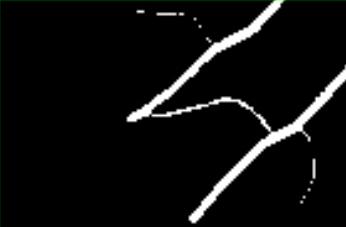}}
				\vspace{-3pt}
			\end{minipage}
				\begin{minipage}[t]{0.191\linewidth}
				\centering
				\centerline{\includegraphics[width=1\linewidth]{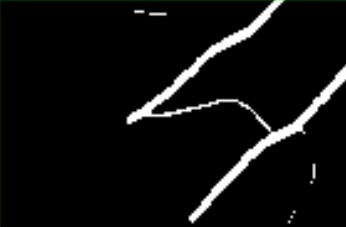}}
				\vspace{-3pt}
			\end{minipage}
                \begin{minipage}[t]{0.191\linewidth}
				\centering
				\centerline{\includegraphics[width=1\linewidth]{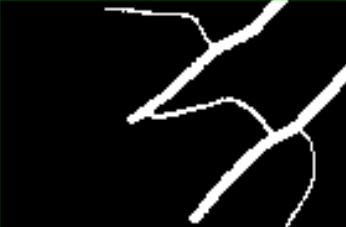}}
				\vspace{-3pt}
			\end{minipage} \\
  \vspace{-6mm}
  \tcblower 
  \vspace{-2mm}
  \scriptsize \textbf{Segmentation in the presence of central vessel reflex} \\
  \begin{minipage}[t]{0.191\linewidth}
			\centering
			\centerline{\includegraphics[width=1\linewidth]{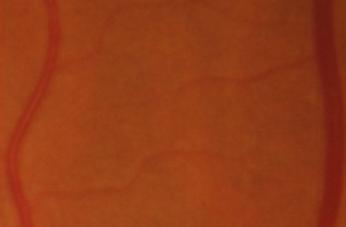}}
		\end{minipage}
			\begin{minipage}[t]{0.191\linewidth}
			\centering
			\centerline{\includegraphics[width=1\linewidth]{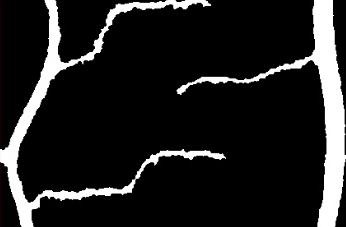}}
		\end{minipage}
			\begin{minipage}[t]{0.191\linewidth}
				\centerline{\includegraphics[width=1\linewidth]{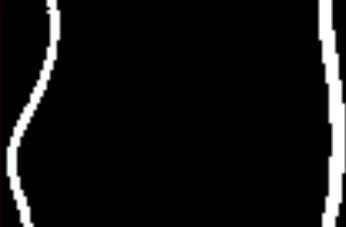}}
			\end{minipage}
				\begin{minipage}[t]{0.191\linewidth}
				\centering
				\centerline{\includegraphics[width=1\linewidth]{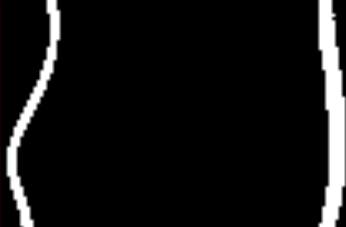}}
			\end{minipage}
                \begin{minipage}[t]{0.191\linewidth}
				\centering
				\centerline{\includegraphics[width=1\linewidth]{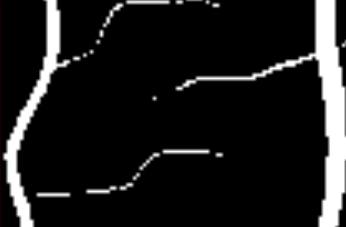}}
			\end{minipage}
   \vspace{-3mm} 
   \\ 
   \begin{minipage}[t]{0.191\linewidth}
			\centering
			\centerline{\includegraphics[width=1\linewidth]{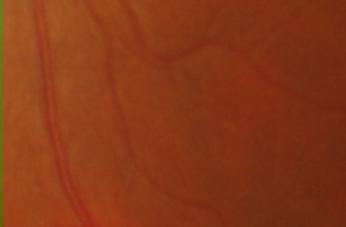}}
			\vspace{-3pt}
                \centerline{\footnotesize Images}
		\end{minipage}
			\begin{minipage}[t]{0.191\linewidth}
			\centering
			\centerline{\includegraphics[width=1\linewidth]{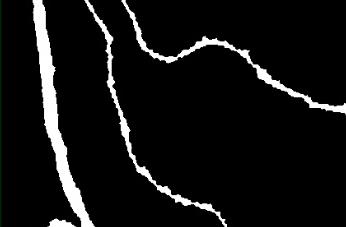}}
			\vspace{-3pt}
                \centerline{\footnotesize GT}
		\end{minipage}
			\begin{minipage}[t]{0.191\linewidth}
			\centerline{\includegraphics[width=1\linewidth]{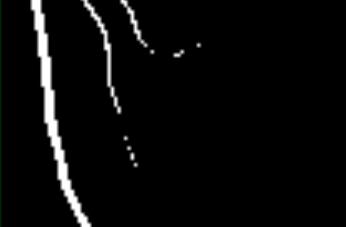}}
				\vspace{-3pt}
                \centerline{\footnotesize DUNet}
			\end{minipage}
				\begin{minipage}[t]{0.191\linewidth}
				\centering
			\centerline{\includegraphics[width=1\linewidth]{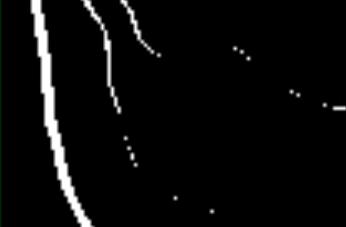}}
				\vspace{-3pt}
                    \centerline{\footnotesize BSEResU-Net}
			\end{minipage}
                \begin{minipage}[t]{0.191\linewidth}
				\centering
			\centerline{\includegraphics[width=1\linewidth]{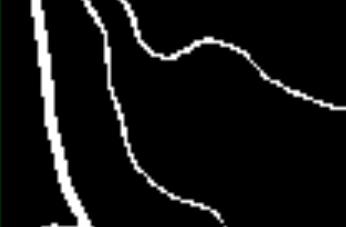}}
				\vspace{-3pt}
                \centerline{\footnotesize Ours}
			\end{minipage}
    \vspace{-3mm}
\end{tcolorbox}
\caption{Detailed comparison of fundus vessel segmentation on the HRF dataset for challenging cases. The first to fifth columns show: fundus image patches, the ground truth, segmentation results generated by DUNet, segmentation results generated by BSEResU-Net, and our results}
\label{img: visualization2}
\end{figure}

Visual comparisons in Figure \ref{img: visualization} highlight our method's superior capability in detecting tiny vessels and resolving topological issues like under-segmentation, outperforming benchmark models like U-Net and addressing limitations in other methods such as R2U-Net and CS-Net. Figure \ref{img: visualization2} further illustrates our method's effectiveness in challenging scenarios, including low-contrast vessels and central vessel reflex, where it accurately identifies intricate vascular details that other methods struggle to capture. Overall, our approach exhibits robust performance across multiple datasets, confirming its efficacy and broad applicability in retinal vessel segmentation.

\begin{table*}[htbp]
\centering
    \caption{The best and second-best results are bold-faced and underlined, respectively.}
    \begin{tabular}{lcccccccc}
    \toprule
        Methods & Year & Param (M) & ACC & SE & SP & F1 & AUC \\ \midrule
        U-Net & 2015 & 31.52 & 0.9628 & 0.7533 & 0.9836 & 0.7943 & 0.9836 \\
        DUNet & 2018 & 0.88 & 0.9641 & 0.7595 & \underline{0.9878} & 0.8143 & 0.9832 \\ 
        Yan et al. & 2018 & 31.35 & 0.9612 & 0.7581 & 0.9846 & - & 0.9801 \\ 
        CS-Net & 2019 & 8.92 & 0.9752 & \textbf{0.8816} & 0.9840 & - & \underline{0.9932} \\ 
        R2U-Net & 2019 & 1.07 & \textbf{0.9862} & 0.8298 & 0.9813 & \underline{0.8475} & 0.9914 \\
        SA-UNet & 2020 & 0.54 & 0.9523 & 0.7328 & 0.9830 & 0.7816 & 0.9612 \\
        CS$^2$-Net & 2021 & 5.96 & 0.9752 & \underline{0.8816} & 0.9840 & - & 0.9932 \\
        FR-UNet & 2022 & 5.72 & 0.9752 & 0.8327 & 0.9869 & 0.8330 & 0.9914 \\
        EXP-Net & 2023 & 43.37 & 0.9677 & 0.8274 & 0.9805 & 0.8313 & 0.9897 \\
        Ours & 2024 & 0.84 & \underline{0.9805} & 0.8695 & \textbf{0.9897} & \textbf{0.8692} & \textbf{0.9933} \\
    \bottomrule
    \end{tabular}
\label{tab: result2}
\end{table*}

\begin{table*}[htbp]
    \caption{The best and second-best results are bold-faced and underlined, respectively.}
    \centering
    \begin{tabular}{lcccccccc}
    \toprule
    Methods & Year & Param (M) & ACC & SE & SP & F1 & AUC \\ \midrule
    DUNet & 2018 & 0.88 & 0.9651 & 0.7464 & \textbf{0.9874} & - & 0.9831 \\ 
    U-Net++ & 2020 & 9.00 & 0.9654 & 0.7939 & 0.9836 & \underline{0.7993} & 0.9823 \\ 
    BSEResU-Net & 2021 & 1.47 & 0.9637 & 0.8067 & 0.9796 & 0.8044 & 0.9837 \\ 
    Li et al. & 2022 & 26.08 & \underline{0.9698} & \underline{0.8178} & 0.9828 & - & \underline{0.9853} \\ 
    Ours & 2024 & 0.84 & \textbf{0.9709} & \textbf{0.8195} & \underline{0.9838} & \textbf{0.8123} & \textbf{0.9885} \\
    \bottomrule
    \end{tabular}
\label{tab: result3}
\end{table*}

\textbf{Comparisons With NAS Methods}.\label{sec4-4}
We analyze our results in comparison with recent NAS models on the same retina vessel segmentation task. To the best of our knowledge, the existing literature on NAS for vascular segmentation remains limited, encompassing only a few models such as \citep{popat2020ga}, HNAS~\citep{houreh2021hnas}, BTU-Net~\citep{rajesh2023evolutionary}, Genetic U-Net~\citep{wei2021genetic}, and MedUNAS~\citep{kucs2023evolutionary}. Here, we compare with all of them based on results reported in the original literature. Table \ref{tab: result4} summarizes the segmentation performance and model complexity of these methods. Results show that our method achieves superior segmentation performance in SE, F1, and AUC than HNAS and BTU-Net while using significantly fewer parameters-approximately 200 times fewer than HNAS v1 and 3 times fewer than BTU-Net. In terms of the ACC metric, our model ranks just below MedUNAS and Genetic U-Net v2. In brief, our method is capable of producing excellent segmentation results with a relatively low degree of model complexity.

\begin{table*}[htbp]
    \caption{Comparison of NAS studies and the proposed methods on the DRIVE dataset. The best and the second best results are bold-faced and underlined, respectively.}
    \centering
    \begin{tabular}{lccccccc}
    \toprule
        Methods & Param (M) & ACC & SE & SP & F1 & AUC \\ \midrule
        popat et al. v1 & 0.18 & 0.9356 & 0.5967 & \underline{0.9850} & - & 0.9465 \\ 
        popat et al. v2 & 8.10 & 0.9534 & 0.7501 & 0.9831 & - & 0.9751 \\
        HNAS v1 & 157 & 0.9542 & 0.7707 & 0.9810 & 0.8108 & 0.9704 \\ 
        HNAS v2 & 2.90 & 0.9546 & 0.7744 & 0.9809 & 0.8129 & 0.9749 \\
        BTU-Net &16.80 & 0.9689 & - & 0.9844 & 0.8178 & 0.9030 \\
        Genetic U-Net v1 & 0.27 & 0.9577 & 0.8300 & 0.9758 & \underline{0.8314} & 0.9823 \\ 
        Genetic U-Net v2 & 0.27 & \underline{0.9707} & 0.8300 & 0.9843 & 0.8314 & \underline{0.9885} \\
        MedUNAS v1 & 2.32 & \textbf{0.9711} & \underline{0.8454} & \textbf{0.9864} & 0.8206 & - \\ 
        MedUNAS v2 & 1.27 & 0.9706 & 0.8341 & 0.9836 & 0.8218 & - \\
        Ours & 0.84 & 0.9701 & \textbf{0.8489} & 0.9825 & \textbf{0.8334} & \textbf{0.9901} \\
    \bottomrule
    \end{tabular}
\label{tab: result4}
\end{table*}

\begin{figure}[htbp]
    \centering
    \includegraphics[width=1\columnwidth]{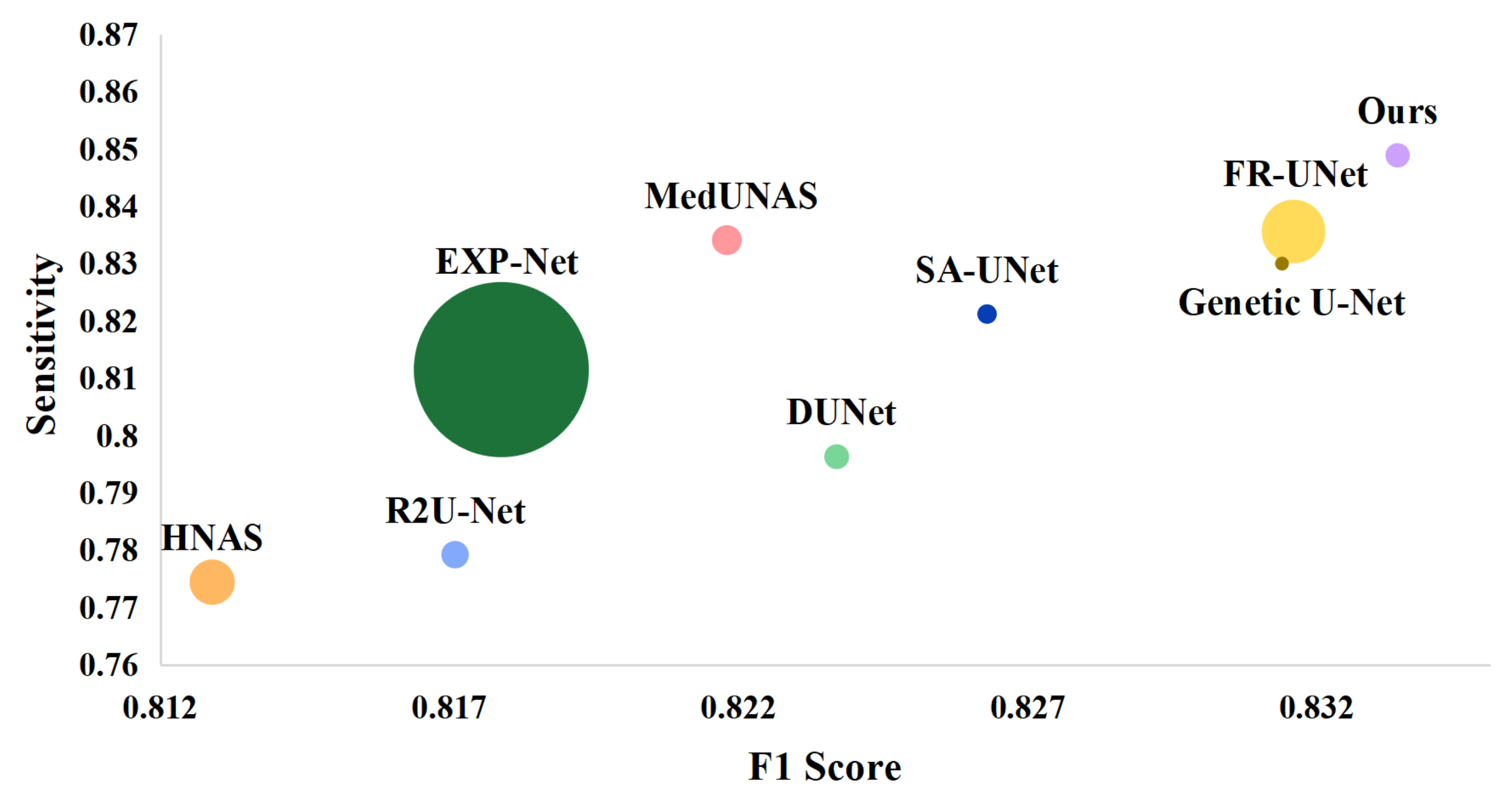}
    \vspace{-0.3cm}
    \caption{Comparison of the number of parameters (size of each colored circle) and performance between our method and other state-of-the-art methods on DRIVE}
    \label{fig:para}
    \end{figure}

\textbf{Parametric Efficiency}.
We compare our method with others on the DRIVE dataset, examining the number of parameters as well as the primary metrics (F1 and SE) used in this study. As shown in Figure \ref{fig:para}, our model demonstrates a clear advantage over other state-of-the-art models in terms of both F1 and SE. Specifically, our model achieves the highest SE, surpassing the nearest competitor FR-UNet by a significant margin, while also maintaining a competitive F1. This balance between SE and F1 highlights the robustness of our approach, particularly when compared to MedUNAS and DUNet, which show lower SE. Additionally, the small size of the colored circle indicates a moderate parameter count, striking a good balance between performance and complexity. Overall, these results highlight the superior generalization capability of our model.

\begin{figure*}[htbp]
			\begin{minipage}[t]{0.33\linewidth}
			\centering
			\vspace{1pt}
			\centerline{\includegraphics[width=1\linewidth]{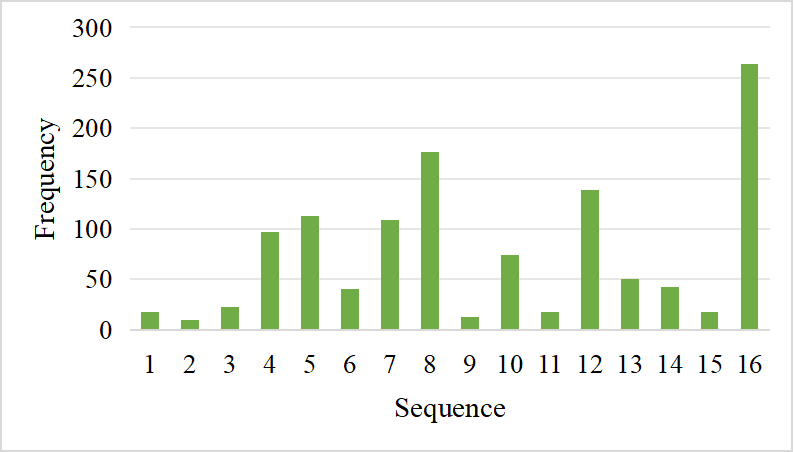}}
			\vspace{-3pt}
			\centerline{\small (a) Depth=2}
		\end{minipage}
			\begin{minipage}[t]{0.33\linewidth}
			\centering
			\vspace{1pt}
			\centerline{\includegraphics[width=1\linewidth]{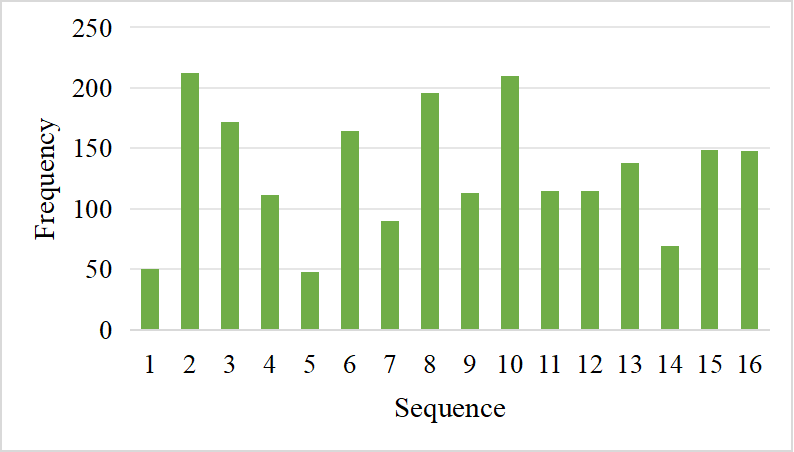}}
			\vspace{-3pt}
			\centerline{\small (b) Depth=3}
		\end{minipage}
			\begin{minipage}[t]{0.33\linewidth}
				\centering
				\vspace{1pt}
				\centerline{\includegraphics[width=1\linewidth]{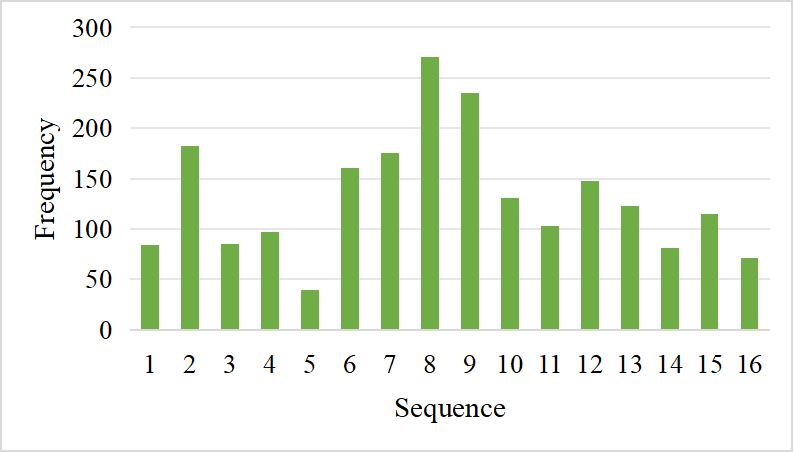}}
				\vspace{-3pt}
				\centerline{\small (c) Depth=4}
		\end{minipage}
		\caption{Frequency of each operation sequence for different network depths}
    \label{fig:fre4}
\end{figure*}

\textbf{Frequency of Operation Sequences}.
We examine a total of 16 node sequence operations, with their detailed compositions and encoding representations presented in Table \ref{tab: operations}. When analyzing the sequences for different U-Net depths, we find that certain configurations are frequently selected during the search, indicating their effectiveness in retinal vessel segmentation tasks. Figure \ref{fig:fre4} shows the top configurations from the search process, along with the frequency of each operation sequence. For a U-Net with a depth of 2, the two most frequent sequences are Sequence 16 (Frequency: 271) and Sequence 8 (Frequency: 235). Sequence 16 employs instance normalization, a 5×5 convolutional kernel, and quadratic neurons. Sequence 8 employs a 3×3 convolutional kernel with quadratic neurons. Both suggest quadratic neurons can capture more spatial information. For U-Net with a depth of 3, the most frequent sequences are Sequence 2 (Frequency: 212) and Sequence 10 (Frequency: 210). Sequence 2 employs a 3×3 convolutional kernel with quadratic neurons and no normalization, utilizing pre-activation. Sequence 10 is similar but uses a larger 5×5 kernel. The presence of quadratic neurons in both of the top sequences strongly suggests their effectiveness in feature learning. For a U-Net with a depth of 4, the most frequent sequences are Sequence 8 (Frequency: 271) and Sequence 9 (Frequency: 235). Notably, Sequence 8 features quadratic neurons, consistent with networks with smaller depths, further highlighting their superiority.

These findings suggest that combining quadratic neurons is a highly effective setup for retinal vessel segmentation tasks, particularly in combination with instance normalization. It is discovered in~\citep{r35} that using quadratic neurons in a deep network has the degree explosion, \textit{i.e.}, given ReLU activation, a quadratic network of $L$ layers is a piecewise polynomial of degree $2^{2^L}$. We think that the reason why quadratic neurons favor instance normalization is instance normalization can inhibit the magnitude of features, thereby facilitating the feature extraction of quadratic neurons.

\begin{figure*}[htbp]
			\begin{minipage}[t]{0.33\linewidth}
			\centering
			\vspace{1pt}
			\centerline{\includegraphics[width=1\linewidth]{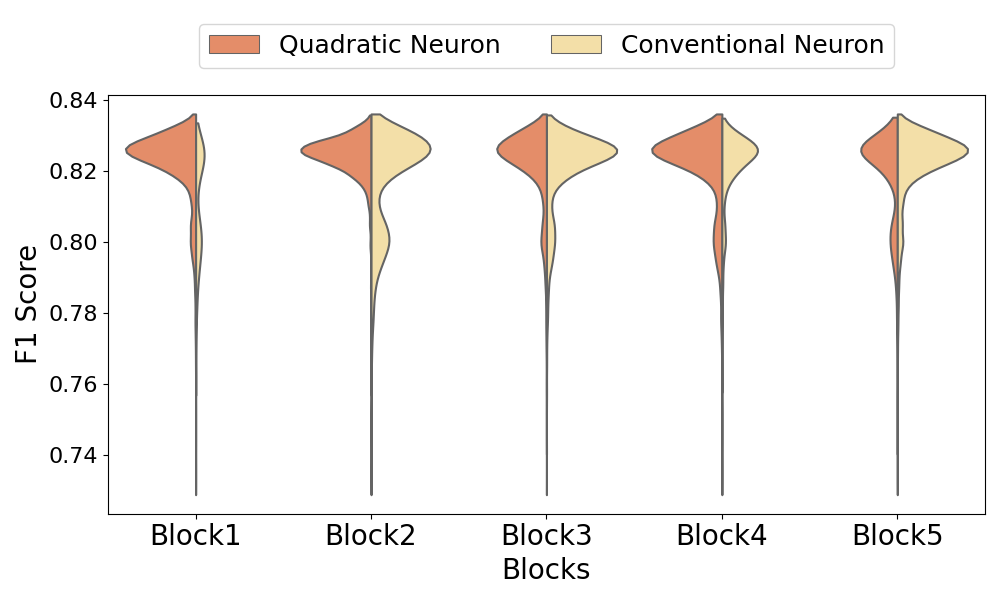}}
			\vspace{-3pt}
			\centerline{\small (a) Depth=2}
		\end{minipage}
			\begin{minipage}[t]{0.33\linewidth}
			\centering
			\vspace{1pt}
			\centerline{\includegraphics[width=1\linewidth]{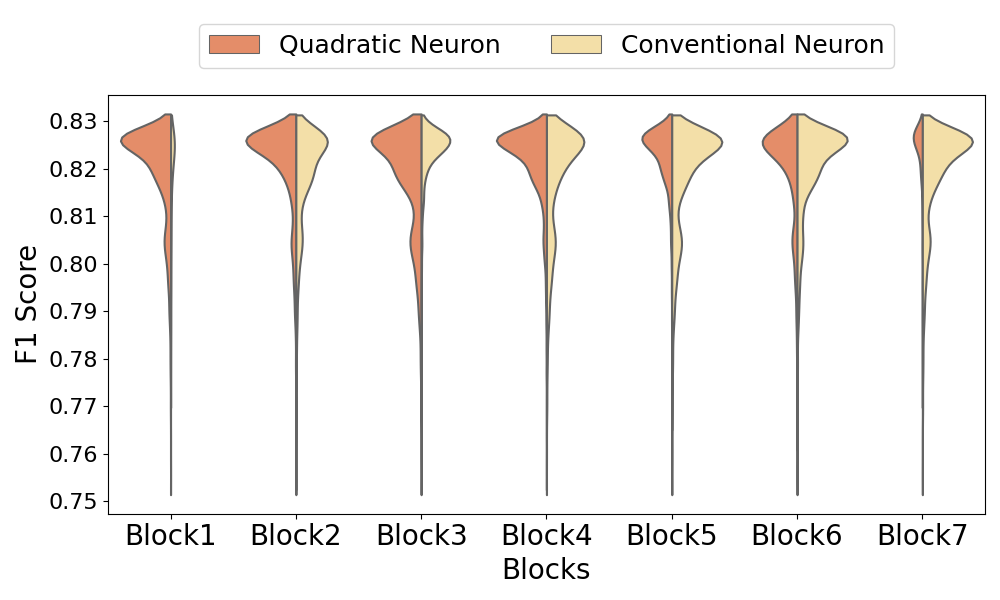}}
			\vspace{-3pt}
			\centerline{\small (b) Depth=3}
		\end{minipage}
			\begin{minipage}[t]{0.33\linewidth}
				\centering
				\vspace{1pt}
				\centerline{\includegraphics[width=1\linewidth]{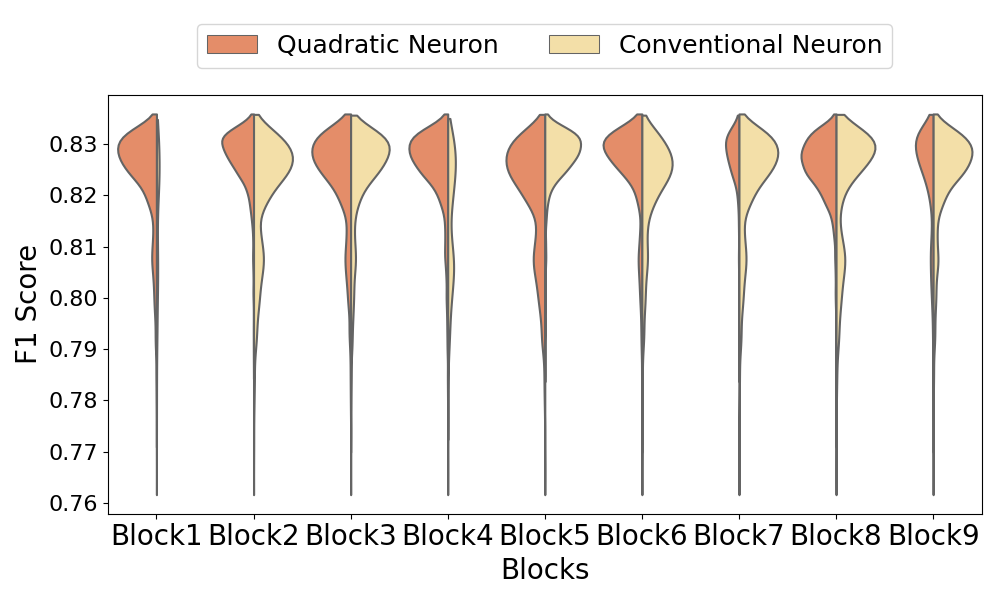}}
				\vspace{-3pt}
				\centerline{\small (c) Depth=4}
		\end{minipage}
		\caption{The impact of neuron type on the F1 for different network depths. The x-axis represents network blocks, with two colors indicating different neuron types. The y-axis shows the F1, reflecting performance across varying neuron distributions}
  \label{img: visualization4}
\end{figure*}

\textbf{The Efficacy of Quadratic Neurons}.
We evaluate the performance of different neuron types across encoder values (2, 3, 4) by analyzing their F1. Figure \ref{img: visualization4} presents the distribution of F1, highlighting the consistent performance of quadratic neurons, with F1 remaining above 0.8 in Block 5 when the encoder value is set to 3. This concentration of high scores indicates a high degree of stability and effectiveness across different network configurations. Additionally, the narrow and high score distribution of quadratic neurons suggests minimal variation and stable performance. In contrast, conventional neurons exhibit a broader distribution range in Block 5. While some configurations achieve high F1, there are also instances with lower scores, indicating higher variability. This suggests that conventional neurons are less stable in performance compared to quadratic neurons in Block 5, with their variability indicating a trade-off between high scores and consistency.

\begin{figure*}[htbp]
    \begin{minipage}[t]{0.33\linewidth}
    \centering
    \vspace{1pt}
    \centerline{\includegraphics[height=4cm]{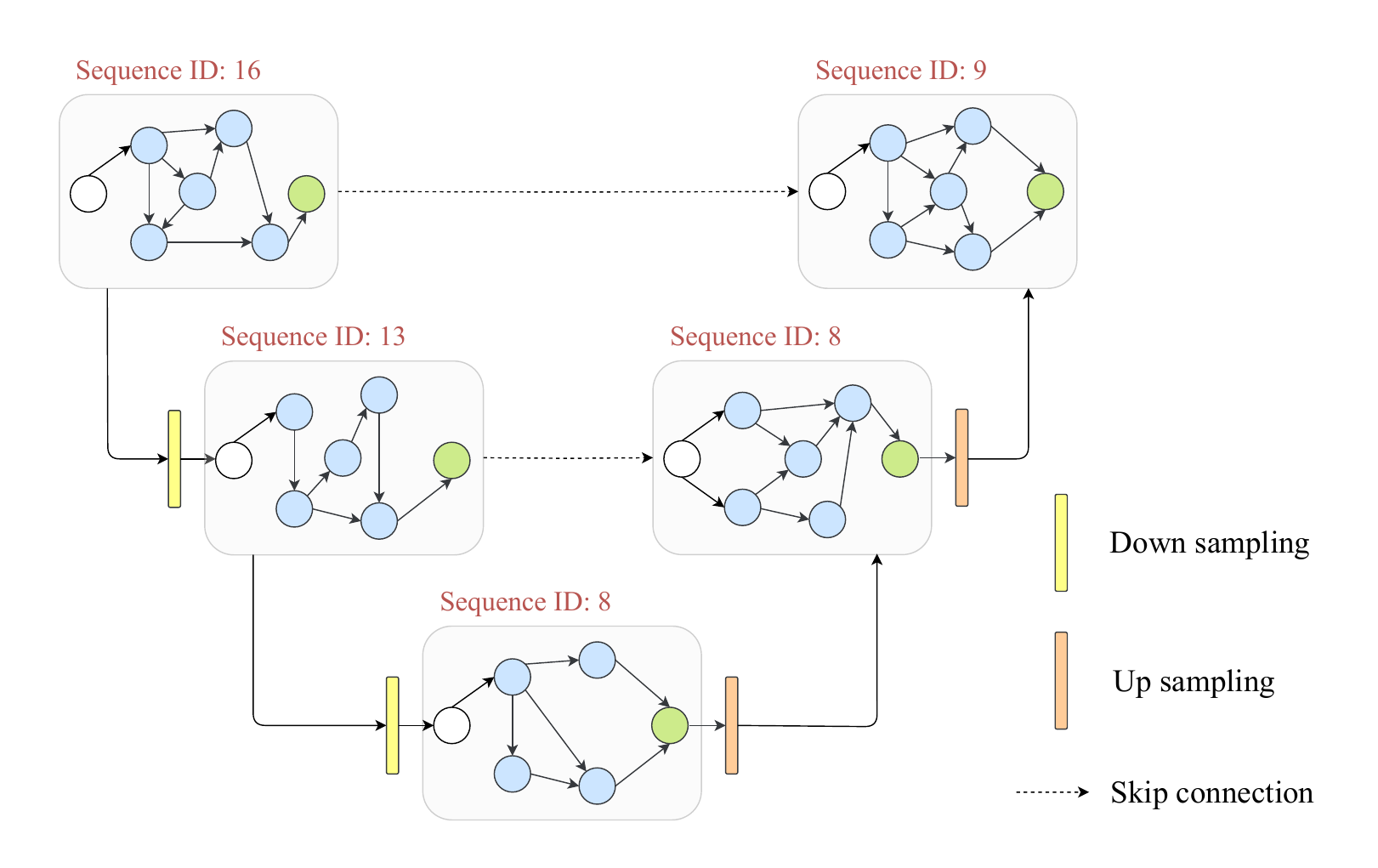}}  
    \vspace{-3pt}
    \centerline{\small (a) Depth=2, Channel=8}
    \end{minipage}
    \begin{minipage}[t]{0.33\linewidth}
    \centering
    \vspace{1pt}
    \centerline{\includegraphics[height=4cm]{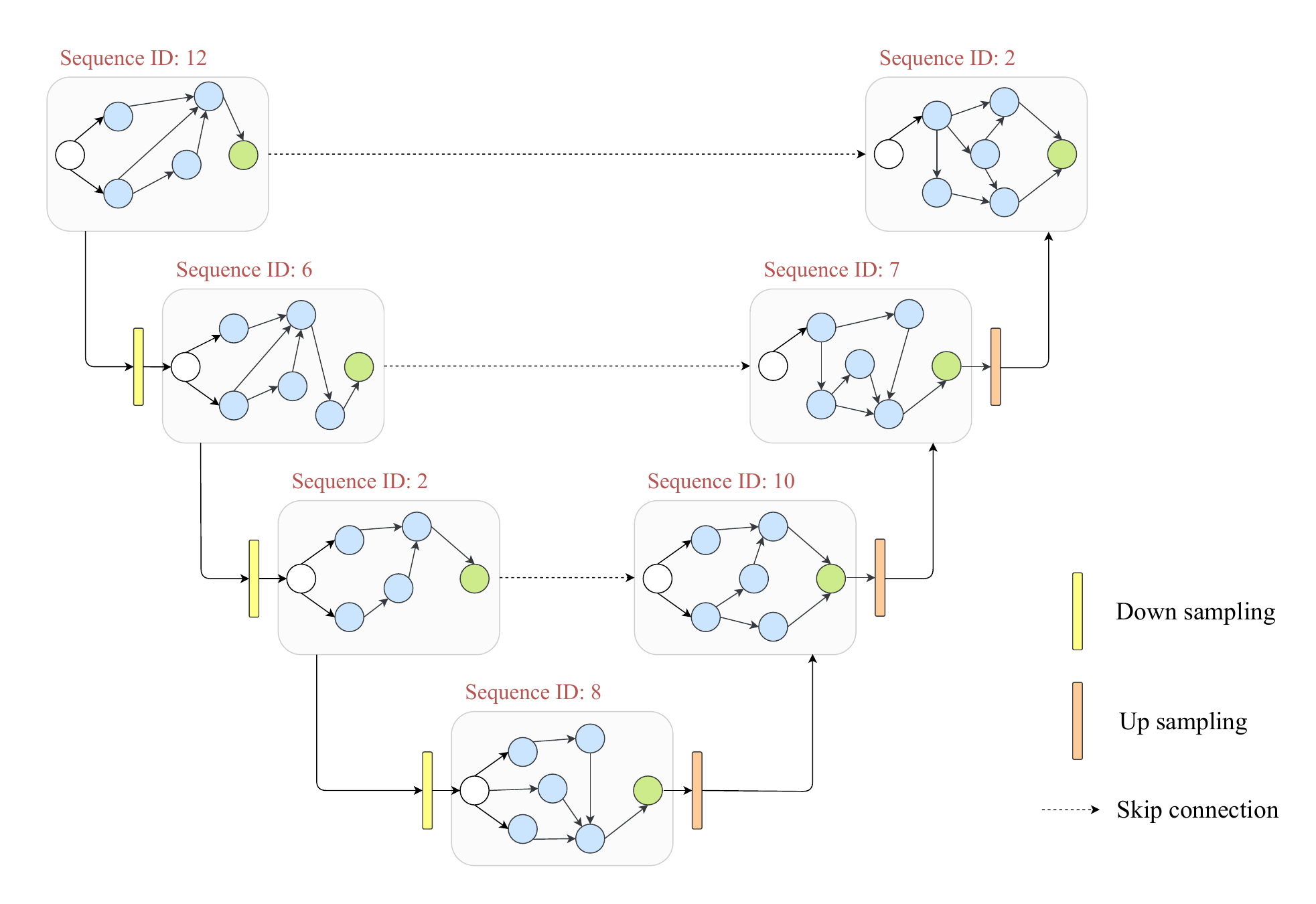}}  
    \vspace{-3pt}
    \centerline{\small (b) Depth=3, Channel=18}
    \end{minipage}
    \begin{minipage}[t]{0.33\linewidth}
    \centering
    \vspace{1pt}
    \centerline{\includegraphics[height=4cm]{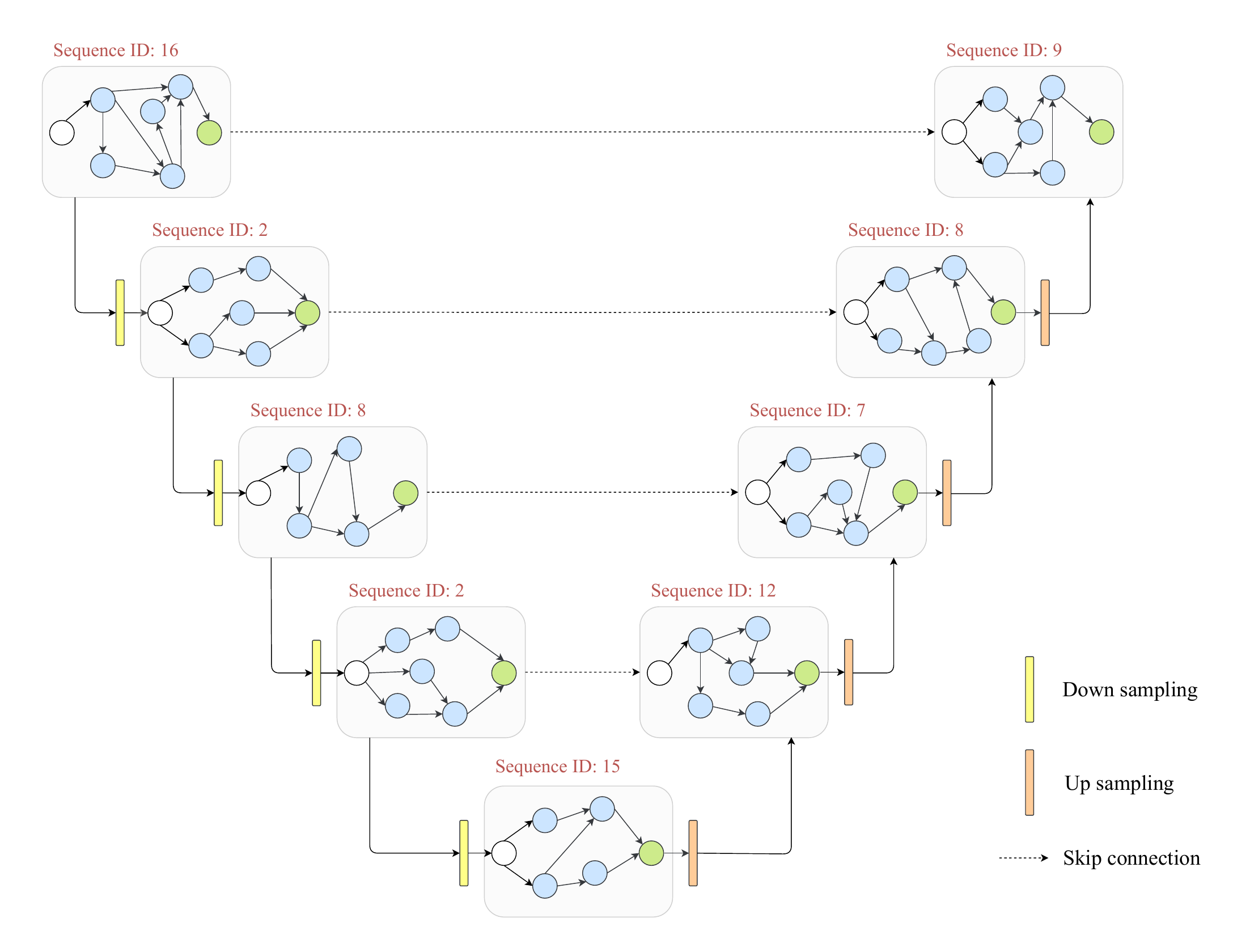}}  
    \vspace{-3pt}
    \centerline{\small (c) Depth=4, Channel=28}
    \end{minipage}
    \caption{Visualization of the best architectures for different network depths obtained from the joint search}
    \label{fig:vis_decode}
\end{figure*}

Here, we present a detailed visualization of the networks discovered through our joint search process for U-Net models with depths of 2, 3, and 4 in Figure \ref{fig:vis_decode}. These visualizations highlight the specific structures and configurations of the top-performing architectures identified during the search, offering insights into the connection and operation types within each network. Particularly, the network structure demonstrates strong integration as the depth increases. At a depth of 2, the neuron types predominantly consist of quadratic neurons, alternating with conventional neurons, which helps the model quickly learn complex features. As the depth increases to 3, almost all layers utilize quadratic neurons, with only the last layer using conventional neurons. At a depth of 4, quadratic neurons still dominate, resulting in a more uniform distribution that enhances inter-layer information transfer and the richness of feature extraction.

\subsection{Plug-and-play Search}
We also explore the plug-and-play neuron programming strategy to assess their impact on the final model performance. In the plug-and-play search, the architecture is optimized first, followed by an independent optimization of the neurons. 

\begin{figure}[htbp]
    \centering
    \includegraphics[width=1\columnwidth]{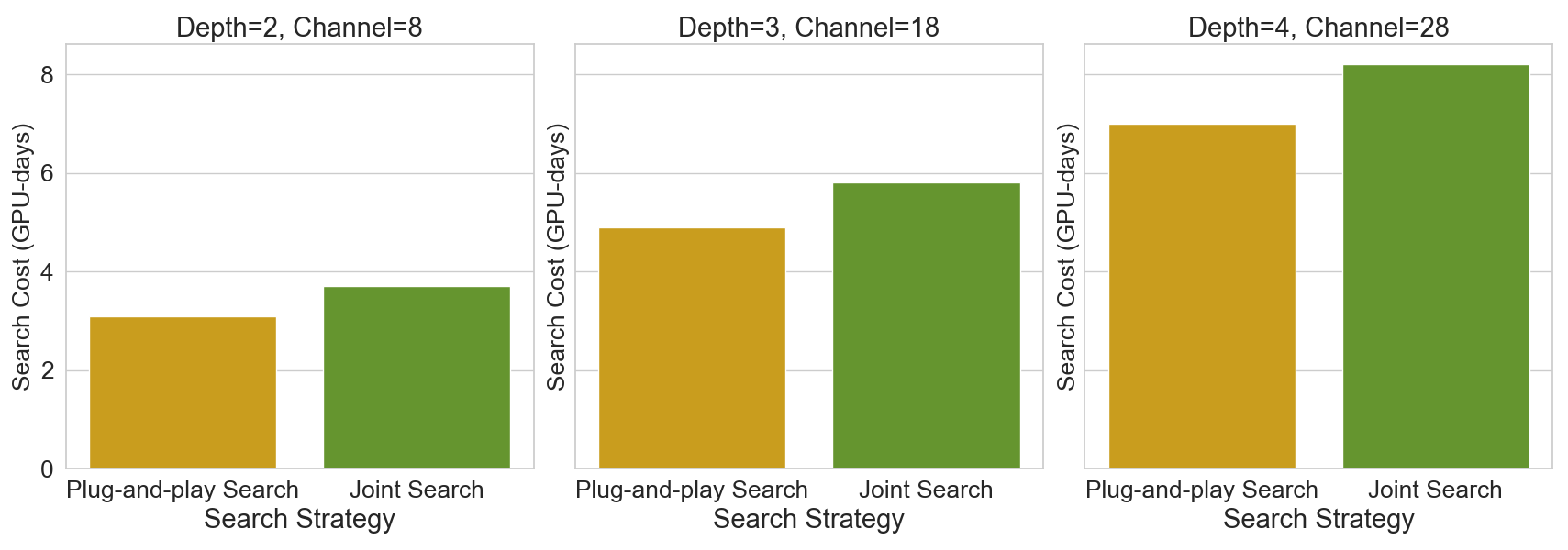}
    \vspace{-0.3cm}
    \caption{Comparison of search costs for different search strategies}
    \label{fig:search_cost}
    \end{figure}
    
\begin{figure*}[htbp]
			\begin{minipage}[t]{0.33\linewidth}
			\centering
			\vspace{1pt}
			\centerline{\includegraphics[width=1\linewidth]{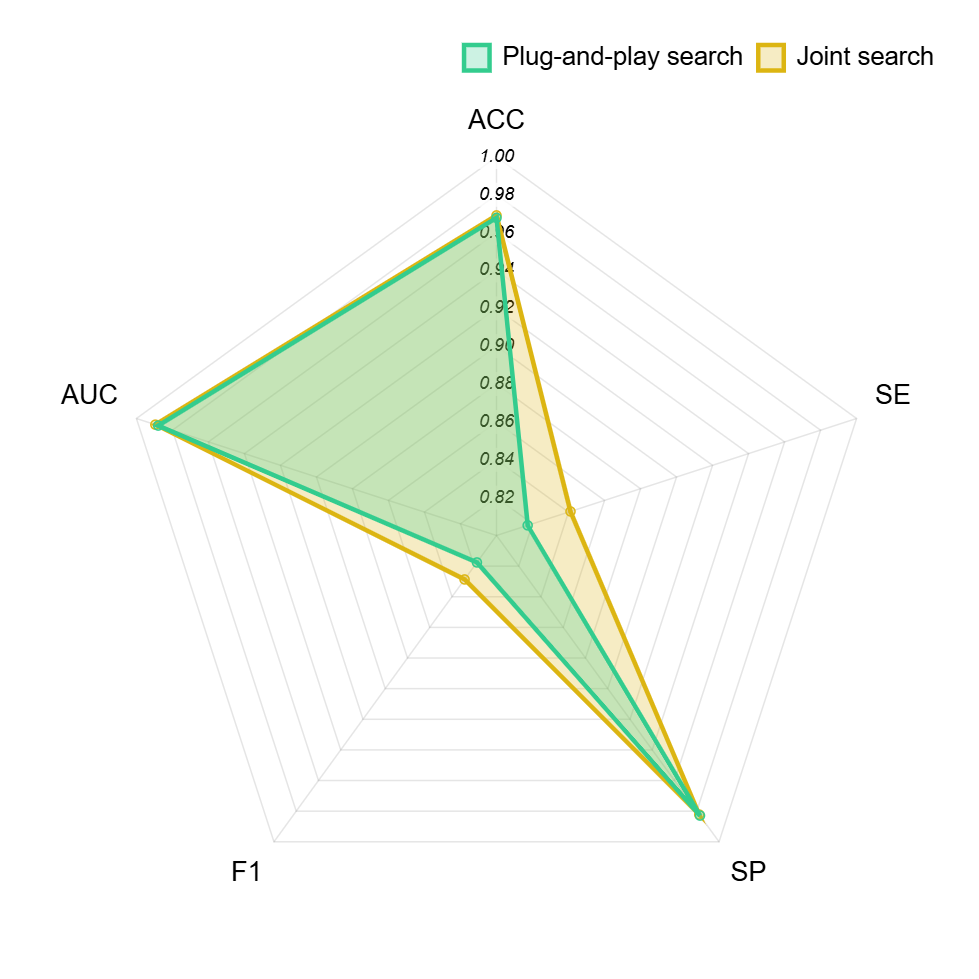}}
			\vspace{-3pt}
			\centerline{\small (a) Depth=2, Channel=8}
		\end{minipage}
			\begin{minipage}[t]{0.33\linewidth}
			\centering
			\vspace{1pt}
			\centerline{\includegraphics[width=1\linewidth]{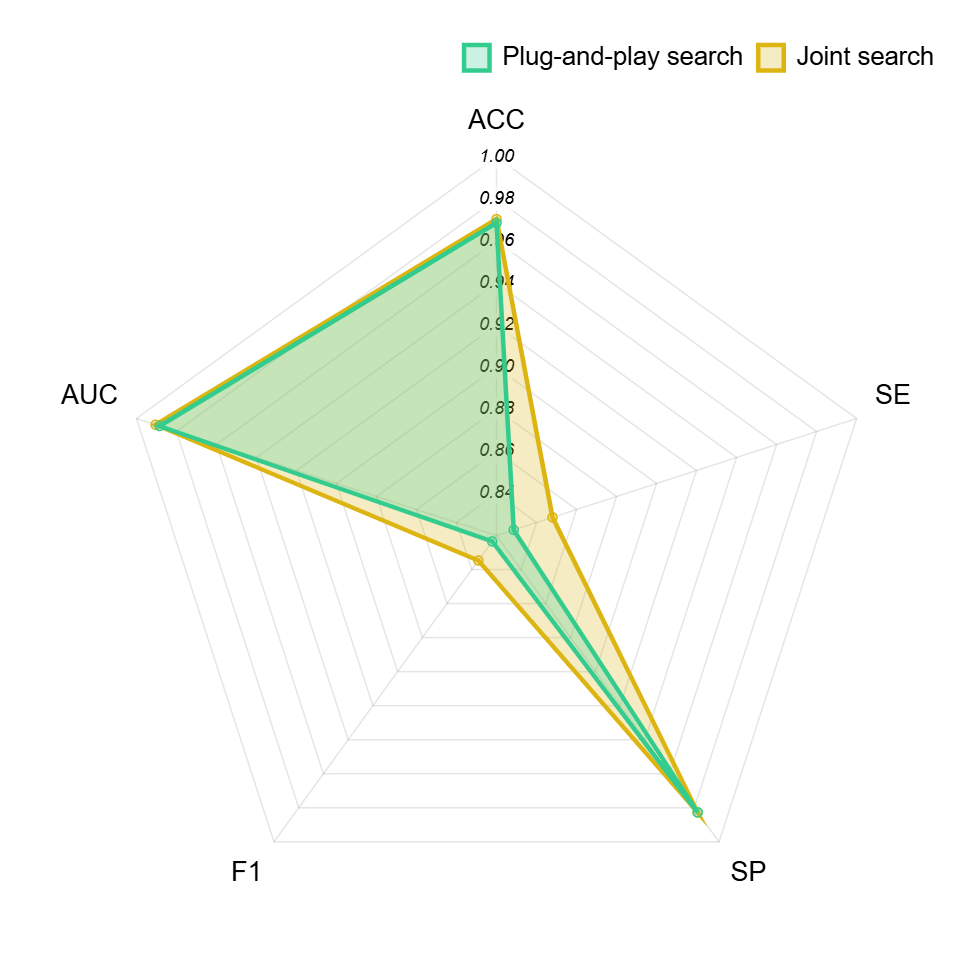}}
			\vspace{-3pt}
			\centerline{\small (b) Depth=3, Channel=18}
		\end{minipage}
			\begin{minipage}[t]{0.33\linewidth}
				\centering
				\vspace{1pt}
				\centerline{\includegraphics[width=1\linewidth]{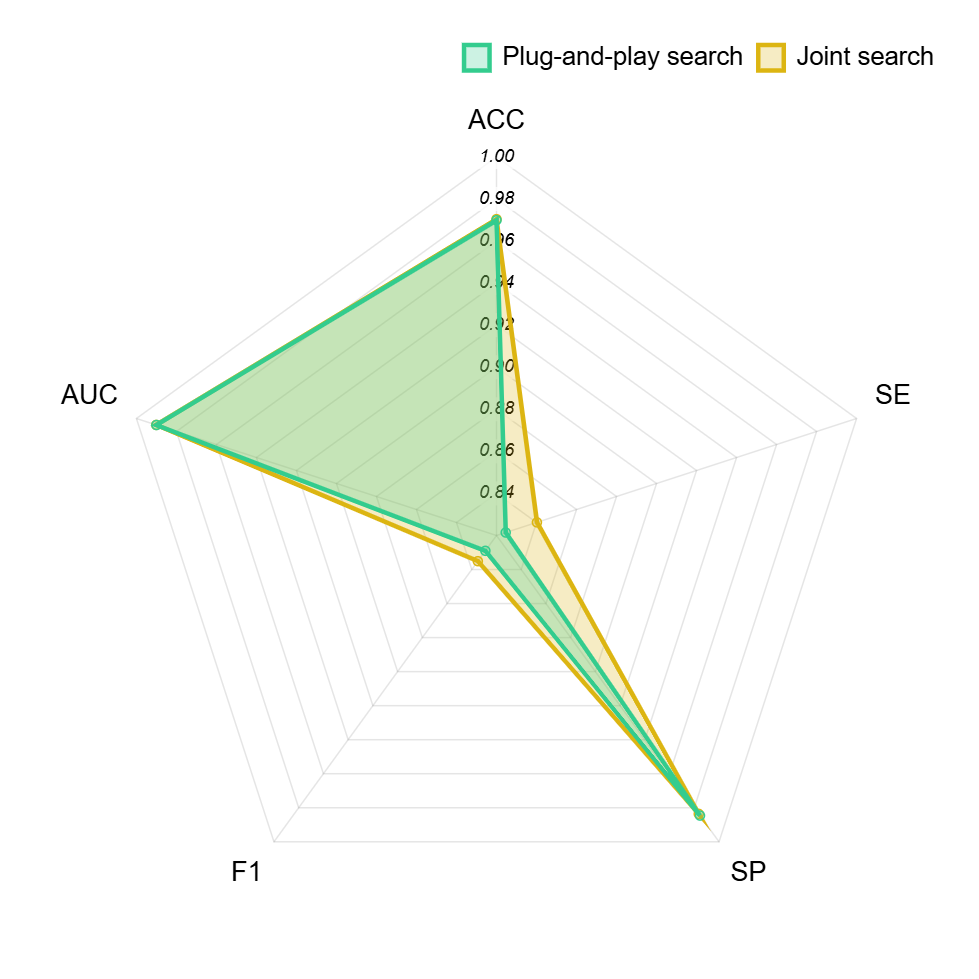}}
				\vspace{-3pt}
				\centerline{\small (c) Depth=4, Channel=28}
		\end{minipage}
		\caption{The performance comparison of the plug-and-play and the joint search strategies in neuron programming}
  \label{fig: nas_compara}
\end{figure*}

Figure \ref{fig:search_cost} presents a comparison of search costs. While plug-and-play search does not achieve the same level of performance as joint search, it requires significantly fewer resources due to its sequential optimization process. As illustrated in Figure \ref{fig: nas_compara}, the final results indicate that joint search offers a distinct advantage. In this approach, the optimization of both architecture and neurons can facilitate a more comprehensive leverage of their interactions. Consequently, this method significantly outperforms the plug-and-play search in key performance metrics, showcasing superior overall performance.

Nevertheless, we don't claim that the plug-and-play search cannot scale. Actually, compared to other methods, the plug-and-play search performs excellently across multiple metrics. In terms of ACC and AUC, it is almost on par with joint search. In SP and F1, it outperforms several compared methods, such as Popat et al. and HNAS, and is close to the performance of Genetic U-Net v2. In brief, 
users can weigh their own needs and available resources when selecting between these methods. For projects prioritizing resource efficiency, the plug-and-play search may be preferable, whereas joint search is ideal for those requiring the highest model performance.

\begin{figure*}[htbp]
    \begin{minipage}[t]{0.33\linewidth}
    \centering
    \vspace{1pt}
    \centerline{\includegraphics[height=4cm]{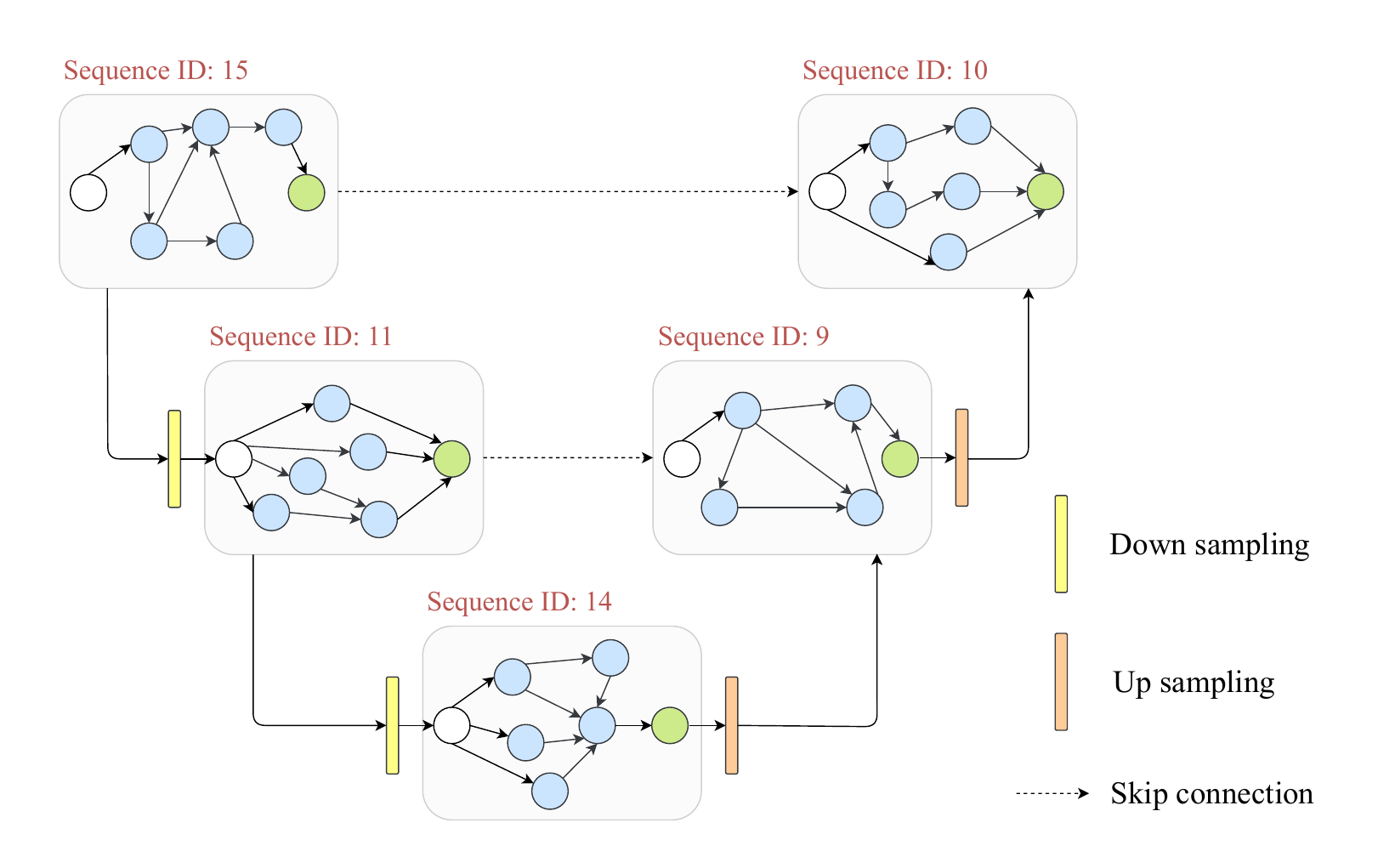}}  
    \vspace{-3pt}
    \centerline{\small (a) Depth=2, Channel=8}
    \end{minipage}
    \begin{minipage}[t]{0.33\linewidth}
    \centering
    \vspace{1pt}
    \centerline{\includegraphics[height=4cm]{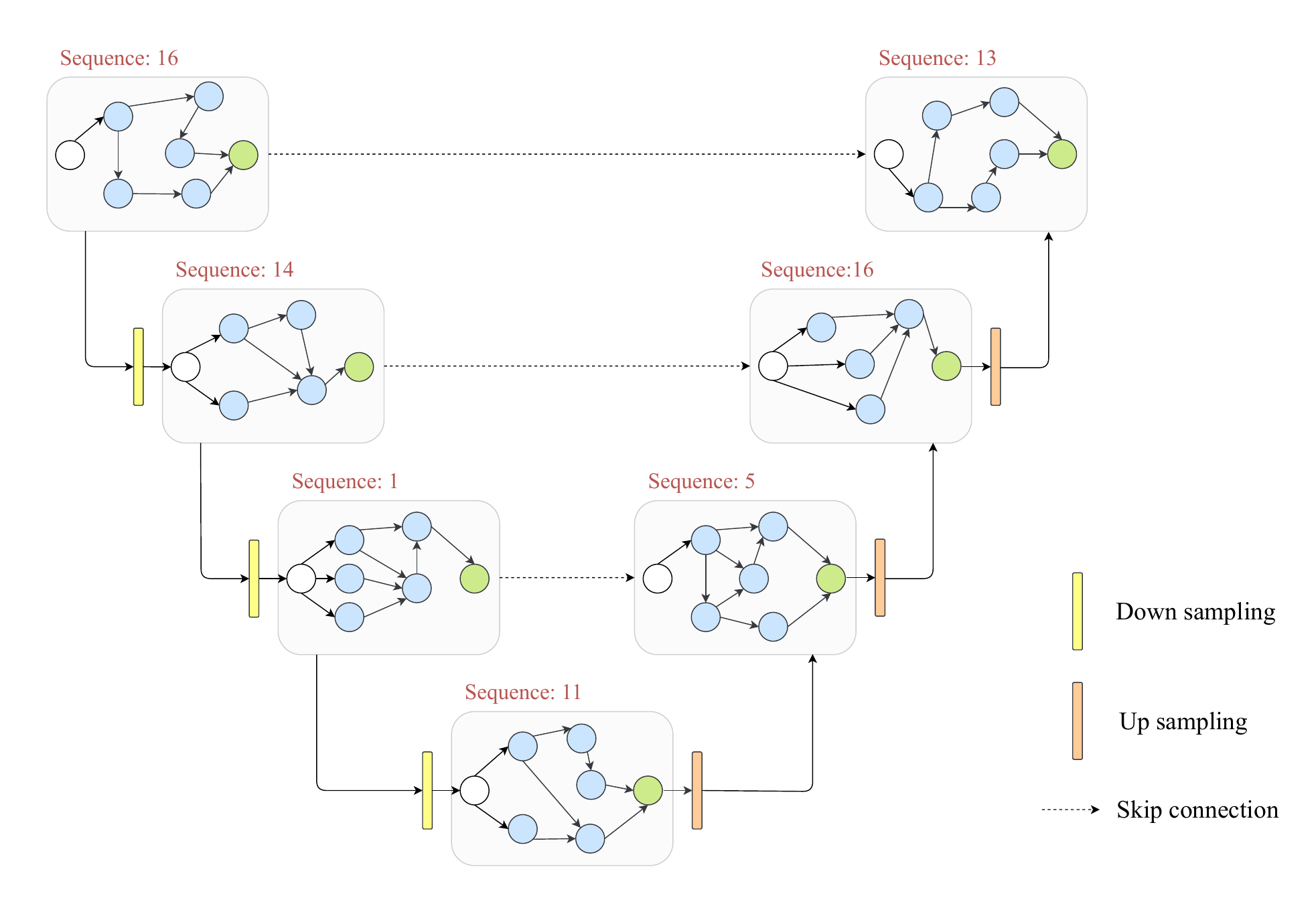}}  
    \vspace{-3pt}
    \centerline{\small (b) Depth=3, Channel=18}
    \end{minipage}
    \begin{minipage}[t]{0.33\linewidth}
    \centering
    \vspace{1pt}
    \centerline{\includegraphics[height=4cm]{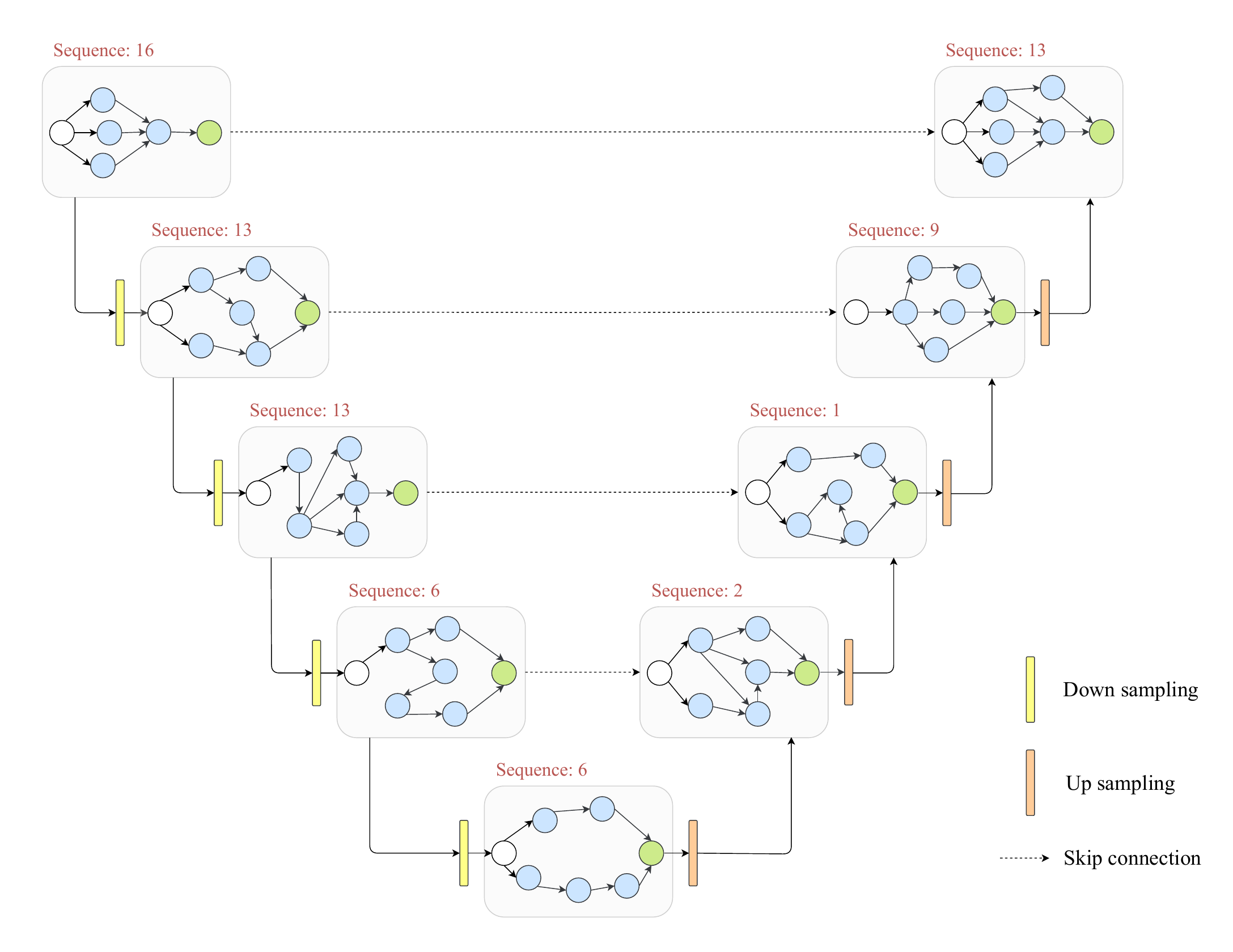}}  
    \vspace{-3pt}
    \centerline{\small (c) Depth=4, Channel=28}
    \end{minipage}
    \caption{Visualization of the architectures for different network depths obtained from the plug-and-play search. Among them, the neuron type is obtained after the architecture is determined}
    \label{fig:visd_decode}
\end{figure*}

\textbf{Visualizing Networks Obtained from Plug-and-Play search}.
Figure \ref{fig:visd_decode} provides three examples of visualized network architectures obtained through the plug-and-play search process. Each block represents a unique architecture characterized by its operation gene and connection gene, detailing the specific combination of convolution operations, activation types, and normalization techniques, as well as the flow of data between nodes. These elements are key in defining the architecture's structure and functionality. The neuron distribution of the plug-and-play search differs significantly from the joint search which is significantly biased towards quadratic neurons. At a depth of 2, the first two layers utilize conventional neurons, while subsequent layers introduce quadratic neurons, resulting in a relatively simple feature extraction process. At a depth of 3, the initial layers use quadratic neurons, but the proportion of conventional neurons increases in the later layers. At a depth of 4, although there is an alternating use of neuron types, the proportion of conventional neurons remains high.

\subsection{Neuron Programming via Hypernetwork}\label{sec4-6}

In the joint search of NAS and neuron programming, we conduct extensive experiments by training nearly 20,000 distinct configurations across various encoder and channel parameters on this retinal vessel segmentation task. This comprehensive dataset allows us to analyze performance characteristics and inform the optimal network designs in a dedicate manner. Here, we present the analysis and results demonstrating hypernetwork’s role in predicting and optimizing neural network architectures.

\textbf{The Training of Hypernetwork}.\label{sec4-6-2}
A core contribution of this work is the development of a hypernetwork to efficiently predict neural network configurations, focusing on key parameters like kernel size and neuron types. This approach has the advantage of reducing the computational burden associated with joint search, as well as serving as a plug-and-play use to the existing good architectures. To ensure the accuracy of the predictions, we employ a selective strategy, using only the top 10 well-performing networks from each parameter group to train a hypernetwork, which is then used to predict neurons with varying depths and channels. The inputs of the hyper-network are the network depth and channel size. 

\begin{figure*}[htbp]
			\begin{minipage}[t]{0.33\linewidth}
			\centering
			\vspace{1pt}
			\centerline{\includegraphics[width=1\linewidth]{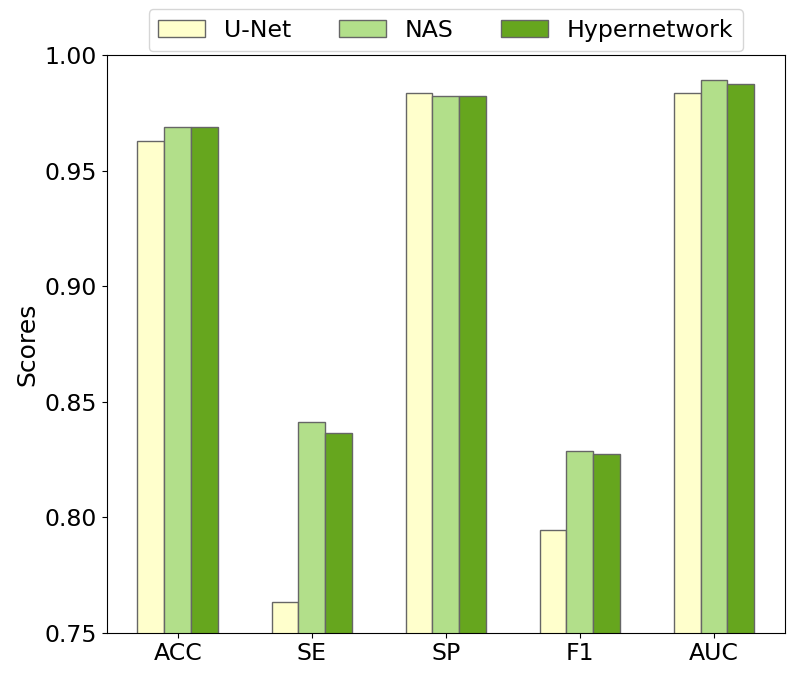}}
			\vspace{-3pt}
			\centerline{\small (a) Depth=2, Channel=8}
		\end{minipage}
			\begin{minipage}[t]{0.33\linewidth}
			\centering
			\vspace{1pt}
			\centerline{\includegraphics[width=1\linewidth]{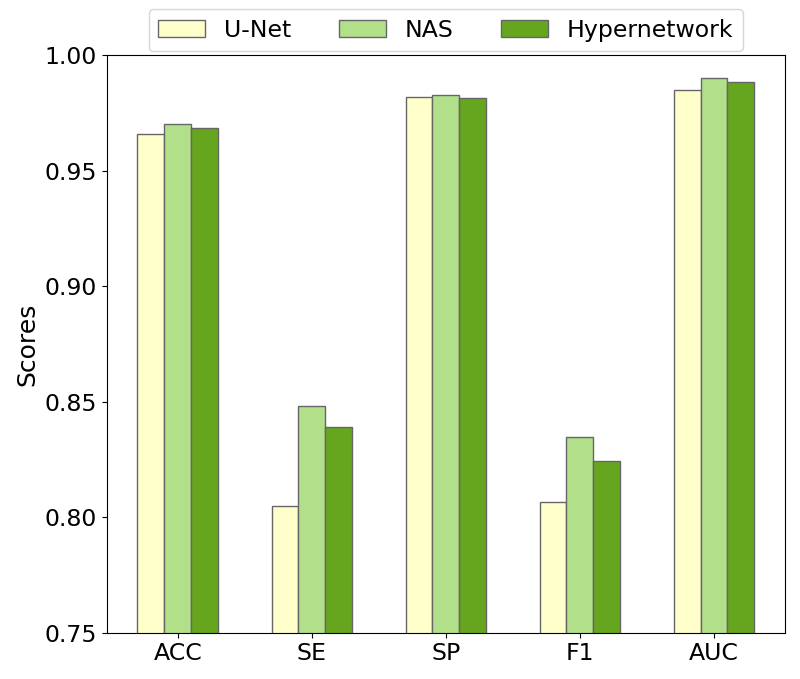}}
			\vspace{-3pt}
			\centerline{\small (b) Depth=3, Channel=18}
		\end{minipage}
			\begin{minipage}[t]{0.33\linewidth}
				\centering
				\vspace{1pt}
				\centerline{\includegraphics[width=1\linewidth]{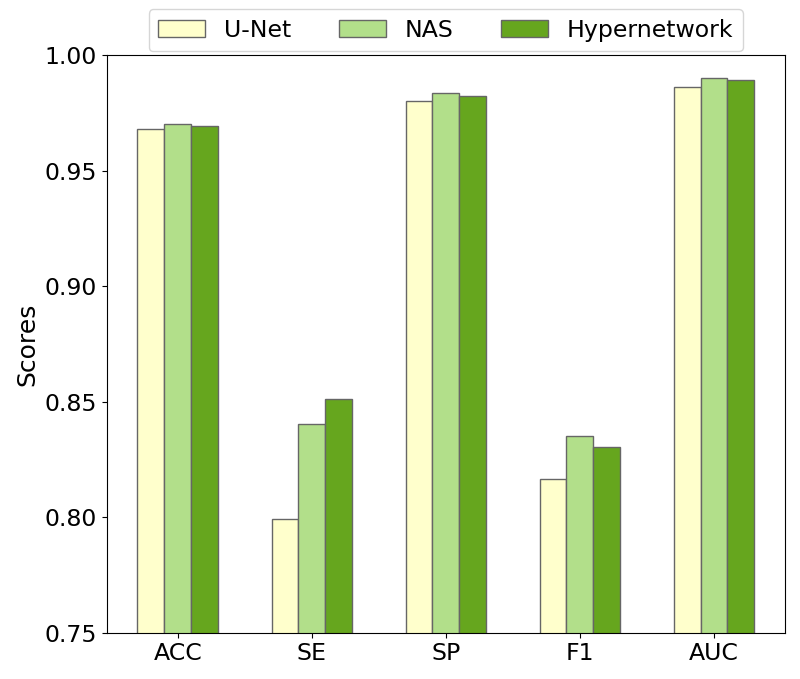}}
				\vspace{-3pt}
				\centerline{\small (c) Depth=4, Channel=28}
		\end{minipage}
		\caption{Comparative performance analysis of network architectures predicted by the hypernetwork}
  \label{img: visualization5}
\end{figure*}

After performing NAS using various combinations of network depth and channel parameters, we obtained approximately 20,000 different network architectures. These architectures feature a range of configurations, including distinct operation genes, which correspond to the different operation types defined in the search space, as well as various connection types. Each network configuration is associated with its corresponding performance metrics, allowing for a comparative evaluation. The training data for the hypernetwork is derived from the subset of these networks that demonstrate the best performance.

\begin{figure}[htbp]
    \centering
    \includegraphics[width=1\columnwidth]{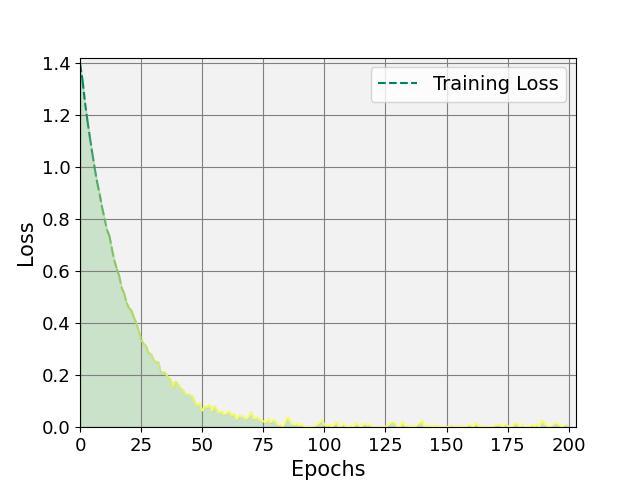}
    \vspace{-0.3cm}
    \caption{Training loss curve during the training process}
    \label{fig: training_loss}
    \end{figure}

The hypernetwork is trained using the Adam optimizer with a learning rate of 0.001 and categorical cross-entropy as the loss function. The training takes 200 epochs and utilizes a batch size of 4. Figure \ref{fig: training_loss} displays the training curves, showing that, as expected, the training loss decreases significantly during the early stages and then stabilizes as the model approaches convergence.

\textbf{The Effectiveness of Hypernetwork}. For a given parameter combination (e.g., encoder=3, channel=18), the hypernetwork predicts a customized network structure, such as 16-3-13-12-9-14-13, where these numbers denote specific parameter sequences within the network. We evaluate three situations (depth=2, channel=8), (depth=3, channel=18), and (depth=4, channel=28). The effectiveness of these predictions is validated by training them and testing the performance, which compares them with networks obtained through NAS methods and the original U-Net. Figure \ref{img: visualization5} compare results from the three sets of predictions. The results demonstrate that the hypernetwork-predicted network performs comparably to optimal networks derived through NAS and outperforms the original U-Net. Notably, the hypernetwork prediction process significantly reduces the computational time, thus lowering the overall computational costs. These findings highlight the hypernetwork's effectiveness in guiding network design.

\section{Ablation Studies}\label{sec4-5}

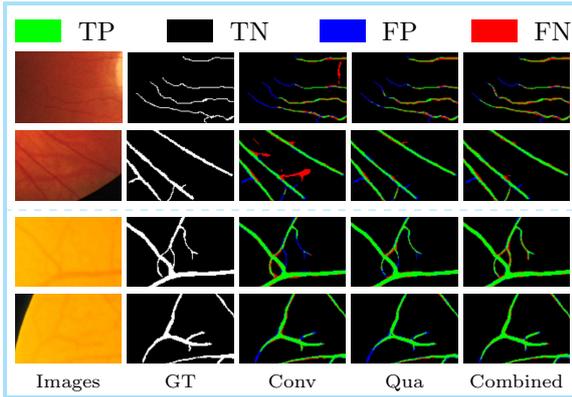
\begin{figure}[ht]
\begin{tcolorbox}[colback=white, colframe=cyan!30, colbacktitle=white, coltitle=black, boxrule=0.5mm, left=0.05mm, right=0.05mm, sharp corners]
  \vspace{-2mm}
    \begin{tikzpicture}
        \node[fill=green, minimum width=0.6cm, minimum height=0.3cm] at (0,0) {};
        \node[anchor=west] at (0.4,0) {TP};

        \node[fill=black, minimum width=0.6cm, minimum height=0.3cm] at (2,0) {};
        \node[anchor=west] at (2.4,0) {TN};

        \node[fill=blue, minimum width=0.6cm, minimum height=0.3cm] at (4,0) {};
        \node[anchor=west] at (4.4,0) {FP};

        \node[fill=red, minimum width=0.6cm, minimum height=0.3cm] at (6,0) {};
        \node[anchor=west] at (6.4,0) {FN};
    \end{tikzpicture}\\
  \begin{minipage}[t]{0.193\linewidth}
			\centering
			\centerline{\includegraphics[width=1\linewidth]{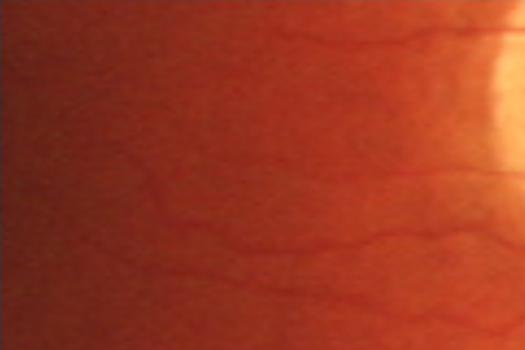}}
		\end{minipage}
			\begin{minipage}[t]{0.191\linewidth}
			\centering
			\centerline{\includegraphics[width=1\linewidth]{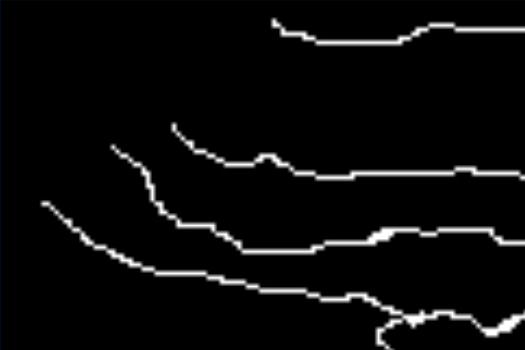}}
		\end{minipage}
			\begin{minipage}[t]{0.191\linewidth}
				\centerline{\includegraphics[width=1\linewidth]{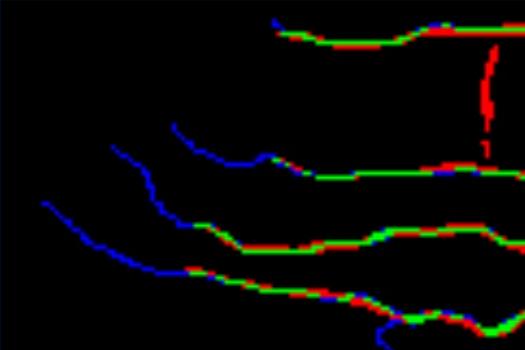}}
			\end{minipage}
				\begin{minipage}[t]{0.191\linewidth}
				\centering
				\centerline{\includegraphics[width=1\linewidth]{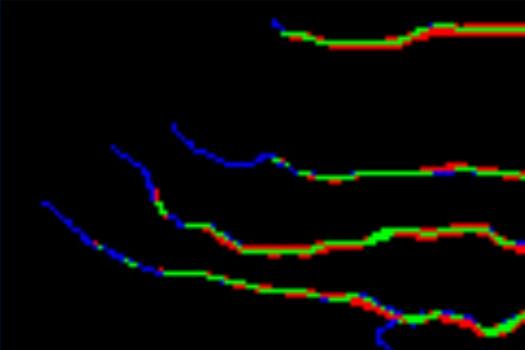}}
			\end{minipage}
                \begin{minipage}[t]{0.191\linewidth}
				\centering
				\centerline{\includegraphics[width=1\linewidth]{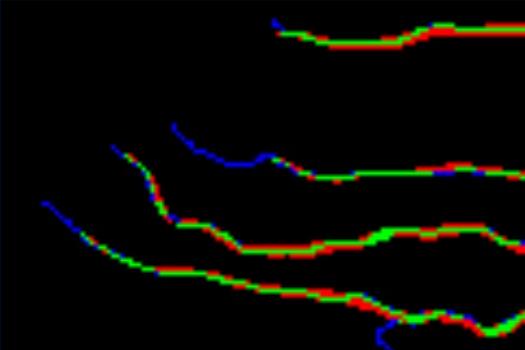}}
			\end{minipage}
   \vspace{-3.5mm} 
   \\ 
   \begin{minipage}[t]{0.191\linewidth}
			\centering
			\centerline{\includegraphics[width=1\linewidth]{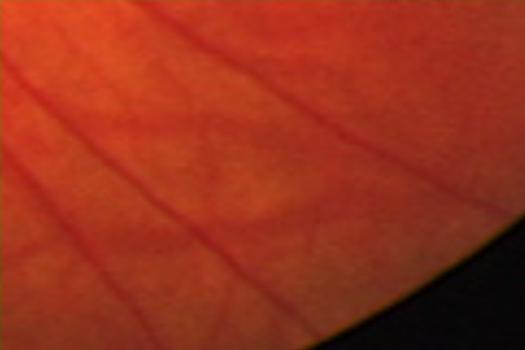}}
			\vspace{-3pt}
		\end{minipage}
			\begin{minipage}[t]{0.191\linewidth}
			\centering
			\centerline{\includegraphics[width=1\linewidth]{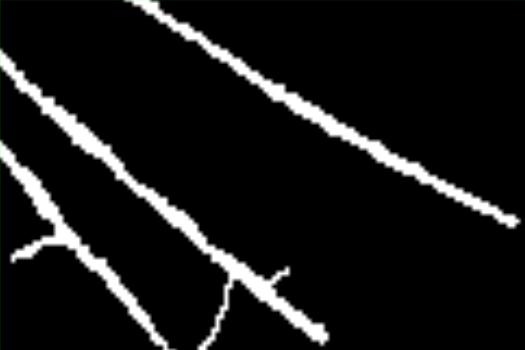}}
			\vspace{-3pt}
		\end{minipage}
			\begin{minipage}[t]{0.191\linewidth}
				\centerline{\includegraphics[width=1\linewidth]{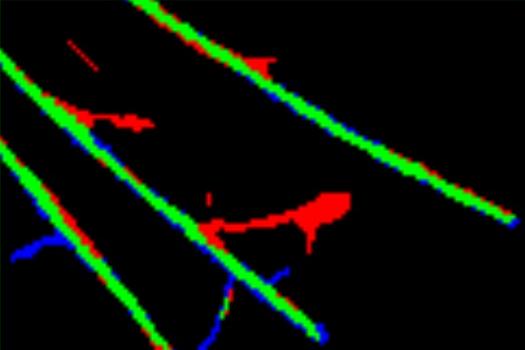}}
				\vspace{-3pt}
			\end{minipage}
				\begin{minipage}[t]{0.191\linewidth}
				\centering
				\centerline{\includegraphics[width=1\linewidth]{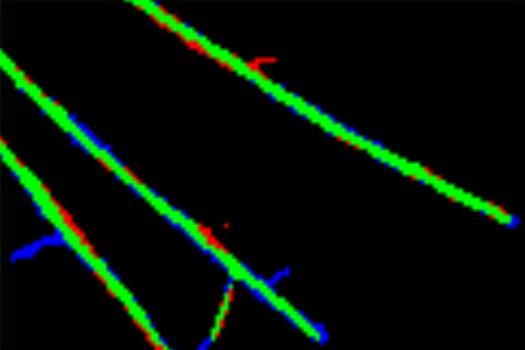}}
				\vspace{-3pt}
			\end{minipage}
                \begin{minipage}[t]{0.191\linewidth}
				\centering
				\centerline{\includegraphics[width=1\linewidth]{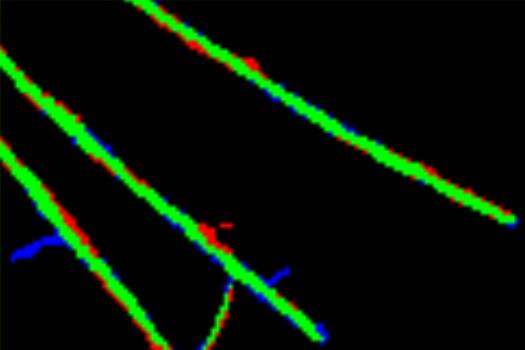}}
				\vspace{-3pt}
			\end{minipage} \\
  \vspace{-6mm}
  \tcblower 
  \vspace{-2mm}
  
  \begin{minipage}[t]{0.191\linewidth}
			\centering
			\centerline{\includegraphics[width=1\linewidth]{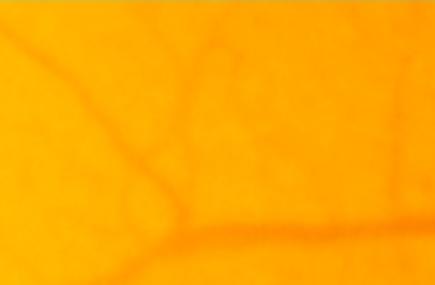}}
		\end{minipage}
			\begin{minipage}[t]{0.191\linewidth}
			\centering
			\centerline{\includegraphics[width=1\linewidth]{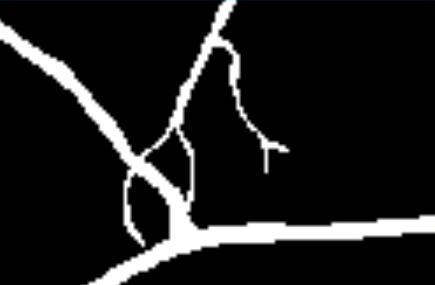}}
		\end{minipage}
			\begin{minipage}[t]{0.191\linewidth}
				\centerline{\includegraphics[width=1\linewidth]{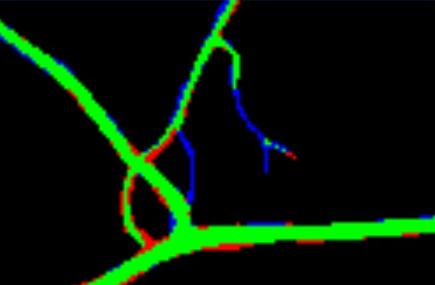}}
			\end{minipage}
				\begin{minipage}[t]{0.191\linewidth}
				\centering
				\centerline{\includegraphics[width=1\linewidth]{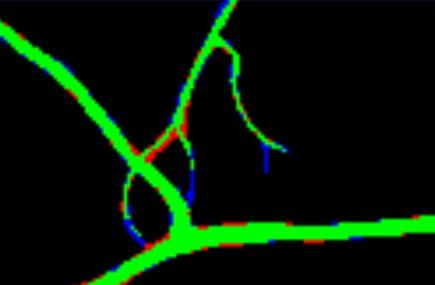}}
			\end{minipage}
                \begin{minipage}[t]{0.191\linewidth}
				\centering
				\centerline{\includegraphics[width=1\linewidth]{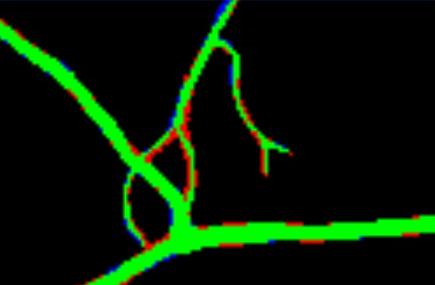}}
			\end{minipage}
   \vspace{-3.5mm} 
   \\ 
   \begin{minipage}[t]{0.191\linewidth}
			\centering
			\centerline{\includegraphics[width=1\linewidth]{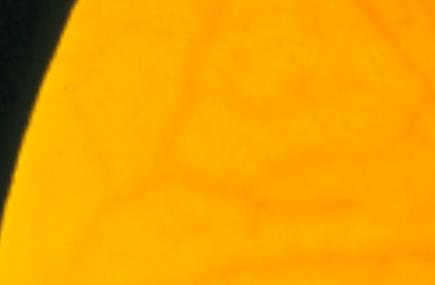}}
			\vspace{-3pt}
                \centerline{\footnotesize Images}
		\end{minipage}
			\begin{minipage}[t]{0.191\linewidth}
			\centering
			\centerline{\includegraphics[width=1\linewidth]{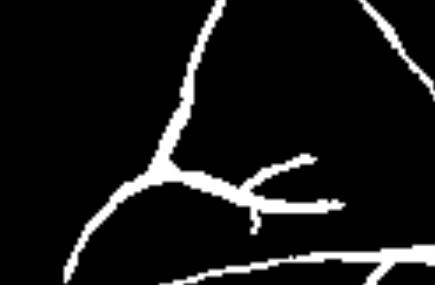}}
			\vspace{-3pt}
                \centerline{\footnotesize GT}
		\end{minipage}
			\begin{minipage}[t]{0.191\linewidth}
			\centerline{\includegraphics[width=1\linewidth]{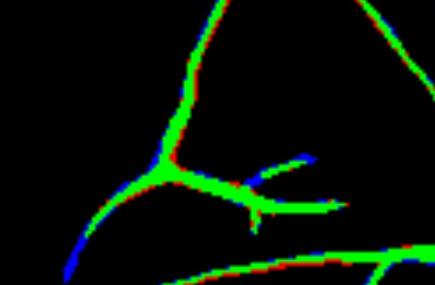}}
				\vspace{-3pt}
                \centerline{\footnotesize Conv}
			\end{minipage}
				\begin{minipage}[t]{0.191\linewidth}
				\centering
			\centerline{\includegraphics[width=1\linewidth]{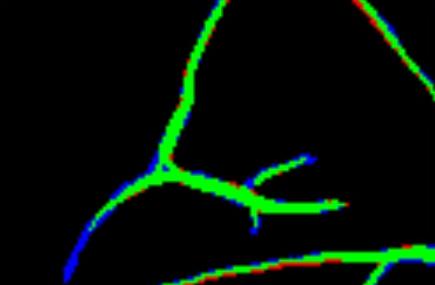}}
				\vspace{-3pt}
                \centerline{\footnotesize Qua}
			\end{minipage}
                \begin{minipage}[t]{0.191\linewidth}
				\centering
			\centerline{\includegraphics[width=1\linewidth]{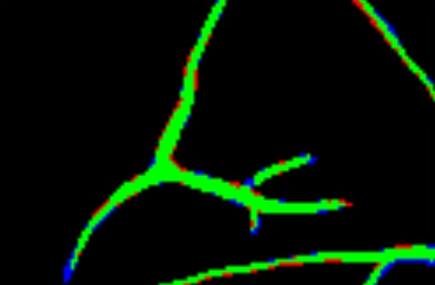}}
				\vspace{-3pt}
                    \centerline{\footnotesize Combined}
			\end{minipage}
    \vspace{-3mm}
\end{tcolorbox}
\caption{Segmentation results using different types of neurons in network architectures on the DRIVE (two rows above) and STARE (two rows below) datasets. From left to right: magnified view, ground truth, segmentation results with conventional neurons, segmentation results with quadratic neurons, segmentation results combining both neuron types}
\label{img: visualization3}
\end{figure}

\textbf{Effect of Neuronal Diversity.}\label{sec4-5-1}
As discussed in Section \ref{subsec2-2}, integrating multiple neurons can enhance the network's expressive capacity. To validate this hypothesis, we replace all neurons in the networks obtained by search with either conventional or quadratic neurons, and train them under identical conditions on the DRIVE dataset. This allows us to analyze how different neuron types affect the network performance. Table \ref{tab: result5} highlights the significant drop in network performance after replacement. The introduction of neuronal diversity enhances the network performance. Figure \ref{img: visualization3} visualizes FP pixels in blue and FN pixels in red. Ideally, a superior model should have fewer FN and FP pixels. The fifth column displays the results of combining both conventional and quadratic neurons, showing significantly fewer blue and red pixels, which reflects superior accuracy in the detailed segmentation of vascular and non-vascular classification.	

\begin{table}[htbp]
    \caption{Segmentation results of different neuron types on the DRIVE dataset. The best and the second best results are highlighted using bold and underline, respectively.}
    \setlength{\tabcolsep}{2pt}
    \begin{tabularx}{\linewidth}{lccccccc}
    \toprule
        \textbf{Neuron type} & \textbf{ACC} & \textbf{SE} & \textbf{SP} & \textbf{F1} & \textbf{AUC} \\ \midrule
        Conv & 0.9684 & 0.8266 & 0.9799 & 0.8265 & 0.9879  \\
        Qua & \underline{0.9687} & \underline{0.8458} & \underline{0.9817} & \underline{0.8300} & \underline{0.9892}  \\
        Conv+Qua & \textbf{0.9701} & \textbf{0.8489} & \textbf{0.9825} & \textbf{0.8334} & \textbf{0.9901}  \\
    \bottomrule
    \end{tabularx}
    \footnotetext{Conv refers to conventional neurons, while qua refers to quadratic neurons}
\label{tab: result5}
\end{table}

\textbf{Effect of Loss Function.}\label{sec4-5-2}
Choosing an appropriate loss function is pivotal for training neural networks in segmentation tasks, as it directly influences the network's training to learn and generalize from the provided training data. In this study, we analyze the segmentation results on the DRIVE dataset using three commonly used loss functions: Jaccard Loss, Dice Loss, and Focal Loss. The findings, shown in Table \ref{tab: result6}, provide a clear comparison of the segmentation outcomes using these different loss functions. Notably, the Focal Loss stands out as the top performer across most evaluation metrics except for SE, highlighting its efficacy in enhancing segmentation quality. By accentuating subtle features and addressing the class imbalance, we conclude that the Focal Loss is a suitable loss function for retinal vessel segmentation.

\begin{table}[htbp]
    \caption{Segmentation results of different loss functions on DRIVE dataset. The best and the second best results are bold-faced and underlined, respectively.}
    \setlength{\tabcolsep}{2pt}
    \begin{tabularx}{\linewidth}{lccccccc}
    \toprule
        \textbf{Loss Function} & \textbf{ACC} & \textbf{SE} & \textbf{SP} & \textbf{F1} & \textbf{AUC} \\ \midrule
        Jaccard Loss & \underline{0.9692} & 0.8393 & \underline{0.9823} & 0.8276 & \underline{0.9712}  \\
        Dice Loss & 0.9690 & \textbf{0.8511} & 0.9809 & \underline{0.8287} & 0.9648  \\
        Focal Loss & \textbf{0.9701} & \underline{0.8489} & \textbf{0.9825} & \textbf{0.8334} & \textbf{0.9901}  \\
    \bottomrule
    \end{tabularx}
\label{tab: result6}
\end{table}

\section{Conclusion}\label{sec5}
In this paper, complementary to NAS, we have introduced neuron programming to retinal vessel segmentation by integrating quadratic neurons into the U-Net architecture. The proposed neuron programming determines the best neuronal type from conventional and quadratic neurons, which fundamentally enhances the network's ability to capture the complex and nonlinear structures inherent in retinal blood vessels. To reduce the computational demand of neuron programming, we have developed a hypernetwork that predicts the optimal network architecture under various parameter configurations. The experimental results demonstrate that our proposed method is competitive compared to the existing state-of-the-art techniques in retinal vessel segmentation. This work advances the field of medical image analysis and opens new avenues for applying diverse neuron types and efficient architecture search methods in other complex image segmentation tasks. Future research could extend this approach to other medical imaging challenges for broader applicability.

\backmatter

\section*{Declarations}
\begin{itemize}
\item \textbf{Funding} The research was supported by ITF, Hong Kong, the Natural Science Foundation of China (Grant No. 61971234), the Scientific Research Foundation of NUPT (Grant No. NY223008) and the Postgraduate Research $\&$ Practice Innovation Program of Jiangsu Province (Grant No. SJCX23\_0252).
\item \textbf{Data Availability} DRIVE dataset is available at \url{https://drive.grand-challenge.org/}. CHASE\_DB1 dataset is available at \url{https://researchdata.kingston.ac.uk/96/}. STARE dataset is available at \url{https://cecas.clemson.edu/~ahoover/stare/}. HRF dataset is available at \url{https://www5.cs.fau.de/research/data/fundus-images/}.
\item \textbf{Competing interests} The authors have no relevant financial or non-financial interests to disclose.
\end{itemize}

\bibliography{sn-bibliography}

\end{document}